\documentclass[12pt]{article}
\usepackage{amsfonts,amssymb,latexsym}
\usepackage[dvips]{graphics}

\renewcommand{\thesection}{\arabic{section}}
\newcommand{\divider}{\rule{13cm}{0.2pt}}
\newcommand{\ie}{i.e.}
\newcommand{\eg}{e.g.}
\newcommand{\pin}{{\rm Pin}}
\newcommand{\spin}{{\rm Spin}}
\newcommand{\diag}{{\rm diag}}
\newcommand{\tr}{{\rm tr}}
\newcommand{\CPT}{$C\!PT$\ }
\newcommand{\sla}{\slash\!\!\!}
\newcommand{\phit}{{\mathop{\varphi}^{\triangle}}}
\newcommand{\phitinv}{{\mathop{\varphi}^{\triangle}}\,{}^{-1}}

\newcommand{\id}{1 \hspace{-1.1mm} {\rm I}}
\newcommand{\be}{\begin{equation}}
\newcommand{\ee}{\end{equation}}
\newcommand{\bea}{\begin{eqnarray*}}
\newcommand{\eea}{\end{eqnarray*}}
\newcommand{\ba}{\begin{array}}
\newcommand{\ea}{\end{array}}
\newcommand{\Rset}{{\mathbb R}}
\newcommand{\Cset}{{\mathbb C}}
\newcommand{\Hset}{{\mathbb H}}
\newcommand{\Zset}{{\mathbb Z}}
\newcommand{\half}{{\scriptstyle {1 \over 2}}}
\newcommand{\qed}{$\Box$}
\newenvironment{pf}{\ \\ \ \\ {\it Proof:} }{\ \\ }
\newenvironment{rem}{\ \\ \ \\ {\it Remark: }}{\ \\ }
\newenvironment{exmp}{\ \\ \ \\ {\it Example: }}{\ \\ }

\begin{document}

\thispagestyle{empty}
\ \\

\hfill IHES-P/00/42
\vspace{3.5cm}

\begin{center}
{\LARGE 
The Pin Groups in Physics: 
   $C$, $P$, and $T$ }\ \ \\ \ \\ \ \\
{\large Marcus Berg${}^{\rm a}$ ,
    C\'{e}cile DeWitt-Morette${}^{\rm a}$,\\[2mm]
   Shangjr Gwo${}^{\rm b}$  and
   Eric Kramer${}^{\rm c}$} \ \\ \ \\ \ \\
   ${}^{\rm a}$Department of Physics and Center for Relativity, \\
   University of Texas, Austin, TX 78712, USA \\[2mm]
   ${}^{\rm b}$Department of Physics, National Tsing Hua University, \\
   Hsinchu 30034, Taiwan \\[2mm]
   ${}^{\rm c}$290 Shelli Lane, Roswell, GA 30075 USA 
\end{center}

\newpage
\thispagestyle{empty}
\begin{center}
{\bf Abstract}
\end{center}
       A simple, but not widely known, mathematical fact concerning the
       coverings of the full Lorentz group sheds light on parity and time
       reversal transformations of fermions. Whereas there is, up to
       an isomorphism, only one \spin\ group which double covers the
       orientation preserving Lorentz
       group, there are two essentially different groups, called \pin\ groups,
       which cover the full Lorentz group. \pin(1,3) is to O(1,3)
       what \spin(1,3) is to SO(1,3).
       The existence of two \pin\ groups
       offers a classification of fermions based on their
       properties under space or time reversal finer than the classification
       based on their properties under orientation preserving Lorentz
       transformations --- provided one can design experiments that
       distinguish the two types of fermions.
       Many promising experimental
       setups give, for one reason or another, identical results for both
       types of fermions. These negative results are reported here because
       they are instructive. Two notable positive results show that the
       existence of two \pin\ groups is relevant to physics:
       \begin{itemize}
           \item
           In a neutrinoless double beta decay, the neutrino emitted and
           reabsorbed in the course of the interaction can only be described in
           terms of \pin(3,1).
           \item
           If a space is topologically nontrivial, the vacuum
           expectation values
           of Fermi currents defined on this space can be totally
           different when
           described in terms of \pin(1,3) and \pin(3,1).
       \end{itemize}
       Possibly more important than the two above predictions, the \pin\
       groups provide a simple framework for the study of fermions; it makes
       possible clear definitions of intrinsic parities and time reversal;
       it clarifies colloquial, but literally meaningless, statements.
       Given the difference between the \pin\ group and the \spin\ group
       it is useful to distinguish their representations, as groups
       of transformations on ``pinors'' and ``spinors'', respectively.
       \ \\ \ \\
       The \pin(1,3) and \pin(3,1) fermions are twin-like particles
       whose behaviors differ only under space or time reversal.
       \ \\ \ \\
       A section on \pin\ groups in arbitrary spacetime dimensions is included.

\setcounter{page}{0}
\newpage

\begin{center}
{\Large {\bf The Pin Groups in Physics: \\[2mm]
$C$, $P$, and $T$}}\\
\end{center}
\ \\

\noindent{\large {\bf Table of contents}}

\begin{itemize}
\item[0.]
Dictionary of Notation
\item[1.]
Introduction
\item[2.]
Background
\begin{itemize}
   \item[2.1] As seen by physicists
   \begin{itemize}
       \item[$\bullet$] Superselection rules
       \item[$\bullet$] Parity violation
       \item[$\bullet$] Standard Model
   \end{itemize}
   \item[2.2] As seen by Wigner
   \item[2.3] As seen by mathematicians
\end{itemize}
\item[3.] The Pin groups in 3 space, 1 time dimensions
       \begin{itemize}
           \item[3.1] The Pin groups
           \begin{itemize}
               \item[$\bullet$] Pin(1,3)
               \item[$\bullet$] Pin(3,1)
           \end{itemize}
           \item[3.2] A Spin group is a subgroup of a Pin group
           \item[3.3] Pin group and Spin group representations on
                     finite-dimensional spaces; classical fields.
           \begin{itemize}
               \item[$\bullet$] Pinors
               \item[$\bullet$] Spinors
               \item[$\bullet$] Helicity
               \item[$\bullet$] Massless spinors, massive pinors
               \item[$\bullet$] Copinors, Dirac and Majorana adjoints
               \item[$\bullet$] Charge conjugate pinors in Pin(1,3)
               \item[$\bullet$] Majorana pinors
               \item[$\bullet$] Unitary and antiunitary transformations
               \item[$\bullet$] Invariance of the Dirac equation under
                                 antiunitary transformations
               \item[$\bullet$] $C\!PT$ invariance
               \item[$\bullet$] Charge conjugate pinors in Pin(3,1)
           \end{itemize}
           \item[3.4]   Pin group and Spin group representations on
                          infinite-dimensional spaces; quantum fields.
           \begin{itemize}
               \item[$\bullet$] Particles, antiparticles
               \item[$\bullet$] Fock space operators, unitary and antiunitary
               \item[$\bullet$] Intrinsic parity
               \item[$\bullet$] Majorana field operator
               \item[$\bullet$] Majorana classical field vs.
                                Majorana quantum field
               \item[$\bullet$] $C\!PT$ transformations
           \end{itemize}
           \item[3.5] Bundles; Fermi fields on manifolds
           \begin{itemize}
               \item[$\bullet$] Pinor coordinates
               \item[$\bullet$] Dirac adjoint in \pin(1,3)
               \item[$\bullet$] Dirac adjoint in \pin(3,1)
               \item[$\bullet$] Pin structures
               \item[$\bullet$] Fermi fields on topologically nontrivial
                                manifolds
           \end{itemize}
           \item[3.6] Bundle reduction
     \end{itemize}
 \item[4.] Search For Observable Differences
   \begin{itemize}
       \item[4.1] Computing observables with Pin(1,3) and Pin(3,1)
       \item[4.2] Parity and the Particle Data Group publications
           \begin{itemize}
               \item[$\bullet$] Parity conservation
           \end{itemize}
       \item[4.3] Determining parity experimentally
             \begin{itemize}
                 \item[$\bullet$] Selection rules: Pion decay
                 \item[$\bullet$] Selection rules: Three-fermion decay
                 \item[$\bullet$] Selection rules: Positronium
                 \item[$\bullet$] Decay rates; cross sections
           \end{itemize}
       \item[4.5] Interference, reversing magnetic fields, reflection
       \item[4.5] Time reversal and Kramer's degeneracy
       \item[4.6] Charge conjugation
           \begin{itemize}
               \item[$\bullet$] Positronium
               \item[$\bullet$] Neutrinoless double beta decay
           \end{itemize}
   \end{itemize}
\item[5.] The Pin groups in $s$ space, $t$ time dimensions
         \begin{itemize}
               \item[5.1] The difference between $s+t$ even and $s+t$ odd
           \begin{itemize}
               \item[$\bullet$] The twisted map
           \end{itemize}
               \item[5.2] Chirality
               \item[5.3] Construction of gamma matrices;
                                         periodicity modulo 8
           \begin{itemize}
               \item[$\bullet$] Onsager construction of gamma matrices
               \item[$\bullet$] Majorana pinors, Weyl-Majorana spinors
           \end{itemize}
               \item[5.4] Conjugate and complex gamma matrices
               \item[5.5] The short exact sequence $\id \rightarrow$ Spin$(t,s)
                         \rightarrow $ Pin$(t,s) \rightarrow {\mathbb Z}_2 \rightarrow 0$
               \item[5.6] Grassman (superclassical)
                                         pinor fields
               \item[5.7] String theory and pin structures
   \end{itemize}
\item[6.] Conclusion
           \begin{itemize}
               \item[6.1] Some facts
               \item[6.2] A tutorial
           \begin{itemize}
              \item[$\bullet$] Parity
               \item[$\bullet$] Time reversal
               \item[$\bullet$]  Charge conjugation
               \item[$\bullet$] Wigner's classification
                                and classification by \pin\ groups
              \item[$\bullet$] Fock space and one-particle states
           \end{itemize}
               \item[6.3] Avenues to explore
           \end{itemize}
\end{itemize}
\newpage
\noindent $\!\!$ Appendices
\begin{itemize}
   \item[A] Induced transformations
   \item[B] The isomorphism of $M_4({\mathbb R})$
                     and ${\mathbb H} \otimes {\mathbb H}$
   \item[C] Other double covers of the Lorentz group
   \item[D] Collected calculations
   \item[E] Collected references
\end{itemize}

\newpage

\setcounter{section}{-1}
\section{Dictionary of Notation}
\label{sec:notation}
The article proper begins with section 1.
\ \\ \ \\
We have occasionally changed some of
our earlier notations to conform to the majority of users.
As much as possible, we have tried to use the usual notation -- but
introducing different symbols for different objects when it is
essential to distinguish them.   For
example, we distinguish the Spin group and the two Pin groups
(sometimes still known as Spin groups) but we speak globally of spin
1/2 particles (lower case ``s'') for all particles, whether they are
represented by spinors or pinors of either type.
This agrees with intuitive notion of spin
as ``the behavior of a field or a state under rotations'' \cite{Polchinski}.
\ \\ \ \\
Our primary references are Peskin and Schroeder \cite{Peskin},
Weinberg \cite{Wein} and Choquet-Bruhat et al. \cite{Yellow,Blue}.
\ \\ \ \\
\divider
\ \\ \ \\
{\em Groups}\\
\begin{itemize}
   \item[]
   {\bf Lorentz group}
   \begin{itemize}
       \item[]
       \begin{itemize}
           \item[]
               O(1,3) leaves $\eta_{\alpha \beta} x^{\alpha} x^{\beta}$
               invariant, $\eta_{\alpha \beta}= \diag(1,-1,-1,-1)$ \\
               \mbox{with} $\alpha, \beta \in \{0,1,2,3\}$, $x^0=t$
           \item[]
               O(3,1) leaves $\hat{\eta}_{\alpha \beta} x^{\alpha} x^{\beta}$
               invariant, $\hat{\eta}_{\alpha \beta}= \diag(1,1,1,-1)$ \\
               \mbox{with} $\alpha, \beta \in \{1,2,3,4\}$, $x^4=t$
       \end{itemize}
       \item[]
       Examples: \\
       \begin{tabular}{l|ll}
           Reverse & $({L^{\alpha}}_{\beta}) \in \mathrm{O}(1,3)$ &
           $(\hat{L}^{\alpha}{}_{\beta}) \in \mathrm{O}(3,1)$   \\ \hline
           1 space axis & $P(1)=\diag(1,1,1,-1)$ &
           $\hat{P}(1)=\diag(-1,1,1,1)$ \\
           3 space axes & $P(3)=\diag(1,-1,-1,-1)$ &
           $\hat{P}(3)=\diag(-1,-1,-1,1)$ \\
           time axis & $T=\diag(-1,1,1,1)$ &
           $\hat{T}=\diag(1,1,1,-1)$
       \end{tabular}
   \end{itemize}
   \[
   {L^{\alpha}}_{\beta}
   {(L^{-1})^{\beta}}_{\gamma}=\delta^{\alpha}_{\beta} \; ,
   \qquad LL^T=\id \mbox{ hence }
   {(L^{-1})^{\beta}}_{\gamma} = {L_{\gamma}}^{\beta} \; .
   \]
\newpage
   \item[]
   {\bf Real Clifford algebra} 
   \begin{itemize}
       \item[]
       The Clifford algebra is a graded algebra
       ${\mathcal C}={\mathcal C}_+ \oplus {\mathcal C}_- \; .$\\
       ${\mathcal C}_+$ is generated by even products of
       $\gamma_{\alpha}$'s, $\Lambda^{\rm even} \in {\mathcal C}_+$   \\
       ${\mathcal C}_-$ is generated by odd products of
       $\gamma_{\alpha}$'s, $\Lambda^{\rm odd} \in {\mathcal C}_-$.

       \item[]
       We choose\footnote{Another choice is
       $\{\gamma_{\alpha}, \gamma_{\beta}\}=-2\eta_{\alpha \beta}$,
       $\{\hat{\gamma}_{\alpha}, \hat{\gamma}_{\beta}\}=
       -2\hat{\eta}_{\alpha \beta}$. Our choice implies that the norm
       $N(v^{\alpha} \gamma_{\alpha})=\eta(v,v)=:||v||^2_{s,t}$.
       With the other choice
       $N(\Lambda)=\Lambda(\alpha(\Lambda))^{\tau}$, where
       $\alpha(\Lambda^{\rm even})=\Lambda^{\rm even}$,
       $\alpha(\Lambda^{\rm odd})=-\Lambda^{\rm odd}$.}
       \ \\
       $\{\gamma_{\alpha}, \gamma_{\beta}\}=2\eta_{\alpha \beta} \qquad
       \{\hat{\gamma}_{\alpha}, \hat{\gamma}_{\beta}\}=
       2\hat{\eta}_{\alpha \beta}$\\

       \item[]
       We could also fix the signature of the metric
       and have $\{\hat{\gamma}_{\alpha}
       \hat{\gamma}_{\beta}\}=-2\eta_{\alpha \beta}$,
       but we prefer to associate $\hat{\gamma}_{\alpha}$
       with $\hat{\eta}_{\alpha \beta}$ rather than with~
       $-\eta_{\alpha\beta}$.\\
  \end{itemize}
   \item[]
   {\bf Pin groups}
   \begin{itemize}
       \item[]
       \begin{tabular}{ccccc}
           $\Lambda_L \in \pin{}(1,3)$ & $\Leftrightarrow$ &
           $\Lambda_L \gamma_{\alpha} \Lambda_L^{-1} =
           \gamma_{\beta} {L^{\beta}}_{\alpha}$ & or &
           $\Lambda_L \gamma^{\alpha} \Lambda_L^{-1} =
           {(L^{-1})^{\alpha}}_{\beta} \gamma^{\beta}$ \\

           $\hat{\Lambda}_L \in \pin{}(3,1)$ & $\Leftrightarrow$ &
           $\hat{\Lambda}_L \hat{\gamma}_{\alpha} \hat{\Lambda}_L^{-1} =
           \hat{\gamma}_{\beta} { \left. \hat{L}^{\beta}
           \right.}_{\alpha}$ & or &
           $\hat{\Lambda}_L \hat{\gamma}^{\alpha} \hat{\Lambda}_L^{-1} =
           {(\hat{L}^{-1})^{\alpha}}_{\beta} \hat{\gamma}^{\beta}$
       \end{tabular} \\
       \item[]
       Examples:
       \[
       \begin{array}{lclclcl}
       \Lambda_{P(1)}&=& \pm \gamma_0 \gamma_1 \gamma_2 & \qquad &
       \hat{\Lambda}_{P(1)}&=& \pm \hat{\gamma}_2 \hat{\gamma}_3
       \hat{\gamma}_4 \\
       \Lambda_{P(3)}&=& \pm \gamma_0   & &
       \hat{\Lambda}_{P(3)}&=& \pm \hat{\gamma}_4 \\
       \Lambda_{T}&=& \pm \gamma_1 \gamma_2 \gamma_3 & &
       \hat{\Lambda}_{T}&=& \pm \hat{\gamma}_1 \hat{\gamma}_2
       \hat{\gamma}_3   \\
       \Lambda_{P(1)}^2&=&-\id & &
       \hat{\Lambda}_{P(1)}^2&=&+\id   \\
         \Lambda_{P(3)}^2&=&+\id   & &
         \hat{\Lambda}_{P(3)}^2&=&-\id \\
       \Lambda_{T}^2 &=& +\id &   &
       \hat{\Lambda}_{T}^2 &=&-\id
       \end{array}
       \]
   \end{itemize}
   \item[]
   {\bf Spin group} 
   \begin{itemize}
       \item[]
       $\spin(1,3) \subset \pin(1,3) \qquad
       \spin(3,1) \subset \pin(3,1)$ \\
       \item[]
       A \spin{} group consists of elements $\Lambda_L$ for $L$ such that
       $\det({L^{\beta}}_{\alpha})=1$. It consists of even elements
       (even products of $\gamma_{\alpha}$) of a \pin{} group. \\
       \item[]
       \[
           \pin(m,n) = \spin(m,n) \ltimes \Zset_2 \qquad
           \mbox{for $m+n>1$}
       \]
        ($\ltimes$ is a semidirect
           product, defined in sec.\ \ref{sec:sequence}).
   \end{itemize}
\end{itemize}
\newpage
\noindent
{\em Group representations, unitary and antiunitary}
\begin{itemize}
   \item[]
   {\bf On finite-dimensional vector spaces, real or complex:}
   \begin{itemize}
       \item[]
       $\Gamma_{\alpha}$ is a real or complex matrix representation
       of $\gamma_{\alpha}$.
       \item[]
       $\hat{\Gamma}_{\alpha}$ is a real or complex matrix representation
       of $\hat{\gamma}_{\alpha}$,
       \[
           \hat{\Gamma}_{\alpha}=i\Gamma_{\alpha} \; .
       \]
   \item[]
   Chiral representation
   \[
   \begin{array}{rclrcl}
       \Gamma_0 &=&
       \left(
       \begin{array}{cccc}
         \; 0 \; & \; 0 \; & \; 1 \; & \; 0 \; \\
         0 & 0 & 0 & 1 \\
         1 & 0 & 0 & 0 \\
         0 & 1 & 0 & 0
         \end{array}
         \right)   \; , \qquad &
       \Gamma_1 &=&
       \left(
       \begin{array}{cccc}
         0 & 0 & \; 0 \; & \; 1 \; \\ 
         0 & 0 & 1 & 0 \\
         0 & -1 & 0 & 0 \\
         -1 & 0 & 0 & 0
         \end{array}
         \right)   \\[10mm]
       \Gamma_2 &=&
       \left(
       \begin{array}{cccc}
         0 & \; 0 \; & \; 0 \; & -i \\ 
         0 & 0 & i & 0 \\
         0 & i & 0 & 0 \\
         -i & 0 & 0 & 0
         \end{array}
         \right)   \; , \qquad &
       \Gamma_3 &=&
       \left(
       \begin{array}{cccc}
         0 & \; 0 \; & \; 1 \; & 0 \\ 
         0 & 0 & 0 & -1 \\
         -1 & 0 & 0 & 0 \\
         0 & 1 & 0 & 0
         \end{array}
         \right)
   \end{array}
   \]
   \[
       \Gamma_5 = i\Gamma_0\Gamma_1\Gamma_2\Gamma_3 =
       \left(
       \begin{array}{cccc}
           -1 & 0 & \; 0 \; & \; 0 \; \\ 
         0 & -1 & 0 & 0 \\
         0 & 0 & 1 & 0 \\
         0 & 0 & 0 & 1
         \end{array}
         \right) \; .
   \]
   \item[]
   Dirac representation, exchange $\Gamma_0$ above for
   \[
   \Gamma_0 =
       \left(
       \begin{array}{cccc}
         \;1 \; & \; 0 \; & 0 & 0 \\ 
         0 & 1 & 0 & 0 \\
         0 & 0 & -1 & 0 \\
         0 & 0 & 0 & -1
         \end{array}
         \right) \; .
   \]
   $\Gamma_1$, $\Gamma_2$, $\Gamma_3$ are the same as in the
   chiral representation.
   \ \\
   \item[]
   A Majorana (real) representation in \pin(3,1):
   \[
   \begin{array}{rclrcl}
       \hat{\Gamma}_1 &=&
       \left(
       \begin{array}{cccc}
         0 & \; 0 \; & \; 0 \; & -1 \\ 
         0 & 0 & 1 & 0 \\
         0 & 1 & 0 & 0 \\
         -1 & 0 & 0 & 0
         \end{array}
         \right)   \; , \qquad &
       \hat{\Gamma}_2 &=&
       \left(
       \begin{array}{cccc}
         \; 1 \; & \; 0 \; & 0 & 0 \\ 
         0 & 1 & 0 & 0 \\
         0 & 0 & -1 & 0 \\
         0 & 0 & 0 & -1
         \end{array}
         \right)   \\[10mm]
       \hat{\Gamma}_3 &=&
       \left(
       \begin{array}{cccc}
         0 & 0 & -1 & 0 \\ 
         0 & 0 & 0 & -1 \\
         -1 & 0 & 0 & 0 \\
         0 & -1 & 0 & 0
         \end{array}
         \right)   \; , \qquad &
       \hat{\Gamma}_4 &=&
       \left(
       \begin{array}{cccc}
         0 & 0 & \; 0 \; & \; 1 \; \\ 
         0 & 0 & 1 & 0 \\
         0 & -1 & 0 & 0 \\
         -1 & 0 & 0 & 0
         \end{array}
         \right)
   \end{array}
   \]
   \item[]
   \label{page:pauli}
   Pauli matrices
   \[
       \sigma_1 =
       \left( \begin{array}{cc} \; 0 \; & \; 1 \; \\
           1 & 0 \end{array} \right)
       \; , \quad
     \sigma_2 =
         \left( \begin{array}{cc} \; 0 \; & -i \\
             i & 0 \end{array} \right)
       \; , \quad
     \sigma_3 =
         \left( \begin{array}{cc} \; 1 \; & 0 \\   
         0 & -1 \end{array} \right)
       \; , \quad
   \]
   \item[]
   Dirac equation for a massive, charged particle
       \begin{eqnarray*}
           && (i\Gamma^{\alpha}\nabla_{\alpha}-m)\psi(x)=0 \qquad \qquad
           (\hat{\Gamma}^{\alpha}\nabla_{\alpha}-m)\hat{\psi}(x)=0 \\
           && \mbox{ with }\nabla_{\alpha}=\partial_{\alpha}+iqA_{\alpha}
       \end{eqnarray*}
   \item[]
   Dirac adjoint (see \ref{sec:bundles} for Dirac adjoints on
                  general manifolds)
     \[
       \bar{\psi}=\psi^{\dagger} \, \Gamma_0
         \; , \qquad
     \bar{\hat{\psi}} =  \hat{\psi}^{\dagger}\hat{\Gamma}_4 \; .
   \]
   \item[]
   Charge conjugate
   \[
       (i\Gamma^{\alpha}(\partial_{\alpha}-
       iq A_{\alpha})-m)\psi^{\rm c}(x)=0 \; , \qquad
       (\hat{\Gamma}^{\alpha} (\partial_{\alpha}-
       iqA_{\alpha})-m)\hat{\psi}^{\rm c}(x)=0
   \]
   and $\psi^{\rm c}={\mathcal C}\psi^*$,
   $\hat{\psi}^{\rm c}=\hat{\mathcal C}\hat{\psi}^*$, where
   \[
   {\mathcal C}\,   \Gamma_{\alpha}^*
       {\mathcal C}^{-1}=-\Gamma_{\alpha} \; , \qquad
     \hat{\mathcal C}\,   \hat{\Gamma}_{\alpha}^* \hat{\mathcal C}^{-1}=
     \hat{\Gamma}_{\alpha} \; .
     \]
   In the Dirac representation ${\mathcal
   C}=\pm \Gamma_2$, $\hat{\mathcal
   C}=\pm \hat{\Gamma}_2$   and ${\mathcal C}{\mathcal C}^*=\id$,
   $\hat{\mathcal C}\hat{\mathcal C}^*=\id$.
   \end{itemize}
   \ \\
\newpage
   \item[]
   {\bf Two-component fermions, Weyl fermions}
   \[
       {\mathcal P}_{\pm}=\half(\id \pm \Gamma_5)
   \]
   \bea
       {\mathcal P}_+ \psi &=& \varphi_R \qquad \mbox{chirality $+1$} \\
       {\mathcal P}_- \psi &=& \varphi_L \qquad \mbox{chirality $-1$}
     \eea
   \item[]
       The antiunitary time reversal
       operator ${\mathcal A}_T$ on $\psi$
       acts on the complex conjugate $\psi^*$, 
       \[
          {\mathcal A}_T {\mathcal C}^{-1} = \Lambda_T \; .
       \]
       \ \\
   \newpage
   \item[]
   {\bf On infinite-dimensional Hilbert spaces of state vectors:}
   \begin{itemize}
       \item[]
       Quantum fields (see \ref{sec:pinrepq})
       \bea
       \Psi(x)&=&{1 \over \sqrt{\Omega}} \sum_{p, s}\left(
       a({\bf p},s) \psi(p,s)+
       b^{\dagger}({\bf p},s) \psi^{\rm c}(p,s)\right)
       \\
         \Psi^{\rm c}(x)&=&{\xi \over \sqrt{\Omega}} \sum_{p, s}\left(
         b({\bf p},s) \psi(p,s)+
         a^{\dagger}({\bf p},s) \psi^{\rm c}(p,s) \right)
         \; .
         \eea
       where $\sum_{p,s}$ combines
       $\sum_s$ and $\int d^4 p \, \delta(p^2-m^2)$, and $\xi$ is a
       phase, $|\xi|=1$.
       \ \\
       \item[]
       States
       \begin{itemize}
       \item[]
           States are created by applying creation and annihilation
           operators on the vacuum state, e.g.
           \[
             \left\{ 
               \begin{array}{l}
               \Psi(x)|0 \rangle = {1 \over \sqrt{\Omega}}
                 \sum_{p,s} b^{\dagger}({\bf p},s) \psi^{\rm
                     c}(p,s)|0\rangle \\
               \langle b | \Psi(x) |0 \rangle = \psi^{\rm c}(p,s)
             \end{array}  \right.
           \]
       $\psi$, $\psi^{\rm c}$ are called classical fields even when they
       are fermionic; the space of classical fields is the domain of the
       classical action (the classical action is not to be confused with
       its minimum value). \\
       \item[]
       $\Psi(x)|0\rangle$ is a linear superposition of one-antiparticle
       states of well-defined momentum $p$ and spin polarization $s$.
   \end{itemize}
\end{itemize}
\end{itemize}

\newpage

\section{Introduction}
A simple, but not widely known, mathematical fact concerning the
coverings of the full Lorentz group sheds light on parity and time
reversal transformations of fermions. Whereas there is, up to
an isomorphism, only one \spin\ group which double covers the
orientation preserving Lorentz
group, there are two essentially different groups, called \pin\ groups,
which cover the full Lorentz group. The name \pin\ is gaining
acceptance because it is a useful name: \pin(1,3) is to O(1,3)
what \spin(1,3) is to SO(1,3). The existence of two \pin\ groups
explains several issues which we discuss in section
\ref{sec:summary}. It offers a classification of fermions based on their
properties under space or time reversal finer than the classification
based on their properties under orientation preserving Lorentz
transformations.
\ \\ \ \\
For the convenience of the reader, we have divided this report into
two parts: in the first part we present the \pin\ groups for three
space dimensions and one time dimension; in the second part we present
the \pin\ groups for $s$ space dimensions and $t$ time dimensions
(with emphasis on the hyperbolic case $t=1$).
In appendix \ref{app:ref} we have collected, by
topic, references of related articles which we have consulted.

\section{Background}
\label{sec:summary}
\subsection{As seen by physicists}
\label{sec:physicist}
Racah, in 1937 \cite{Racah}, and Yang and Tiomno, in 1950
\cite{YangTiomno}, pointed out that
under a space inversion four different transformations for fields of
spin 1/2 are possible.   Yang and Tiomno added, ``The types of
transformation properties to which the various known spin-1/2 fields belong
are
physical observables and could in principle be determined experimentally
from their mutual interactions and their interactions with fields of integral
spin.''. They wrote down a list of all possible spinor interactions
using the four types of spinors, and attempted to exclude some interactions
based on the guiding principle of parity conservation.
Fermi even scheduled a special session at a conference he organized in
Chicago (September 1951) devoted to these ideas and to the
experimental distinction between the different kinds of spinors
\cite{Wightman}.
\ \\ \ \\
Under the impact of superselection rules \cite{Wick},
the discovery of parity violation \cite{LeeYang,Wu}, and the
success of the Standard Model \cite{Salam1,Salam2,Wein67},
the Yang and Tiomno paper fell by the wayside;
its goal was rendered obsolete.
{\it Nevertheless the fact remains that there are four
different kinds of spin-1/2 particles.}
Why has this fact, noted already in 1937, been largely ignored?
\ \\ \ \\
\noindent{\em a) The impact of superselection rules
in the 1960s\/}\\
In 1952, Wick, Wightman, and Wigner discussed limitations
of the concept of intrinsic parity of elementary particles,
in a paper affectionately known as W${}^3$ \cite{Wick}.
These limitations follow
from super\-selection rules --- rules which
restrict the nature and scope
of possible measurements.
More precisely, there is a superselection rule if the following
conditions are satisfied: {\it i)}
there is an exact conservation law, and {\it ii)} 
the Hilbert space can be decomposed into orthogonal subspaces $\{{\mathcal
H}_1, {\mathcal H}_2,\ldots \} $
such that there are no observables which
contain matrix elements between any pair of those subspaces, thus
making the relative phases between components of the state vector in
different subspaces ${\mathcal H}_{i}$ and ${\mathcal H}_{j} (i \ne j)$
irrelevant. 
\ \\ \ \\
In particular, these restrictions cast a shadow on the
possibility of identifying Dirac fields by their transformation properties
under space inversion.   But it has been shown \cite{DeWittDeWitt}
that the four choices of \pin\ structure {\em are\/}
observable---admittedly in an exotic setting, but even such an
example 
is sufficient to rule out a \pin\ superselection rule operating in general.
Moreover, as pointed out by Weinberg
\cite{Wein}
``the issue
of superselection rules is a bit of a red herring; it may or may not be
possible to prepare physical systems in arbitrary superpositions of
states, but one cannot settle the question by reference to symmetry
principles, because whatever one thinks the symmetry group of nature may be,
there is always another group whose consequences are identical except for the
absence of superselection rules.''
\ \\ \ \\
Weinberg proceeds to give the concrete example
(p.\ 62, p.\ 90) of the galilean group which
introduces a superselection rule forbidding the superposition of
states of different masses. However, one can add to the galilean Lie
algebra one generator which commutes with all the other generators and
whose eigenvalues are the masses of the various states.   In the
enlarged galilean group, there is no 
need for a mass superselection rule. 
\ \\ \ \\
In 1967, Aharanov and Susskind provided an interesting new angle on
W${}^3$'s
claim that there is a superselection rule between states of
half-odd integer spin and states of integer spin. They
proposed a slow neutron-beam experiment for ruling
out a fermion-boson
superselection rule \cite{Aharanov2}. 
In the proposed experiment, one part of a system
is rotated relative to another.
These experiments are now classic (a
review can be found in \cite{Badurek}). Nevertheless, 
as summarized  for example by Wightman
in his very readable 1994 account \cite{Wightman},
these experimental setups do not
rotate the entire system --- in fact the main premise
of the experiments is precisely
to separate the system into two parts --- hence do not
directly apply to the W${}^3$ fermion-boson superselection rule. Again
the fact that the superselection rule stated by W${}^3$ survives
is of no direct concern within the scope of our paper,
for the reasons mentioned in the  two previous paragraphs.
Indeed in section \ref{sec:interference} we investigate an experimental setup
suggested by the Aharonov-Susskind experiment.
Superselection rules apply neither in the Aharonov-Susskind nor in our
considerations.
\ \\ \ \\
In addition, arguments based heavily (be it implicitly or explicitly)
on ``exact'' conservation laws have often later been 
subject of revision, the most obvious example being the following
about parity. We will have some more to say about ``exact'' conservation
laws in section 4.2.
\ \\ \ \\
Finally, we are aware of the fact that much progress has been
made in investigating superselection rules since the 1960s; 
we merely wish to recall one of the reasons for the earlier lack 
of interest in the \pin\ groups.
\ \\ \ \\
\noindent {\em b) The impact of parity violation.
Right-left asymmetry.\/}\\
The angular distribution of electrons from the beta decay of the
polarized Co${}^{60}$ nucleus, as well as other experiments
involving weak interactions, are
best interpreted in a theory of two-component spinors (see for
instance T.D. Lee \cite{Lee}). This theory distinguishes neutrinos, whose
spins are antiparallel to their momenta (left-handed), 
from antineutrinos, whose spins are
parallel to their momenta (right-handed). 
Neutrinos are emitted in $\beta^+$ decay, and antineutrinos in
$\beta^-$ decays, such as Co${}^{60}$ decays.
In a true (massless) two-component theory, 
antineutrinos whose spins are antiparallel to
their momenta do not exist, so the theory is ``maximally''
parity-violating. The two-component versus
the four-component fermion theory is central to the discussion of
\spin\ and \pin, and to the analysis of parity, which can be found in
section \ref{sec:Pin13}. 
With the experimental evidence of
at least one neutrino being massive, the massless two-component 
``maximally parity-violating'' formalism
has lost some of its absolute character in the Standard Model;
it is therefore like conservation laws such as strangeness
which were thought to be exact
but are nonetheless useful in their range of validity.
\ \\ \ \\
\noindent {\em c) The impact of the Standard Model\/}\\
In the Standard Model, all spin-1/2 particles are chiral particles
defined by the Weyl representation (the two-component theory) and
the concept of intrinsic
parity does not apply to a chiral
particle. Stated in other words, since a left particle becomes a right
particle under space inversion, how does one define its parity? 
Indeed, the Particle Data Group publications \cite{PDG}
do not attribute parity to leptons, presumably for this same reason.
However, quarks and leptons are not ``truly'' two-component, since
mass terms mix left and right. The mere circumstance
that the quarks and leptons of the
Standard Model are written in terms of chiral fermions 
certainly does not rule out the existence of two \pin\ groups.
\ \\ \ \\
\subsection{As seen by Wigner}
In a fundamental paper \cite{Wigner1}, Wigner established the fact that
relativistic invariance implies that physical states are represented
by unitary representations of the Poincar\'{e} group, and simple
systems by irreducible ones. In an article published in 1964 
\cite{Wigner2} Wigner, using an unpublished manuscript written much
earlier with Bargmann and Wightman, analyzes the representations of
the full Lorentz group (see fig.~\ref{fig:Lorentz}). He recalls
first the representations of the proper orthochronous Lorentz group
(labelled $\id$ in fig.~\ref{fig:Lorentz}), then includes space, time,
and spacetime reflections.
\ \\ \ \\
His analysis is anchored on $SL(2,\Cset)$ which is a covering group of
the proper orthochronous Lorentz group. The group $SL(2,\Cset)$ is
isomorphic, but not
identical, to $\spin^\uparrow(3,1) \in \spin(3,1)$ (these 
covering groups are
defined in section \ref{sec:spinsub}). Adding
reflections to $SL(2,\Cset)$, Wigner constructs four {\em distinct}
covering groups. In the process of examining all the possibilities
offered by
adding reflections, Wigner constructs a multiplication table of
reflection operators (Table I in \cite{Wigner2}). Additional
considerations eliminate some unwanted entries of the multiplication
table. Wigner's results explain the observations of Racah, Yang and
Tiomno. Wigner contemplates the existence of a ``whole group'' as opposed
to four distinct groups, but notes that it is not uniquely defined.
\ \\ \ \\
Wigner's work in the formalism of one-particle states has been
extended to Fock space (see in particular 
\cite{Moussa2,MoussaStora,MoussaStora2} and references
therein). An excellent presentation of Wigner's work and its quantum
field theory extension can be found in Moussa's lecture notes
\cite{Moussa2}, in which elementary methods for describing
representation of the the Poincar\'{e} group are used, 
and the aim is to describe spin in particle physics in a natural way.
\ \\ \ \\
Is there anything to be added to Wigner's analysis? The answer is
yes. Wigner's interest in a ``whole group'' and his concern about it
lacking a unique definition is taken care of in this report: there are
two well-defined ``whole groups'', namely the two \pin\ groups. In
comparing our work with Wigner's, one should keep in mind two facts:
\begin{itemize}
\item
Superselection rules are no longer viewed the same way Wigner 
thought of them, as mentioned in section \ref{sec:physicist}a. 
\item
Wigner works with quantum mechanical operators and their projective
representations on one-particle states. We work with operators on Fock
space. Therefore the phases in this report are not Wigner's, 
but of course the
phases in quantum mechanics and the phases in quantum field theory are
related, since a representation on Fock space dictates a representation
on a given one-particle state.
\end{itemize}
Our discussion is structured as follows:
\begin{itemize}
\item
The Pin groups
\item
Their representations on classical fields (not their projective
representations on quantum one-particle states)
\item
Their projective representations in quantum field theory.
\end{itemize}
We can nevertheless establish correspondences between Wigner's results
and ours, namely
\begin{itemize}
\item
Wigner's multiplication table (his Table I) corresponds to the eight
double covers of the Lorentz group listed here in appendix
\ref{app:covers}. Wigner's Table I includes unwanted possibilities
which he eventually excludes,
the eight double covers include double covers other than the \pin\
groups. The double covers which are the \pin\ groups are called
Cliffordian. See Chamblin \cite{Chamblin2} for discussion 
of the non-Cliffordian
double covers.
\item
After elimination of unwanted entries, Table I (modulo Wigner's
phases) corresponds to \pin\ group multiplications (our
eq.\ (\ref{eq:sqrtable}), which includes multiplication of elements
$\Lambda \in \pin(1,3)$ and multiplication of elements $\hat{\Lambda}
\in \pin(3,1)$). 
\end{itemize}
In brief Wigner works in quantum mechanics with four distinct groups
based on $SL(2,\Cset)$; we work in quantum field theory with two
groups \pin(1,3) and \pin(3,1).

\subsection{As seen by mathematicians}
The earliest reference to \pin\ groups we know of is in the 1964 paper
\cite{Atiyah}
of Atiyah, Bott and Shapiro on Clifford modules --- a paper not likely
to have come to the attention of physicists in those days.
Moreover, the authors label both groups \pin($k$), rather than
$\pin(k,0)$ and $\pin(0,k)$, so that the differences between the two
groups is noticed only by a careful and motivated reader.
\ \\ \ \\
Possibly detracting from the difference between the two \pin\ groups
is Car{\nolinebreak}tan's book, {\em Le\c{c}ons sur la Th\'{e}orie
des Spineurs I\/} \cite{Cartan}. We quote from the translation:
\ \\ \ \\
Page 3.
``Let $\phi$ be a quadratic form, $\phi(n-h,h)$,
\begin{equation}
 \phi := \ x_1^2+x_2^2+ \ldots + x_{n-h}^2-x_{n-h+1}^2-
\ldots - x_n^2 \quad
\end{equation}
we shall assume, without any loss of generality, that $n-h\geq h$''.
\ \\ \ \\
There is no loss
of generality in considering only O$(s,t)$, with $s \geq t$, but there
{\em is\/} loss of generality in considering only \pin($s,t$) with $s
\geq t$. It is little known that Cartan did distinguish spinors of the
first and second kind, here identified as the two different kinds of
{\it pinors}. 
\ \\ \ \\
Shortly after the Atiyah, Bott and Shapiro paper, Karoubi published
in {\it Annales Scientifiques de l'Ecole Normale Sup\'{e}rieure} a
long article on ``Alg\`{e}\-bres de Clifford et K-th\'{e}orie''
which contains
a careful study of $\pin(t,s)$ and $\pin(s,t)$. However,
it is not surprising
that physicists did not relate Karoubi's mathematical analysis
to the experimental question of parity.
\ \\ \ \\
When one of us (CD) could not figure out why there are different
obstruction criteria for characterizing the manifolds which admit a
\pin\ bundle, a letter from Y. Choquet-Bruhat paved the way for
identifying not one but two \pin\ groups; a letter from S. Gutt gave
us the construction of the two non-isomorphic \pin\ groups and a
reference to Karoubi's article. The reason for the different criteria
became obvious; two groups, two bundles, each with its own criterion.
\ \\ \ \\
The goal of this report is to clarify parity and related topics by
defining them in terms of the \pin\ groups (sec. \ref{sec:pingroup}),
and to investigate the
physical consequences of the fact that there are two \pin\ groups.

\newpage

\section{The Pin groups in 3 space, 1 time dimensions}
\label{sec:Pin13}
\begin{figure}[h]
   \begin{center}
     \resizebox{10cm}{!}{\includegraphics{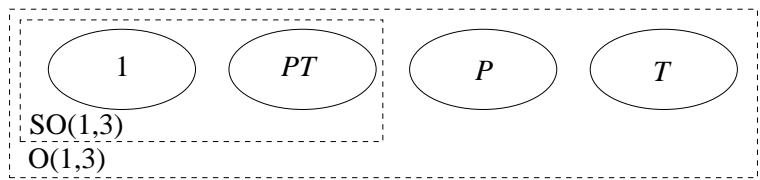}}
   \end{center}
\caption{Components of the Lorentz group}
\label{fig:Lorentz}
\end{figure}

The title of this section could be ``Basic Mathematics''. Here, we explain
why there are two \pin\ groups, and we analyze their differences. For
more information, see for instance \cite{Yellow}.
\ \\ \ \\
Let O$(1,3)$ be the Lorentz group of transformations of
$(\Rset^4,\eta)$ which leaves invariant the quadratic form
$\eta_{\alpha \beta} x^{\alpha}x^{\beta}$, where
\begin{equation}
(\eta_{\alpha \beta}):=\diag(1,-1,-1,-1)
\end{equation}
and let O$(3,1)$ be the Lorentz group of transformations of
$(\Rset^4,\hat{\eta})$, where
\begin{equation}
(\hat{\eta}_{\alpha \beta}):=\diag(1,1,1,-1)
\end{equation}
The Lorentz groups O(1,3) and O(3,1) are isomorphic; nevertheless we
shall use different symbols for their elements because
$({L^{\alpha}}_{\beta}) \in {\rm O(1,3)}$ is not identical to
$(\hat{L}^{\alpha}{}_{\beta}) \in {\rm O(3,1)}$
(see examples in the section on notation).
\ \\ \ \\
The full Lorentz group consists of four components (fig. \ref{fig:Lorentz}).
 \\
Each component is labelled by a representative element:\\
\ \\
\begin{tabular}{ll}
   $\id$ & the unit element \\
   $P$ & the reversal of one or three space axes \\
   $T$ & the reversal of the time axis
\end{tabular} \\
\ \\
The component connected to $\id$ is called the
{\em proper orthochronous\/}
Lorentz group. The two components connected respectively to $\id$
and $PT$ make up the subgroup of the Lorentz group consisting
of orientation
preserving transformations; \ie\ the matrix of the transformation has
determinant 1. If it changes the time orientation, it also changes the
space orientation.
\ \\ \ \\
The \pin\ groups entered physics by the requirement that the Dirac
equation be invariant under Lorentz transformations. For the sake
of clarity and brevity we proceed in the following order:
\ \\ \ \\
\begin{enumerate}
\item[\ref{sec:pingroup}]
 The Pin groups
\item[\ref{sec:spinsub}]
 A Spin group as a subgroup of a Pin group
\item[\ref{sec:pinreps}]
 Pin group and Spin group representations on
 finite-dimensional spaces; classical fields.
\item[\ref{sec:pinrepq}]
 Pin group and Spin group representations on
 infinite-dimensional spaces; quantum fields.
\item[\ref{sec:bundles}]
Bundles; Fermi currents on topologically nontrivial manifolds
\item[\ref{sec:reduction}]
Bundle reduction; massless and massive neutrinos
\end{enumerate}
\ \\ \ \\
The distinction between \pin\ and \spin\ is not always recognized. A
\spin\ group is a subgroup of a \pin\ group, but the expression ``\spin\
group'' is unfortunately still often used to mean the full group. The word
``\pin''
was originally a joke\footnote{The joke has
been attributed to J.P. Serre \cite{Atiyah} but
upon being asked, he did not confirm this.}: \pin($n$) is
to O($n$) what $\spin(n)$ is to SO($n$).

\subsection{The Pin groups}
\label{sec:pingroup}
\begin{figure}[h]
   \begin{center}
     \resizebox{10cm}{!}{\includegraphics{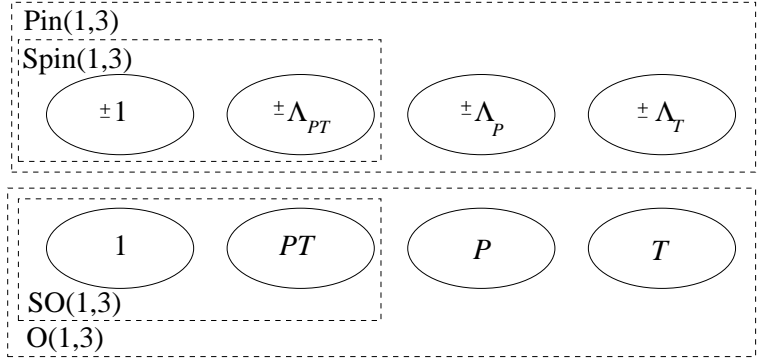}}
   \end{center}
\caption{Double cover of the Lorentz group}
\label{fig:cover}
\end{figure}

\begin{center}
{\em Pin(1,3) \/}
\end{center}
\ \\
Let $\{\gamma_{\alpha}\}$ be the generators of a {\em real\/} Clifford
algebra, such that
\begin{equation}
\{\gamma_{\alpha}, \gamma_{\beta}\}=2\eta_{\alpha \beta} \id
\; , \qquad \eta_{\alpha \beta}=\diag(1,-1,-1,-1)
\end{equation}
and let $({L^{\alpha}}_{\beta}) \in {\rm O(1,3)}$.
\ \\ \ \\
Pin(1,3) consists of the invertible elements $\Lambda_L$ of the Clifford
algebra such that
\begin{equation}
\Lambda_L \gamma_{\alpha}\Lambda_L^{-1}=\gamma_{\beta}{L^{\beta}}_{\alpha}
\qquad \mbox{or equivalently } \qquad
\Lambda_L \gamma^{\alpha}\Lambda_L^{-1}=
{(L^{-1})^{\alpha}}_{\beta}\gamma^{\beta}
\label{eq:homo}
\end{equation}
and such that
\[
\Lambda \Lambda^{\tau}=\pm \id \; .
\]
Here $\tau$ is the reversion, e.g. $(\gamma_0 \gamma_1
\gamma_2)^{\tau}=\gamma_2 \gamma_1 \gamma_0$.
\ \\ \ \\
The two elements $\pm \Lambda_L$ of the \pin\ group
are said to {\em cover\/} the single
element $L$ of the Lorentz group (see fig. \ref{fig:cover})
\ \\ 
\ \\ 
For future reference we solve eq. (\ref{eq:homo})
in a few cases. The solution is readily obtained when $L$ is
diagonal. For example, the reflection of 3 space axes in
$O(1,3)$ is $P=\diag(1,-1,-1,-1)$. Hence
\[
\left\{
\begin{array}{l}
\Lambda_{P(3)} \, \gamma_0 \, \Lambda_{P(3)}^{-1} = \gamma_0 \\
\Lambda_{P(3)} \, \gamma_i \, \Lambda_{P(3)}^{-1} = -\gamma_i
\quad , \quad \mbox{for } i \in \{1,2,3\}
\end{array}
\right.
\]
and the solution is
\[
\Lambda_{P(3)}=\pm \gamma_0 \; .
\]
If $L$ is an element of the proper orthochronous Lorentz group,
\[
\Lambda_L = \exp({\scriptstyle{1 \over 8}} [\gamma_{\alpha},
\gamma_{\beta}]\, \theta^{\alpha \beta})
\]
where $\theta^{\alpha \beta}$ is an antisymmetric tensor made of boost
and rotation generators.
\ \\
\newpage
\begin{center}
{\em Pin(3,1) \/}
\end{center}
\ \\
Let $\{\hat{\gamma}_{\alpha}\}$ be the generators of a {\em real\/} Clifford
algebra, such that
\begin{equation}
\{\hat{\gamma}_{\alpha}, \hat{\gamma}_{\beta}\}=2\hat{\eta}_{\alpha \beta} \id
\; , \qquad \hat{\eta}_{\alpha \beta}= \diag(1,1,1,-1)
\end{equation}
and let $(\hat{L}^{\alpha}{}_{\beta}) \in {\rm O(3,1)}$.
\ \\ \ \\
Pin(3,1) consists of the invertible elements $\hat{\Lambda}_L$ of the Clifford
algebra such that
\begin{equation}
\hat{\Lambda}_L \hat{\gamma}_{\alpha}\hat{\Lambda}_L^{-1}=
\hat{\gamma}_{\beta}\hat{L}^{\beta}{}_{\alpha}
\qquad \mbox{or equivalently } \qquad
\hat{\Lambda}_L \hat{\gamma}^{\alpha}\hat{\Lambda}_L^{-1}=
{(\hat{L}^{-1})^{\alpha}}_{\beta}\hat{\gamma}^{\beta}
\label{eq:homo2}
\end{equation}
and such that
\[
\hat{\Lambda} \hat{\Lambda}^{\tau}=\pm 1 \; .
\]
We summarize in the following table
some results from solving eq.\ (\ref{eq:homo}) and
eq.\ (\ref{eq:homo2}):
\begin{equation}
   \begin{array}{lclclcl}
       \Lambda_{P(1)}&=& \pm \gamma_0 \gamma_1 \gamma_2 & &
       \hat{\Lambda}_{P(1)}&=& \pm \hat{\gamma}_2 \hat{\gamma}_3
         \hat{\gamma}_4 \\
       \Lambda_{P(3)}&=& \pm \gamma_0   & &
       \hat{\Lambda}_{P(3)}&=& \pm \hat{\gamma}_4 \\
       \Lambda_{T}&=& \pm \gamma_1 \gamma_2 \gamma_3 & &
       \hat{\Lambda}_{T}&=& \pm \hat{\gamma}_1 \hat{\gamma}_2 \hat{\gamma}_3   \\
   \end{array}
   \label{eq:lambdatable}
\end{equation}
It follows that
\begin{equation}
   \begin{array}{lclclcl}
       \Lambda_{P(1)}^2&=&-\id & &
       \hat{\Lambda}_{P(1)}^2&=&+\id   \\
         \Lambda_{P(3)}^2&=&+\id   & &
         \hat{\Lambda}_{P(3)}^2&=&-\id \\
       \Lambda_{T}^2 &=& +\id &   &
       \hat{\Lambda}_{T}^2 &=&-\id
   \end{array}
   \label{eq:sqrtable}
\end{equation}
equations which can be used to distinguish \pin(3,1) and \pin(3,1).
We note that $P(3)$ is $P(1)$ followed by a rotation;
$\Lambda^2_{P(3)}$ is $\Lambda^2_{P(1)}$ followed by the effect of a
$2\pi$ rotation on the pinor, hence
$\Lambda^2_{P(3)}=-\Lambda^2_{P(1)}$.

\subsection{A Spin group is a subgroup of a Pin group}
\label{sec:spinsub}
If $({L^{\alpha}}_{\beta}) \in SO(1,3)$ then $\Lambda_L \in
\spin(1,3)$.
\ \\ \ \\
$\spin(1,3)$ consists of even elements (products of
an even number of $\gamma_{\alpha}$) of \pin(1,3).
\begin{center}
\fbox{
   \begin{tabular}{l}
       \spin(1,3) is isomorphic to \spin(3,1), but\\
       \pin(1,3) is not isomorphic to \pin(3,1).
   \end{tabular}
}
\end{center}
A simple but convincing argument
that the two Pin groups are not isomorphic
consists in writing the
multiplication tables of the four generators of $\pin(1,0)$ and
$\pin(0,1)$:
\ \\ \ \\
\begin{tabular}{llll}
$(\pm 1, \pm \gamma)$ & with $\gamma^2=1$ & is isomorphic to &
$\Zset_2 \times \Zset_2$ \\
$(\pm 1, \pm \hat{\gamma})$ & with $\hat{\gamma}^2=-1$ & is isomorphic to &
$\Zset_4$ \\
\end{tabular}\\
\ \\
The proof for 1 time, 3 space dimensions is easier to carry out in
terms of representations of the groups (see section \ref{sec:pinreps}).
\ \\ \ \\
We prove in section \ref{sec:sequence} that 
\ \\ \ \\
\begin{tabular}{lcll}
\pin(1,3)&=&$\spin(1,3) \ltimes
\Zset_2$ & (where $\ltimes$ is a semidirect product) \\
\pin(3,1)&=&$\spin(3,1) \ltimes \Zset_2$
\end{tabular}\\
\ \\
and nevertheless \pin(1,3) is not isomorphic to \pin(3,1).
\ \\ \ \\
The following is true for the Lie algebras of the \pin\ and \spin\
groups. 
\ \\ \ \\
\begin{tabular}{lcccl}
The Lie algebras & $\mathcal{L}(\pin(t,s))$ & and
& $\mathcal{L}(\spin(t,s))$ & are identical.\\
The Lie algebras & $\mathcal{L}(\spin(t,s))$ & and
& $\mathcal{L}(\spin(s,t))$ & are isomorphic.
\end{tabular}\\
\ \\ \ \\
Since \spin(1,3) and \spin(3,1) are isomorphic, the differences
between \pin(1,3) and \pin(3,1) appear only in discussions
of space or time
reversals.
$\Lambda_P$, $\Lambda_T$ are not in \spin(1,3) but
$\Lambda_P^2$ and $\Lambda_T^2$ are in \spin(1,3) and
can be used to identify a \spin\ group as a subgroup
of either \pin(1,3) or \pin(3,1).
\ \\ \ \\
We have come across confusion between the properties of parity and the
properties of $2\pi$ and $4\pi$ rotations. Table \ref{table:rot} should
clarify this confusion.
\begin{table}[h]
\[
   \begin{array}{|@{\hspace{2mm}}cclclcrlcr@{\hspace{2mm}}|}\hline
       \mbox{Let}\; L & = & R(2\pi) \;\mbox{be a}\; 2\pi \; \mbox{rotation}
         &   & \mbox{then}\; \Lambda_{R(2\pi)}
         & = & & \!\!\!\!\hat{\Lambda}_{R(2\pi)} & = & -\id   \\
         \mbox{Let}\; L & = & R(4\pi) \;\mbox{be a}\; 4\pi \; \mbox{rotation}
         &   & \mbox{then}\; \Lambda_{R(4\pi)}
         & = & & \!\!\!\!\hat{\Lambda}_{R(4\pi)} & = & \id   \\
         \mbox{Let}\; L & = & P(1) \;\mbox{reverse 1 space axis}\;
         &   & \mbox{then}\; \Lambda^{2}{_{P(1)}}
         & = & -\id &, \; \hat{\Lambda}^{2}{_{P(1)}}
         & = & \id   \\
         \mbox{Let}\; L & = & P(3) \;\mbox{reverse 3 space axes}\;
         &   & \mbox{then}\; \Lambda^{2}{_{P(3)}}
         & = & \id &, \; \hat{\Lambda}^{2}{_{P(3)}}
         & = & -\id   \\ \hline
       \end{array}
\]
\caption{Parity and rotations in \pin(1,3) vs. \pin(3,1).
Note that $P(3)$ is the reversal of one axis $P(1)$
together with a $\pi$
rotation.}
\label{table:rot}
\end{table}
\ \\
$\Lambda_{R}$ belongs to the Spin group and {\em does not distinguish}
$\psi$ and $\hat{\psi}$ particles, whereas $\Lambda_{P}$ belongs to
a Pin group without belonging to the Spin group.
In a nonorientable space there is no fundamental difference between
rotation and reflection, but in an orientable space there is.
\ \\ \ \\
In brief:
\begin{center}
\begin{tabular}{llll}
$\spin ^\uparrow(s,t)$ & double covers & Proper
orthochronous Lorentz group
& $SO^\uparrow(s,t)$\\ 
$\spin (s,t)$ &double covers &Orientation preserving Lorentz
group &$SO(s,t)$\\
$\pin^\uparrow(s,t)$ & double covers &Orthochronous
Lorentz group &$O^\uparrow(s,t)$   \\
$\pin (s,t)$ & double covers &Full Lorentz group &$O(s,t)$
\end{tabular}
\end{center}
\begin{rem}
It is the $Spin^{\uparrow}$
group which
can be written in a $2 \times 2$ complex matrix representation:
\[
 \spin^{\uparrow} (1,3) \simeq SL(2,\Cset) \; .
\]
\end{rem}
\subsection{Pin group and Spin group representations on finite
dimensional spaces. Classical fields}
\label{sec:pinreps}
We use only real Clifford algebras because we are interested in real
spacetimes, but we use real or complex matrix representations.
Let
\begin{itemize}
   \item[]
   $\Gamma_{\alpha}$ be a real or complex matrix representation
   of $\gamma_{\alpha}$.
   \item[]
   $\hat{\Gamma}_{\alpha}$ be a real or complex matrix representation
   of $\hat{\gamma}_{\alpha}$.
\end{itemize}
We can set $\hat{\Gamma}_{\alpha}=i\Gamma_{\alpha}$ but this bijection
does not define an algebra isomorphism. Indeed, let
$\phi:\pin(1,3) \rightarrow \pin(3,1)$ ; define
\[
\phi(\Gamma_{\alpha})=\hat{\Gamma}_{\alpha} \qquad
\mbox{by} \qquad \hat{\Gamma}_{\alpha}=i\Gamma_{\alpha}
\]
then
\[
\phi(\Gamma_{\alpha})\, \phi(\Gamma_{\beta})
\neq \phi(\Gamma_{\alpha}\Gamma_{\beta}) \; .
\]
On the other hand, the elements of the Spin subgroups consist of even
products of gamma matrices; the mapping
\[
\phi(\Gamma_{\alpha}\Gamma_{\beta})=\hat{\Gamma}_{\alpha}
\hat{\Gamma}_{\beta} \qquad
\mbox{by} \qquad \hat{\Gamma}_{\alpha}\hat{\Gamma}_{\beta}=
-\Gamma_{\alpha}\Gamma_{\beta}
\]
maps \spin(1,3) into itself; \spin(1,3) and \spin(3,1) are identical.
\begin{center}
\ \\
{\em Pinors \/}\\
\end{center}
A representation of a \pin\ group on a vector space defines a
pinor. For example, a fermion of mass $m$ which satisfies the Dirac
equation in an electromagnetic potential is a Dirac pinor $\psi$
or $\hat{\psi}$:
\begin{eqnarray}
 (i\Gamma^{\alpha}\nabla_{\alpha}-m)\, \psi(x)&=&0
 \qquad \psi(x) \in \Cset^4 \label{eq:Dirac} \\
 (\hat{\Gamma}^{\alpha}\nabla_{\alpha}-m)\,
   \hat{\psi}(x)&=&0
\qquad \hat{\psi}(x) \in \Cset^4 \label{eq:Dirachat}
\end{eqnarray}
where $\nabla_{\alpha}=\partial_{\alpha}+iqA_{\alpha}$.
Although they obey the same equation, $\psi$ and $\hat{\psi}$ are
different objects because they transform differently under space or
time reversal.
\ \\ \ \\
We use the same notation $\Lambda_L$ for an element of a \pin\
group and its matrix representation. For instance we write the pinor
transformation $\psi \mapsto \psi'$
induced by a Lorentz transformation
\be
\psi'({L^{\alpha}}_{\beta} \, x^{\beta})=\Lambda_L \psi(x^{\alpha})
\label{eq:Lorentztr}
\ee
\ \\
\begin{center}
{\em Spinors \/}
\end{center}
The space of linear representations of \spin(1,3) on $\Cset^4$
(similar property for \spin(3,1)) splits into two spaces:
\[
S=S_+ \oplus S_-
\]
$S_+$ and $S_-$ are eigenspaces of the chirality operator
\[
\Gamma_5=i\Gamma_0\Gamma_1\Gamma_2\Gamma_3 \; .
\]
$\Gamma_5$
commutes with even elements of a \pin\ group, and
anti-commutes with the odd elements. Let $\varphi$ be an eigenspinor of
$\Gamma_5$, let $\Gamma_+$ be an even element of \pin(1,3) and
$\Gamma_-$ be an odd element of \pin(1,3).
Since $\Gamma_5^2=\id$, the eigenvalues of $\Gamma_5$ are $\pm 1$:
\[
\Gamma_5\, \varphi=\lambda \varphi \;,   \qquad
\mbox{where $\lambda \in \{1,-1\}$,}
\]
thus
\[
\Gamma_5\Gamma_+ \, \varphi=\lambda \Gamma_+ \, \varphi
\qquad \mbox{ and } \qquad \Gamma_5\Gamma_- \, \varphi=
-\lambda \Gamma_- \, \varphi \; .
\]
Hence
\[
\Gamma_+:S_+ \rightarrow S_+ \;
\mbox{ and } \;
\Gamma_- : S_- \rightarrow S_- \; .
\]
Since the eigenvalues of $\Gamma_5$ are $\pm 1$,
the projection matrices
\[
{\mathcal P}_{\pm}=\half(\id \pm \Gamma_5)
\]
project a 4-component $\psi$ into two 2-component Weyl spinors
$\varphi_{\rm L}$ and $\varphi_{\rm R}$:
\bea
\varphi_{\rm L}&=&\half(\id - \Gamma_5)\psi \\
\varphi_{\rm R}&=&\half(\id + \Gamma_5)\psi \; ,
\eea
here L and R stand for left and right;
the use of the words left and right is justified in the paragraph on
helicity below.
\ \\ \ \\
A representation adapted to the splitting $S_+ \oplus S_-$ is called a
chiral representation; in the chiral representation $\Gamma_5$ is
block-diagonal:
\[
\Gamma_5 = \left( \begin{array}{cc}
-\id & \;\; 0 \;\; \\
0 & \id \end{array} \right) \; .
\]
\ \\
\begin{center}
{\em Helicity}
\end{center}
\ \\
In terms of the momentum operator $p_{\mu}=-i \, \partial/\partial
x^{\mu}$, the Dirac equation (\ref{eq:Dirac})
with $m=0$ is
\[
 (\Gamma^0 p_0+ \Gamma^i p_i)\, \psi = 0
\qquad \qquad i \in \{1,2,3\} \; .
\]
When multiplied by $\Gamma_1\Gamma_2\Gamma_3$, this equation reads
in the chiral representation
\[
(\Gamma_5 \, p_0 - (\sigma^i \otimes \id_2) \, p_i) \, \psi =0
\; ,
\]
where $\sigma^i$ are the Pauli matrices (see p.\ \pageref{page:pauli}).
\ \\ \ \\
If $\psi$ is a plane wave
\[
\psi(x,s)=u(p,s) \exp(-i p\cdot x)
\]
the spinor $u(p,s)$ satisfies the equation
\[
(\Gamma_5 - \sigma^i p_i / p_0) u(p,s) = 0 \; .
\]
The helicity operator $h=\half \sigma^i \hat{p}_i$ (where
$\hat{p}_i=p_i/|{\bf p}|=p_i/|p_0|$)
tells us if the spin of the particle is
\begin{itemize}
\item[]
oriented along the direction
of motion
\begin{itemize}
\item[]
(``right-handed'', helicity eigenvalue +1/2),   or
\end{itemize}
\item[]
oriented opposite to
the direction of motion
\begin{itemize}
\item[](``left-handed'', helicity eigenvalue
$-$1/2).
\end{itemize}
\end{itemize}
One often hears the phrase ``a Weyl spinor cannot correspond to an
eigenstate of parity'', but this is a meaningless statement, because the
parity operator $\Lambda_P$ does not act on (2-component)
spinors. In other words, only products of an even number of gamma matrices
can be block-diagonalized; since $\Lambda_P$ is
made of an odd number of gamma
matrices, it cannot be block-diagonalized.
Thus $\Lambda_P$ does not preserve the
splitting $S_+ \oplus S_-$, and is not an operator on the space of
Weyl spinors.
\ \\ \ \\
The fact that only left-handed neutrinos are emitted in Co${}^{60}$
disintegration is referred to as ``parity is not conserved in beta
decay''. Here ``parity is not conserved'' means that the interaction
Hamiltonian does not commute with the space reversal operator.
\ \\
\begin{center}
{\em   Massless spinors, massive pinors\/}
\end{center}
The massless Dirac operator
$\Gamma^{\alpha}\nabla_{\alpha}$ changes the helicity of a Weyl
fermion. The massive Dirac operator $\Gamma^{\alpha}\nabla_{\alpha}+m$
is the sum of a helicity-changing and a helicity-conserving operator;
therefore a massive fermion can only be defined by the 4-component
Dirac representation.
\ \\
\begin{center}
{\em Copinors, Dirac and Majorana
adjoints in Pin(1,3)\/}
\end{center}
\label{sec:copinor}
The representation $\rho$ of \pin(1,3) on ${\mathbb C}^4$ by
$\rho(\gamma_{\alpha})=\Gamma_{\alpha}$ which defines pinors as
contravariant vectors is not the only useful representation. In order
to make tensorial objects from spinorial objects, one needs to
introduce covariant pinors, also called copinors. In the copinor
representation $\rho(\gamma_{\alpha})=\Gamma_{\alpha}^{-1}$ is a right
action on copinors. Let $\psi$ be a pinor with components
$\{\psi^A\}$ in a basis $\{ e_A \}$,
\[
\psi(x)=\psi^A(x) \, e_A \; ;
\]
by definition, the adjoint $\bar{\psi}$ of the pinor $\psi$ is the
copinor
\[
\bar{\psi}(x)=\bar{\psi}_A(x) \, e^A
\]
such that the duality pairing
\[
\langle \bar{\psi} , \psi \rangle :=
\int_{\rm spacetime} \hspace{-4mm} dv(x) \bar{\psi}_A(x)\psi^A(x)
\; , \qquad \mbox{$dv(x)$ a volume element,}
\]
is invariant under Lorentz transformations:
\[
\langle \bar{\psi} , \psi \rangle =
\langle \overline{\Lambda \psi} , \Lambda \psi \rangle
\; .
\]
We are therefore interested in the solutions of the equation
\be
\overline{\Lambda \psi}=\bar{\psi}\Lambda^{-1} \; .
\label{eq:barpsi}
\ee
For $\Lambda$ covering the proper orthochronous Lorentz group, 
the solutions of (\ref{eq:barpsi})
most frequently used are the Dirac adjoint $\bar{\psi}$ and the
Majorana adjoint $\bar{\psi}_M$ of a spinor $\psi$.
The Dirac adjoint is
\be
\bar{\psi}=\psi^{\dagger} \, \Gamma_0
\; ,
\label{eq:Diracadj}
\ee
where a dagger stands for the complex conjugate transposed.
\ \\ \ \\
There is another solution of eq.\ (\ref{eq:barpsi}), namely
$\bar{\psi}_M$, called the
Majorana adjoint or Majorana conjugate of $\psi$; it
is defined by
\[
\bar{\psi}_M := \tilde{\psi}\, {\mathcal T}
\]
where $\tilde{\psi}$ is the transpose of ${\psi}$, and ${\mathcal T}$
defines the isomorphism of the group $\{\tilde{\Lambda}\}$ with the
group $\{\Lambda\}$. The qualifier ``Majorana'' has been used for
different purposes:
\begin{itemize}
\item
A Majorana adjoint as defined above.
\item
A Majorana representation is a representation by matrices all
real or purely imaginary (see Notation,
and section \ref{sec:mod8}).
\item
A Majorana particle is identical to its antiparticle (see
next section).
\end{itemize}
The Majorana adjoint was introduced by
Van Nieuwenhuizen \cite{VanNieu} and
we refer the reader to his article
for the definition and uses of the Majorana adjoint in arbitrary
dimensions.
\ \\ \ \\
The Dirac adjoint for $\Lambda \in \pin(1,3)$ and for $\hat{\Lambda}
\in \pin(3,1)$ is treated in \ref{sec:bundles}, after we have
introduced pinor coordinates.
\ \\
\begin{center}
{\em Charge conjugate pinors in Pin(1,3)\/}
\end{center}
The Dirac pinor $\psi(x)$ satisfies the equation
\be
(i\Gamma^{\alpha}(\partial_{\alpha}+iq A_{\alpha})-m)\psi(x)=0
\label{eq:Diracpin}
\ee
The complex conjugate of this equation is
\be
(-i\Gamma^{\alpha \, *}(\partial_{\alpha}-iq A_{\alpha})-m)\psi^*(x)=0
\label{eq:Diraccompl}
\ee
therefore we introduce
a map ${\mathcal C} : \Cset^4 \rightarrow \Cset^4$ such that
\be
{\mathcal C}\,   \Gamma_{\alpha}^* {\mathcal C}^{-1}
= - \Gamma_{\alpha}
\label{eq:Cdef}
\ee
and we define the charge conjugate pinor as
\be
\psi^{\rm c}={\mathcal C}\psi^* \; .
\label{eq:defpsic}
\ee
which is then a solution of
\be
(i\Gamma^{\alpha}(\partial_{\alpha}-iq A_{\alpha})-m)\psi^{\rm c}(x)=0
\; .
\label{eq:Diracconj}
\ee
In the Dirac representation ${\mathcal
C}=\pm \Gamma_2$ so ${\mathcal C}{\mathcal C}^*=\id$, which is
necessary for $(\psi^{\rm c})^{\rm c}=\psi$.
\ \\ \ \\
The operation $\psi \rightarrow \psi^{\rm c}$ defined by
(\ref{eq:defpsic}) is an antiunitary operation which consists of two
steps; take the complex conjugate of $\psi$, then apply a unitary matrix.
\ \\ \ \\
A pinor and its charge conjugate have opposite eigenvalues of the
parity operator $\Lambda_P$. Let a pinor
$\psi$ be in an eigenstate of $\Lambda_{P(3)}$, abbreviated to
$\Lambda_P$ (reversal of 3 space axes); we have shown that
$\Lambda_P=\pm \Gamma_0$.
Using eq.\ (\ref{eq:Cdef}) we find
\[
\ba{rcll}
\Lambda_P \psi &=& \lambda \psi \; , \hspace{3cm} &
\Lambda^2_P = \id \; , \qquad \lambda = \pm 1 \\ [3mm]
\Lambda_P \psi^{\rm c} &=& \Lambda_P ({\mathcal C}\psi^*) \\
&=&
-{\mathcal C}(\Lambda_P^* \psi^*) & 
\mbox{by (\ref{eq:Cdef}),} \\
&=& -\lambda \psi^{\rm c} & \mbox{by (\ref{eq:defpsic}).}
\ea
\]
To summarize, if $\psi$ is an eigenpinor of $\Lambda_P$,
\be
\Lambda_P \psi = \lambda \psi \; , \qquad
\Lambda_P \psi^{\rm c} = -\lambda \psi^{\rm c} \; .
\label{eq:paritypsi}
\ee
We now compute $\Lambda_P \, \psi(p)$ which is needed in the section on
intrinsic parity.
Let
$\Lambda({\bf p})$ be a 3-momentum boost, then we can write
\bea
\Lambda_P \, \psi(p,s)&=&\Lambda_P   \Lambda({\bf p}) \Lambda_P^{-1}
\Lambda_P \, \psi(p_0,0) \\
&=&\Lambda(-{\bf p})\Lambda_P \, \psi(p_0,0) \\
&=&\Lambda(-{\bf p})\, \lambda \, \psi(p_0,0)
\qquad \mbox{if the ${\bf p}=0$ pinor is an eigenstate of $\Lambda_P$}\\
\eea
Thus, if the pinor at rest has $\Lambda_P$ eigenvalue $\lambda$, we
may write
\be
\Lambda_P \, \psi(p) = \lambda\,   \psi(p\tilde{P})
\label{eq:boost}
\ee
where $p\tilde{P}=(p_0,-{\bf p})$.
Thus all we need to require for the pinor to transform into something
proportional to itself at the new spacetime point $(x_0,-{\bf x})$
is that the pinor at rest (${\bf p}=0$) is an eigenpinor of
parity. The ``eigenvalue'' $\lambda$ at nonzero momentum
is the same as the eigenvalue for the pinor at rest.
\ \\
\begin{center}
{\em  Majorana pinors \/}
\end{center}
By definition a Majorana pinor $\psi^M$ is such that 
\[
(\psi^M)^{\rm c} = \psi^M \; .
\]
To be meaningful the property must remain 
satisfied under a parity transformation. In \pin(1,3)
\begin{eqnarray*}
\Lambda_P \psi^{\rm c}&=& \Lambda_P {\mathcal C} \psi^* \\
&=&  -{\mathcal C} \Lambda_P^* \psi^* \\
&=& - {\mathcal C}(\Lambda_P \psi)^* \\
&=& -(\Lambda_P \psi)^{\rm c} \; , 
\end{eqnarray*}
therefore the condition $\psi^c =\psi$ does not remain satisfied
for the transformed
pinor. On the other hand, in \pin(3,1) $\hat{\psi}^{\rm c}= 
\hat{\psi}$ is
form invariant under a parity transformation:
\[
\hat{\Lambda}_P \hat{\psi}^{\rm c} = (\hat{\Lambda}_P \hat{\psi})^{\rm
c} \; .
\]
{\bf Conclusion:} The classical field of a Majorana fermion can only be a
section of a \pin(3,1) bundle. Briefly,
\[
\fbox{
a Majorana pinor can
only be a Pin(3,1) pinor.}
\]
Yang and Tiomno \cite{YangTiomno} and
Berestetskii, Lifschitz and Pitaevskii \cite{BLP} have also concluded
that, of the four possible parities $(\pm 1, \pm i)$, a Majorana
pinor could be assigned only two. In these references, which bring out
the particular status of Majorana particles (called ``strictly
neutral'' in \cite{BLP}), the four choices were not related to the
existence of {\rm two} Pin groups.
\begin{rem}
In parity-asymmetric theories
it may not be useful
to require the Majorana condition to be invariant under parity.
\end{rem}
\begin{rem}
P. Van Nieuwenhuizen \cite{VanNieu} defines a Majorana particle such
that its Dirac adjoint is equal to its Majorana adjoint. According to
his definition, a Majorana pinor is such that $\psi^{\rm c}=\pm\psi$,
rather than the more commonly used $\psi^{\rm c}=\psi$,
or $\psi^{\rm c}=-\psi$, where one sticks to one choice.
\end{rem}
\ \\
\begin{center}
{\em   Unitary and antiunitary transformations\/}
\end{center}
Motion reversal (sometimes called ``time reversal'')
is a transformation which changes $t$ into $-t$ but, if there
is an electric charge, does not change its sign. Motion reversal is an
antiunitary transformation. Both the
antiunitary motion reversal and the unitary time reversal are useful,
but they play different roles. For example,
Maxwell's equations in the presence of
charge density $\rho$ and current density ${\bf J}$ read either
\be
\begin{array}{cclcccc}
\nabla \times   {\bf E} + {\displaystyle{\partial {\bf B} \over \partial
t}} &=& 0 &
\qquad &
\nabla \cdot {\bf B} &=& 0\\[3mm]
\nabla \times {\bf B} - {\displaystyle{{\partial {\bf E} \over \partial
t}}} &=& {\bf J} &
\qquad &
\nabla \cdot {\bf E} &=& \rho   \\
\end{array}
\label{eq:Maxw}
\ee
or, if one wants to emphasize their relativistic invariance,
\be
\begin{array}{ccccccc}
{\rm d}F&=&0 &\qquad \mbox{or} \qquad &
\epsilon^{\mu \nu \rho \sigma}\partial_{\nu}F_{\rho\sigma}&=&0 \\
\delta F + J&=&0&\qquad \mbox{or} \qquad &
\partial_{\mu} F^{\mu\nu}&=&J^{\nu}
\end{array}
\label{eq:Maxwcov}
\ee
where $F$ and $J$ without indices are differential forms.
If one is interested in motion reversal, eqs. (\ref{eq:Maxw}) are the
appropriate Maxwell's equations. On the other hand, if
one works with the full Lorentz group, and uses eqs. (\ref{eq:Maxwcov}),
the covariant current 4-vector $J_{\alpha}$ transforms
under a Lorentz transformation $L$ as follows:
\[
J \rightarrow J' \qquad \mbox{such that}
\qquad
J_{\beta}(x) = J'_{\alpha}(Lx) {L^{\alpha}}_{\beta} \; .
\]
If the Lorentz transformation
$L$ is the time reversal $({T^{\alpha}}_{\beta})=\diag(-1,1,1,1)$,
then the charge density changes sign:
\[
J_0({\bf x}, t) = -J_0'({\bf x}, -t) \;  .
\]
\ \\
\begin{center}
{\em Invariance of the Dirac equation under antiunitary
transformations\/}
\end{center}
We recall that the Dirac equation is invariant under a unitary
transformation $\Lambda_L$ induced by a Lorentz transformation $L$ if
the new pinor $\psi$ is related to the original pinor $\psi$ by
\[
\psi'(Lx)=\Lambda_L \psi(x)
\qquad
\mbox{with} \qquad   \Lambda_L \Gamma_{\alpha}
\Lambda_L^{-1} = \Gamma_{\beta} {L^{\beta}}_{\alpha} \; .
\]
Under an antiunitary\footnote{Here, like in eq.\ (\ref{eq:defpsic}),
the composite operation $A$ is antiunitary,
but it is carried out 
by a unitary matrix ${\mathcal A}_L$ which acts on complex
conjugate pinors, so $A: \; \psi \rightarrow {\mathcal A}_L \psi^* $. }
transformation $A$
the new pinor $\psi'$ is related to the original one by
\[
\psi'(Lx)={\mathcal A}_L \psi^{*}(x)
\qquad \mbox{with} \qquad
 {\mathcal A_L } \Gamma_{\alpha}^* {\mathcal A_L}^{-1}=
 \Gamma_{\beta}{L^{\beta}}_{\alpha} \; .
\]
This condition is equivalent to
\[
({\mathcal A}_L {\mathcal C}^{-1})\,   \Gamma^{\alpha}\,
( {\mathcal C} {\mathcal A}_L^{-1})=
\Gamma_{\beta}{L^{\beta}}_{\alpha} \; .
\]
Therefore ${\mathcal A}_L {\mathcal C}^{-1}$ 
carries out a {\em unitary} transformation which acts on
pinors, rather than their complex conjugates.
For example, if ${\mathcal A}_T$ is the motion reversal,
then
\[
{\mathcal A}_T {\mathcal C}^{-1} = \Lambda_T \; .
\]
$\Lambda_T$ is made of an odd number of gamma matrices,
and ${\mathcal A}_T$ is made of an even number of matrices.
\ \\ \ \\
The reason for introducing the antiunitary operation
$A$ performed by complex conjugation and the matrix ${\mathcal A}_T$
is the requirement, necessary in a theory free of negative energy
states, that the fourth component of the energy-momentum vector does
not change sign under time reversal.
\ \\
\newpage
\begin{center}
{\em \CPT invariance}
\end{center}
\CPT invariance means invariance under the combined transformation
of charge, parity
and antiunitary time reversal. It follows from the above relation
between unitary and antiunitary time reversal that \CPT invariance
is simply invariance under $\Lambda_P\Lambda_T$, where we
emphasize that $\Lambda_T$ is
unitary. The combination $\Lambda_P\Lambda_T$ covers the component
$PT$ of the full Lorentz group, which
together with the component connected to unity constitutes
the component
of the orientation preserving
transformations (determinant 1).
Thus, {\em CPT invariance is invariance under orientation
preserving Lorentz transformations.}
\ \\ \ \\
For the consistency of a relativistic formalism it is advisable to
derive first the equations in the framework of Lorentz transformations
before investigating the transformations of interest in a specific
context. With \CPT, for example, it can be easier to work with
$\Lambda_P \Lambda_T$ than with \CPT in the traditional sense. We
will have more to say on \CPT in the quantum field theory section.
\ \\

\begin{center}
{\em Charge conjugate pinors in Pin(3,1)\/}
\end{center}
The Dirac pinor $\hat{\psi}(x)$ in
a \pin(3,1) representation satisfies the equation
\be
(\hat{\Gamma}^{\alpha}(\partial_{\alpha}+iq A_{\alpha})-m)\hat{\psi}(x)=0
\label{eq:Diracpin2}
\ee
The charge conjugate $\hat{\psi}^c$ of $\hat{\psi}$ must satisfy the
equation
\be
(\hat{\Gamma}^{\alpha} (\partial_{\alpha}-
iqA_{\alpha})-m)\hat{\psi}^{\rm c}(x)=0
\label{eq:Diracconj2}
\ee
Therefore the map $\hat{\mathcal C}$ such that
\be
\hat{\mathcal C}\,   \hat{\Gamma}_{\alpha}^* \hat{\mathcal C}^{-1}=
\hat{\Gamma}_{\alpha}
\label{eq:Chatdef}
\ee
defines the charge conjugate $\hat{\psi}^{\rm c}$ of $\hat{\psi}$,
\be
\hat{\psi}^{\rm c}=\hat{\mathcal C}\hat{\psi}^* \; .
\ee
The requirement $\hat{\mathcal C}\hat{\mathcal C}^*=\id$ is
indeed satisfied
in \pin(3,1), and ${\mathcal C}{\mathcal C}^*=\id$ is satisfied in
\pin(1,3).
We also check that in \pin(3,1), like in \pin(1,3), a pinor and its charge
conjugate have opposite eigenvalues of the parity operator
$\Lambda_P$. Let a pinor $\hat{\psi}$ be in an
eigenstate of $\hat{\Lambda}_{P(3)} \equiv \hat{\Lambda}_P = \pm
\hat{\Gamma}_4$.
Using eq.\ (\ref{eq:Chatdef}) we find
\bea
\hat{\Lambda}_P \hat{\psi} &=& \hat{\lambda} \hat{\psi} \; , \hspace{3cm}
\hat{\Lambda}^2_P = -\id \; , \qquad \hat{\lambda} = \pm i \\ [3mm]
\hat{\Lambda}_P \hat{\psi}^{\rm c} &=& \hat{\Lambda}_P (\hat{\mathcal
C}\hat{\psi}^*)=
\hat{\mathcal C}(\hat{\Lambda}_P^* \hat{\psi}^*) \\
&=& -\hat{\lambda} \hat{\psi}^{\rm c} \; ,   \hspace{4cm} \mbox{since
$\hat{\Lambda}^*_P=-\hat{\Lambda}_P$.}
\eea
In conclusion, both $\Lambda_P^2=\id$ and $\hat{\Lambda}_P^2=-\id$ imply
opposite eigenvalues of
the parity operator for a pinor and its charge conjugate.
\ \\ \ \\
We record the following:
\be
\begin{array}{cccccl}
{\mathcal C} \, \Gamma^*_{\alpha} \, {\mathcal C}^{-1} &=&
-\Gamma_{\alpha} & \qquad \mbox{In the
Dirac representation} \; {\mathcal C}&=& \pm \Gamma_2 \\

\hat{\mathcal C} \, \hat{\Gamma}^*_{\alpha} \, \hat{\mathcal C}^{-1} &=&
\hat{\Gamma}_{\alpha} & \qquad \mbox{In the
Dirac representation} \; \hat{\mathcal C}&=& \pm \hat{\Gamma}_2 \; . \\
\end{array}
\label{eq:CChat}
\ee

\subsection{Pin and Spin representations on infinite-dimensional
spaces. Quantum fields.}
\label{sec:pinrepq}

\begin{center}
{\em Particles, antiparticles}
\end{center}

In section \ref{sec:pinreps}, charge conjugation meant electrical
charge conjugation. Here the notion of charge conjugate fields is
extended to ``charges''
other than electric: strong isospin, strangeness, etc., charge
conjugate pairs are called {\em antiparticles\/}. Equations
(\ref{eq:CChat}) used in defining electrical charge conjugation are now
used in defining antiparticles.
\ \\ \ \\
``The reason for antiparticles'' is the title of a lecture given by
Feynman as the first Dirac Memorial lecture in 1986. This is how
Feynman introduced his lecture in honor of Dirac: ``Dirac with his
relativistic equation for the electron was the first to, as he put it,
wed quantum mechanics and relativity together'' and Feynman notes that
the ``crucial idea necessary'' for achieving this is the existence of
antiparticles. In the context of quantum field theory it can be shown
that antiparticles are required by causality; an antiparticle
gives rise to
a contribution to the commutator of two fermion fields which exactly
cancels the contribution from the particle at spacelike separation,
as required by causality. Antiparticles are necessary; their
existence is implied in systems invariant under \CPT. 
\ \\ \ \\
The Dirac field operator $\Psi$ acts on a Fock space of particle and
antiparticle states.
The free field
decomposes into particle and antiparticle
plane wave solutions of the Dirac equation:
\be
\Psi(x)={1 \over \sqrt{\Omega}} \sum_{p, s}\left(
a({\bf p},s) \psi(x,p,s)+b^{\dagger}({\bf p},s) \psi^{\rm c}(x,p,s)\right)
\label{eq:Psidef}
\ee
where $\Omega$ assigns a dimension to $\Psi$, and
$\sum_{p,s}$ combines
$\sum_s$ and $\int d^4 p \, \delta(p^2-m^2)$,
and where the mode functions
\be
\ba{rcll}
\psi(x,p,s)&=&
u(p,s) \; \exp(-ip \cdot x) \qquad&   \mbox{particle field} \\
\psi^{\rm c}(x,p,s)&=&v(p,s) \; \exp( ip \cdot x )&
\mbox{antiparticle field}
\label{eq:uandv}
\ea
\ee
are free particle and antiparticle solutions of the Dirac equation,
with
\[
\psi^c={\mathcal C} \psi^* \qquad
\mbox{and \quad ${\mathcal C}^{-1}\Gamma_{\alpha}
{\mathcal C} = -\Gamma_{\alpha}^*$} \; .
\]

The operators
$a$, $a^{\dagger}$, $b$, $b^{\dagger}$ are annihilation and
creation operators on the Fock space:
\bea
a({\bf p},s) && \qquad \mbox{annihilates a particle of momentum ${\bf
p}$   and spin $s$} \\
b^{\dagger}({\bf p},s) && \qquad \mbox{creates an antiparticle
of momentum ${\bf p}$ and spin $s$} \\
\eea
and obvious definitions for $a^{\dagger}$ and $b$.
The usual form of the 
nonzero commutation relations are
\bea
[a({\bf p},s) , a^\dagger({\bf p}',s') ] &=&\delta({\bf p}-{\bf p}')
\, \delta_{s \, s'} \\
\lbrack b({\bf p},s) , b^\dagger({\bf p}',s') \rbrack
&=&\delta({\bf p}-{\bf p}')
\, \delta_{s \, s'}
\eea
Recall that the relativistically invariant quantity is 
$\delta^{(3)}({\bf p}'-{\bf p}) E_{\bf p}$.
\ \\

\begin{center}
{\em Fock space operators, unitary and antiunitary}
\end{center}
In section \ref{sec:pinreps} we introduced three operators on the
space of pinors $\psi$; the charge conjugation operator
${\mathcal C}$, the space and time reversal unitary operators
$\Lambda_P$ and $\Lambda_T$.
We also introduced the antiunitary operation 
$\psi \rightarrow \psi^{\rm c}={\mathcal A}_T \psi^*$ such that
\[
{\mathcal A}_T {\mathcal C}^{-1} = \Lambda_T \; .
\]
The corresponding operators on Fock space
are introduced below in eqs. (\ref{eq:unitary}), (\ref{eq:Uona}),
(\ref{eq:Uonb}) and (\ref{eq:AT}).
\ \\ \ \\
Wigner has shown that a symmetry operator on the Hilbert space of
states is either linear and unitary, or antilinear and antiunitary. A
detailed proof can be found in Weinberg's book (\cite{Wein}
pp. 91-96).
\ \\ \ \\
If one requires the theory to be free of negative energy states, then
the time reversal operator is antiunitary (see for instance the books
by Lee \cite{Lee} or Weinberg \cite{Wein}).
\ \\ \ \\
Let us start from the beginning.
Let $|\alpha \rangle$ and $|\beta \rangle$ be two state vectors, and
$\xi$, $\eta$ two complex numbers.
An operator $U$ is said to be linear if
\[
U (\xi |\alpha \rangle + \eta |\beta \rangle) =
\xi U |\alpha \rangle + \eta U |\beta \rangle \; ,
\]
isometric if $
\langle \alpha | U^{\dagger} U | \beta \rangle  =
\langle \alpha \, | \, \beta \rangle$, 
and unitary if $ U^{\dagger}U = UU^{\dagger} = \id$.
An operator $A$ is said to be antilinear if
\[
A (\xi |\alpha \rangle + \eta |\beta \rangle) = \xi^* A |\alpha \rangle + \eta^* A |\beta \rangle
\]
and antiunitary if $\langle \alpha | A^{\dagger} A |\beta \rangle  =
\langle \alpha \, | \, \beta \rangle ^*$.
\ \\ \ \\
The charge conjugation operator $U_C$ on Fock space is by definition
the unitary operator, $U_C^{-1}=U_C^{\dagger}$, on Fock
space which exchanges particles and
antiparticles:
\be
\ba{rclcrcl}
U_C \, a({\bf p},s) \, U_C^{\dagger}&=&\xi_a b({\bf p},s)
& \qquad &
U_C \, a^{\dagger}({\bf p},s) \, U_C^{\dagger}
&=&\xi^*_a \, b^{\dagger}({\bf p},s) \\
U_C \, b({\bf p},s) \, U_C^{\dagger}&=&\xi_b a({\bf p},s)
& \qquad &
U_C \, b^{\dagger}({\bf p},s) \, U_C^{\dagger}
&=&\xi^*_b \, a^{\dagger}({\bf p},s)
\ea
\label{eq:unitary}
\ee
where $\xi_a$ and $\xi_b$ are arbitrary phases at this stage.
Hence the charge conjugate field operator $\Psi^{\rm c}$ of $\Psi$ is
\[
\Psi^{\rm c}=U_C \Psi U_C^{\dagger}
\]
and
\[
\Psi^{\rm c}(x)={1 \over \sqrt{\Omega}} \sum_{p, s}\left(
\xi_a b({\bf p},s) \psi(p,s)+\xi_b^*
a^{\dagger}({\bf p},s) \psi^{\rm c}(p,s) \right)
\; .
\]
\begin{tabular}{ll}
$\Psi$ & creates antiparticles and
annihilates particles, \\
$\Psi^{\rm c}$ & creates particles and
annihilates antiparticles.
\end{tabular}\\
\ \\
The matrix elements of the operator $\Psi$ take their values in the
space of classical fields $\psi$.
For example, $b^{\dagger}$ creates an antiparticle and
\[
\langle \, b \, | \Psi(a({\bf p},s),b^{\dagger}({\bf p},s))|
\, 0 \, \rangle =
 \psi^c_b(p,s) \; .
\]
Given $\psi^{\rm c}={\mathcal C} \psi^*$ and the fact that
${\mathcal C}$ (an operator in \pin(1,3)) is different from
$\hat{\mathcal C}$ (an operator in \pin(3,1)) we reexpress $\Psi^{\rm
c}$ in terms of ${\mathcal C}$, and will later on express
$\hat{\Psi}^{\rm c}$ in terms of $\hat{\mathcal C}$.
With arguments suppressed
\bea
\Psi^{\rm c}&=&{1 \over \sqrt{\Omega}} \sum_{p,s}
\left(\xi_a \, b \, ({\mathcal C}^{-1} \psi^{\rm c})^* +
\xi^*_b\,   a^{\dagger}\, {\mathcal C}\psi^* \right) \\
&=& {\mathcal C}\left( {1 \over \sqrt{\Omega}}
\sum_{p,s} \left(\xi_a \, b \, \psi^{{\rm c}\, *} + \xi^*_b \, a^{\dagger}
\, \psi^*\right)\right)
\qquad \mbox{since ${\mathcal{C C}}^*=\id$} \\[2mm]
&=:& \, \xi \, {\mathcal C}\Psi^* \; ,
\eea
if we require $\xi^*_b=\xi_a := \xi$, or equivalently
\be
\xi_a \xi_b = 1 \; .
\label{eq:xiconj}
\ee
We review how this fact is verified experimentally
in section \ref{sec:chargeconj}.
\ \\ \ \\
In the above expression for $\Psi^{\rm c}$,
the operator $\Psi^*$ is defined by
\[
\langle \, {\rm out} \, | \Psi^* | \, {\rm in} \, \rangle =
\left(\langle \, {\rm out} \, | \Psi | \, {\rm in} \, \rangle\right)^*
\]
\begin{rem}
Note the difference between $\Psi^*$ and
$\Psi^{\dagger}$ defined by
\[
\left(\langle \, {\rm out} \, | \Psi | \, {\rm in} \, \rangle
\right)^* =
\langle \, {\rm in} \, | \Psi^{\dagger} | \, {\rm out} \, \rangle
\; .
\]
\end{rem}
\begin{rem}
The operator $U_C$ on Fock space is unitary; it acts only on the
creation and annihilation operators. On the other hand, the operator
${\mathcal C}$ on the space of pinors is antiunitary.
\end{rem}
\ \\
\begin{center}
{\em Intrinsic parity\/}
\end{center}
In section \ref{sec:pinreps} we gave the transformation laws of
classical fermion fields, eq.\ (\ref{eq:Lorentztr}), under Lorentz
transformations $L$, and thus in particular under $L=P(3)$, reflection
of 3 axes; in this section we do not need $P(1)$ and we abbreviate
$P(3)$ to $P=\diag(1,-1,-1,-1)$. We now determine the transformation
law under $P$ of the field operator
\[
\Psi(x)={1 \over \sqrt{\Omega}}\sum_{p,s}\left(
a(p,s) \psi_{p,s}(x) + b^{\dagger}(p,s) \psi^{\rm c}_{p,s}(x)\right)
\]
Let $U_P$ be a unitary operator, $U_P^{\dagger}=U_P^{-1}$, such
that\footnote{Weinberg's   \cite{Wein} convention is $U_P \,
a^{\dagger}(p,s) \,
U_P^{\dagger}=\eta_a \, a^{\dagger}(p\tilde{P}, s)$,
so it differs from ours by a
complex conjugation. Our convention is the same as in Peskin and
Schroeder \cite{Peskin}.}
\begin{eqnarray}
U_P \, a(p,s) \, U_P^{\dagger}&=&\eta_a \, a(p\tilde{P}, s)
\;\; \mbox{(the components of $p\tilde{P}$ are $(p_0,-{\bf p})$ )}
\label{eq:Uona} \\
U_P \, b(p,s) \, U_P^{\dagger}&=&\eta_b \, b(p\tilde{P}, s) \; .
\label{eq:Uonb}
\end{eqnarray}
The requirement
\[
U_P \, | 0 \rangle = | 0 \rangle
\]
fixes the values of $\eta_a$ and $\eta_b$ with respect to the vacuum.
\begin{rem}
It is easy to convince oneself that spins do not change under a parity
transformation. Intuitively one has a picture like
fig. \ref{fig:mirror}.
\begin{figure}[h]
   \begin{center}
     \resizebox{6cm}{!}{\includegraphics{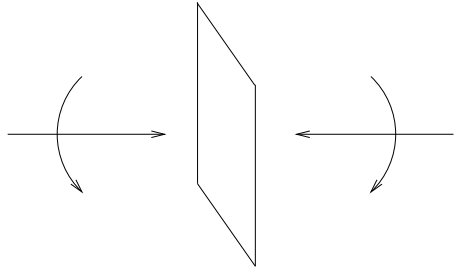}}
   \end{center}
\caption{Spins do not change under reflection.}
\label{fig:mirror}
\end{figure}
Another reason is that we want to add the spin operator {\bf s} to an
orbital angular momentum operator of the form {\bf r} $\times$ {\bf
p}, which does not change sign under parity.
\end{rem}
\begin{rem}
\[
(U \eta_a a \, U^{-1})^*=U \eta_a^* a^{\dagger} \, U^{-1}\; .
\]
\end{rem}
There is obviously a relationship between $\eta_a$ and $\eta^*_b$ so
that $\Psi(x)$ has a well-defined transformation under $P$. In order
to relate $U_P$ to $\Lambda_P$ acting on $\psi$ and $\psi^c$, we
proceed as in the section {\em Particles, antiparticles}; namely we
look for a (complex) constant of proportionality such that
\be
U_P \, \Psi(x,s) \, U_P^{\dagger} \; \propto \; \Lambda_P \Psi(Px,s) \; ,
\label{eq:etaLambda}
\ee
where the components of $Px$ are $(x^0, -{\bf x})$.
We know that $\psi$ and $\psi^{\rm c}$ depend on $x$ only through $p \cdot
x$, so when $x \mapsto Px$, then $p \mapsto p \tilde{P}$;
the components of $p\tilde{P}=p\tilde{P}^{-1}$ are $(p_0, -{\bf p})$.
\ \\ \ \\
Eq. (\ref{eq:etaLambda}) is meaningful only in local quantum field
theory, and so are eqs. (\ref{eq:AT}), (\ref{eq:UC}) and
(\ref{eq:UP}), which all have the same structure as (\ref{eq:etaLambda}).
\ \\ \ \\
Given the definition (\ref{eq:Psidef}, \ref{eq:uandv}) of $\Psi$ and the
Dirac equations (\ref{eq:Diracpin}, \ref{eq:Diracconj},
\ref{eq:Diracpin2}, \ref{eq:Diracconj2}) satisfied by $\psi(x)$,
$\hat{\psi}(x)$ and their charge conjugates,
we have for a particle at rest (omitting the spin label $s$)
\be
\ba{cccccc}
\Gamma_0u({\bf p}=0)&=&u({\bf p}=0) \; ,
\qquad & -\Gamma_0v({\bf p}=0)&=&v({\bf p}=0) \\
i\hat{\Gamma}_4\hat{u}({\bf p}=0)&=&\hat{u}({\bf p}=0)\; ,
\qquad   & -i\hat{\Gamma}_4\hat{v}({\bf p}=0)&=&\hat{v}({\bf p}=0)
\ea
\label{eq:uveigen}
\ee
Therefore for a particle at rest, in momentum space $u$ is an
eigenpinor of $\Gamma_0$ with eigenvalue $1$, and $v$ is an eigenpinor
of $\Gamma_0$ with eigenvalue $-1$. Similarily,
$\hat{u}$ is an eigenpinor of $\hat{\Gamma}_4$ with eigenvalue $-i$
and $\hat{v}$ is an eigenpinor
of $\hat{\Gamma}_4$ with eigenvalue $i$.
\ \\ \ \\
We have established (see the
Dictionary of Notation) that the
space reversal operator
$\Lambda_{P(3)} \in \pin(1,3)$ is $\pm \Gamma_0$, where the choice of
pin structure dictates the sign. We have also computed
in eqs.\ (\ref{eq:paritypsi}, \ref{eq:boost}) the action of $\Lambda_P$ on
$\psi(x)$ when $\psi$ is an eigenpinor
of $\Lambda_P$ at ${\bf p=0}$, with eigenvalue $\lambda$,
\[
\Lambda_P u(p\tilde{P},s) \exp(-ip\cdot x)   =
\lambda u(p,s) \exp (-ip\cdot x) \; .
\]
If we choose $\Lambda_P=\Gamma_0$, it follows from (\ref{eq:uveigen})
that
\bea
\Lambda_P u(p\tilde{P})&=&u(p) = \lambda u(p) \\
\Lambda_P v(p\tilde{P})&=&-v(p) = -\lambda v(p) \\
\eea
We can now compute, omitting reference to $s$ since $s$ is not changed
by $U_P$,
\bea
U_P \Psi(x) U_P^{-1}&=&{1 \over \sqrt{\Omega}}\sum_{p}\left(
\eta_a a(p\tilde{P}) u(p)e^{-ip\cdot x} +
\eta^*_b b^{\dagger}(p\tilde{P})
v(p)e^{ip\cdot x}\right) \\
&=& {\Lambda_P \over \lambda \sqrt{\Omega}}\sum_{p}\left(
\eta_a a(p\tilde{P}) u(p\tilde{P})e^{-ip\cdot x} - \eta_b^*
b^{\dagger}(p\tilde{P}) v(p\tilde{P})e^{ip\cdot x}
\right) \; .
\eea
The sum over $p$ is an integral, under the change of variable
$p\tilde{P} \mapsto p$ we have
\[
u(p\tilde{P}) \exp(-ip\cdot x) \; \mapsto \;
u(p) \exp(-ip\cdot Px)   \; .
\]
Finally,
\[
U_P \Psi(x) U_P^{-1}=
{\Lambda_P \over \lambda \sqrt{\Omega}}\sum_{p}\left(
\eta_a a(p) u(p)e^{-ip\cdot Px}   - \eta_b^*
b^{\dagger}(p) v(p) e^{ip\cdot Px}\right) \; .
\]
For this to be proportional to $\Lambda_P \Psi(Px)$, we must require
$\eta_a = -\eta^*_b =: \eta$, or equivalently
\be
\eta_a \eta_b = \eta_a (-\eta_a^*)=-1 \; .
\label{eq:etaconj}
\ee
This sign has been experimentally verified (see section
\ref{sec:parityexp} on positronium).
Thus we have
\be
U_P \, \Psi(x) \, U_P^{\dagger} = (\eta/ \lambda) \,
\Lambda_P \Psi(Px) \; , \qquad \Lambda_P=\Gamma_0 \; .
\label{eq:etalambda}
\ee
Similarly one finds for \pin(3,1)
\be
U_P \, \hat{\Psi}(x) \, U_P^{\dagger} = (\eta/\hat{\lambda}) \hat{\Lambda}_P
\hat{\Psi}(Px) \; , \qquad \hat{\Lambda}_P=\hat{\Gamma}_0 \; .
\label{eq:etalambdahat}
\ee
\begin{rem}
If we use $\Lambda_P=-\Gamma_0$ or $\hat{\Lambda}_P=-\hat{\Gamma}_0$,
the r.h.s. of (\ref{eq:etalambda}) and (\ref{eq:etalambdahat})
simply change sign.
\end{rem}
We summarize the results for the two \pin\ groups by
\[
\fbox{$U_P \, \Psi(x,s) \, U_P^{\dagger} =
(\eta /\lambda) \, \Lambda_P \Psi(Px,s)\qquad   \mbox{\pin(1,3)}
$}
\]
\[
\fbox{$U_P \, \hat{\Psi}(x,s) \, U_P^{\dagger} =
(\eta / \hat{\lambda}) \, \hat{\Lambda}_P \hat{\Psi}(Px,s)
\qquad \mbox{\pin(3,1)}
$}
\]
\begin{rem}
\ \\
\be
U_P^2 \Psi(x) U_P^{\dagger \, 2} =
(\eta/\lambda)^2 \Lambda_P^2 \Psi(x)
= \eta^2 \lambda^2 \Lambda_P^2 \Psi(x) \; .
\label{eq:Psqr}
\ee
It has been argued
that eq.\ (\ref{eq:Psqr}) must give $\eta^2 \lambda^2
\Lambda_P^2\Psi = -\Psi$ since a
fermion changes sign under a rotation of $2\pi$. However, 
on an orientable space, a
$2\pi$ rotation (successive transformation by infinitesimal angles)
is very different from the discrete symmetry $P$. For example, as
we have shown, $\Lambda_P^2=1$ whereas $\hat{\Lambda}_P^2=-1$, but
still $\Lambda_{R(2\pi)}=\hat{\Lambda}_{R(2\pi)}=-1$.
Thus we leave $\eta$ to be determined.
\end{rem}
\ \\ 
The definition of intrinsic parity has changed throughout the
years. When the canonical reference was Bjorken and Drell
\cite{Bjorken}, the intrinsic parity of a field $\psi$ was the eigenvalue of
$\Lambda_P$:
\[
\Lambda_P \psi = \lambda \psi
\; ,
\qquad \mbox{for $\psi$ in an eigenstate of $\Lambda_P$.}
\]
This is the definition used in the fundamental paper of Tripp \cite{Tripp}
entitled ``Spin and Parity Determination of Elementary
Particles''. A common reference nowadays
is Peskin and Schroeder \cite{Peskin},
where intrinsic parity $\eta$ is defined by the
field operator $\Psi$:
\[
U_P \Psi(x) U_P^{-1} = \eta \Lambda_P \Psi(Px)
\; ,
\qquad \mbox{for $\Lambda_P\, u=u$.}
\]
Since we want to allow for both \pin\ groups and use the modern view
of field theory, we conclude from equation (\ref{eq:etalambda})
that the most general quantity of interest is $\eta/\lambda$, \ie\
\[
\fbox{
\begin{tabular}{l}
$\eta/ \lambda$ is intrinsic parity.
\end{tabular}
}
\]
or for \pin(3,1) it would be denoted $\eta/\hat{\lambda}$.
The fact that we can have four possible parities comes from $(\eta
/\lambda)^4=(\eta/ \hat{\lambda})^4= 1$.
\ \\

\begin{center}
{\em Majorana field operator}
\end{center}
By definition a Majorana field operator $\Psi^{\rm M}$ is such that
\be
\Psi^{\rm M}(x)=
(\Psi^{\rm M}(x))^{\rm c}=\xi \, {\mathcal C} \Psi^{\rm M \, *}(x)
\label{eq:defMaj}
\ee
hence
\[
\Psi^{\rm M}(x)=
{1 \over \sqrt{\Omega}} \sum_{p,s}
\left( a({\bf p},s)\psi_{p,s}(x) +
a^{\dagger}({\bf p},s) \psi^{\rm c}_{p,s}(x)\right)
\]
In other words, the creation/annihilation operators $a$ and $b$ are
the same: $a=b$, thus $\eta_a=\eta_b$ and 
eq.\ (\ref{eq:etaconj}) (which is $\eta_a \eta_b = -1$)
implies $\eta:=\eta_a=\eta_b$ is imaginary.
Putting everything together, we have for the field operator
\ \\ \ \\
Pin(1,3): $\lambda$ is real: $\eta/\lambda=-(\eta/\lambda)^*$.
  ${\mathcal C}\Lambda_P^* =
-\Lambda_P {\mathcal C}$.
\begin{eqnarray*}
U_P (U_C \Psi(t,{\bf x}) U_C^{-1}) U_P^{-1}
&=&(\eta/\lambda)\xi \, \Lambda_P \, {\mathcal C} \, \Psi^*(t,-{\bf x})  \\
&=& -(\eta/\lambda) \xi  {\mathcal C} \Lambda_P^*  \, 
(\Psi(t,-{\bf x}))^* \\
&=& +\xi {\mathcal C} \, ((\eta/\lambda)\Lambda_P \Psi(t,-{\bf
x}))^* \\ 
&=& U_C (U_P \Psi(t,{\bf x}) U_P^{-1})U_C^{-1} \; .
\end{eqnarray*}
\ \\ \ \\
Pin(3,1): $\hat{\lambda}$ is imaginary: $\eta/\hat{\lambda}=
(\eta/\hat{\lambda})^*$.
$\hat{\mathcal C}\hat{\Lambda}_P^* =
\hat{\Lambda}_P \hat{\mathcal C}$.
\begin{eqnarray*}
U_P (U_C \hat{\Psi}(t,{\bf x}) U_C^{-1}) U_P^{-1}
&=&(\eta/\hat{\lambda})\xi \, \hat{\Lambda}_P \, \hat{{\mathcal C}}
\, \hat{\Psi}^*(t,-{\bf x})  \\
&=& (\eta/\hat{\lambda}) \xi  \hat{{\mathcal C}} \hat{\Lambda}_P^*  \, 
(\hat{\Psi}(t,-{\bf x}))^* \\
&=& \xi \hat{{\mathcal C}} \, ((\eta/\hat{\lambda})\hat{\Lambda}_P
\hat{\Psi}(t,-{\bf x}))^* \\
&=& U_C (U_P \hat{\Psi}(t,{\bf x}) U_P^{-1})U_C^{-1} \; .
\end{eqnarray*}
Thus, in both cases, the phase $\eta/\lambda$ makes sure that
the two operations $U_C$ and $U_P$ commute on Fock space.
That is, we {\em can} make
statements about $\Psi^{\rm c}$ (such as the Majorana condition 
$\Psi^{\rm c}=\Psi$) which are invariant under parity for both Pin groups.
\ \\

\begin{center}
{\em Majorana classical field vs. Majorana quantum field}
\end{center}
We have established in section \ref{sec:pinreps} 
that the Majorana condition on a classical Dirac
field 
\[
\psi^{\rm c} = \psi
\]
can be satisfied only by sections $\hat{\psi}$ of a \pin(3,1)
bundle. We have also established that the Majorana condition on a
quantum Dirac field $\Psi$
\[
\Psi^{\rm c} = \Psi
\]
can be satisfied by both types of operators $\Psi$ and $\hat{\Psi}$. 
On the other hand the classical field and field operator are 
of course related;
the matrix elements of an operator $\Psi$ take their values in the
space of classical fields $\psi$. See for instance the subsection on
Fock space operators, the equation
\[
\langle \, b \, | \Psi(a({\bf p},s),b^{\dagger}({\bf p},s))|
\, 0 \, \rangle =
 \psi^c_b(p,s) \qquad \mbox{in \pin(1,3).} 
\]
We also have
\[
\langle \, b \, | \hat{\Psi}(\hat{a}({\bf p},s),\hat{b}^{\dagger}
({\bf p},s))| \, 0 \, \rangle =
 \hat{\psi}^c_b(p,s) \qquad  \mbox{in \pin(3,1).}
\]
The nature of observed particles (i.e. excitations of the field) is
dictated by the annihilation and creation operators. Since we do not
observe the operator but its matrix elements, we observe the classical
fields $\psi$. Hence the Majorana condition ``particle identical to
its antiparticle'' needs to be implemented on the classical field
$\psi$ as well as the field operator $\Psi$ -- and we confirm the
statements made by Yang and Tiomno, Beresteskii, Lifschitz, and
Pitaevskii that a Majorana particle can only be a \pin(3,1) particle.

\begin{rem}
We have seen that
$\eta$ is necessarily imaginary
for a Majorana field operator. This is an example of additional information
which can be used to actually determine the \pin\ group through
$\lambda$.
We already showed above that
a Majorana pinor must be a \pin(3,1) pinor,
\ie\ $\hat{\lambda}$ is imaginary, thus
if we impose both $\Psi^{\rm c}=\Psi$ and $\psi^{\rm c}=\psi$
the total intrinsic parity $\eta/\hat{\lambda}$
of a Majorana particle is real.
\end{rem}
\begin{rem}
Weinberg \cite{Wein} obtains an imaginary parity for
a Majorana field operator. He works with \pin(3,1), so
$\hat{\lambda}$ is imaginary as well as $\eta$, and 
we would expect the parity
$\eta/\hat{\lambda}$ to be real as 
in the previous remark. However, he redefines the parity operator
using other conserved quantities such as baryon number.
See section \ref{sec:search} for our
discussion of parity and conserved quantities.
\end{rem}
\begin{rem}
Majorana fermions may be necessary in
supersymmetric theories, such as eleven-dimensional
$N=1$ supergravity (see \eg\ \cite{Chamblin}). Indeed,
in this theory, if the
superpartner of the graviton -- the gravitino -- were not a Majorana
fermion, the number of bosonic and fermionic degrees of freedom
would not match.
Even in the simplest $N=1$ theory in four dimensions, 
the photino is a Majorana fermion \cite{Sohnius}.
\end{rem}
\ \\

\begin{center}
{\em \CPT transformations\/}
\end{center}
There are different equivalent formulations of the $C\!PT$ theorem,
combining charge conjugation, space and time reversal. See references
in appendix \ref{app:ref}: Collected References. We shall compute the
effect of $U_C U_P A_T$ on the operator $\Psi(x)$ where $U_C$ and $U_P$
are the unitary operators defined by eqs. (\ref{eq:unitary}),
(\ref{eq:Uona}) and (\ref{eq:Uonb}),
and $A_T$ is the antiunitary
operator defined by
\begin{equation}
A_T \Psi(t,{\bf x}) A_T^{-1} = \zeta {\mathcal A}_T \Psi(-t,{\bf x})
\label{eq:AT}
\end{equation}
where ${\mathcal A}_T$ is the operator on the space of pinors defined
in section \ref{sec:pinreps}:
\[
{\mathcal A}_T = \Lambda_T {\mathcal C} =
\pm (\Gamma_1 \Gamma_2 \Gamma_3) (\pm \Gamma_2) =
\mp (\pm \Gamma_1 \Gamma_3)
\qquad \mbox{since } (\Gamma_2)^2 = \id \; .
\]
The choice $\zeta \Lambda_T {\mathcal C} = -\Gamma_1 \Gamma_3$ is the
one used in the book by Peskin and Schroeder \cite{Peskin}.
\ \\ \ \\
Eq. (\ref{eq:AT}) is the bridge connecting the Fock space and the
space of pinors; so are the previously
established relationships
\begin{equation}
U_C \Psi(x) U_C^{-1} = \xi {\mathcal C} \Psi^*(x)
\label{eq:UC}
\end{equation}
\begin{equation}
U_P \Psi(x) U_P^{-1} = (\eta/\lambda) \Lambda_P \Psi(Px) \; .
\label{eq:UP}
\end{equation}
In Pin(1,3), $\Lambda_P = \pm \Gamma_0$ and ${\mathcal C}=\pm
\Gamma_2$, therefore eqs.\ (\ref{eq:AT}), (\ref{eq:UC}) and
(\ref{eq:UP}) give
\begin{eqnarray*}
(U_CU_P A_T) \Psi(t,{\bf x}) (U_CU_P A_T)^{-1}
&=& \pm \zeta
(\eta/\lambda)\xi   \, \Gamma_0 \Gamma_1 \Gamma_2 \Gamma_3 \Psi(t,{\bf x}) \\
&=& {\rm (phases)} \; \Gamma_5 \Psi(t,{\bf x}) \\
&=& {\rm (phases)} \; \Lambda_P \Lambda_T \Psi(t,{\bf x}) \; . 
\end{eqnarray*}
In conclusion, if $C\!PT$ refers to an operator on one-particle
states, we have established in section \ref{sec:pinreps} that $C\!PT$
is the unitary operator $\Lambda_P \Lambda_T$ which corresponds to
orientation preserving Lorentz transformations (transformations of
determinant 1). If $C\!PT$ is an operator on Fock space, it is the
antiunitary transformation $U_C U_P A_T$ carried out by $ {\rm
(phases)} \cdot \Lambda_P \Lambda_T $. The determinant of   $ {\rm
(phases)} \cdot \Lambda_P \Lambda_T $ is 1.
\ \\ \ \\
A similar result is obtained when working with \pin(3,1).
\ \\ \ \\
Invariance of a theory under $C\!PT$ transformations implies the
existence of antiparticles in the theory.

\subsection{Bundles; Fermi fields on manifolds}
\label{sec:bundles}
Given a representation $\rho$ of a \pin\ group on a vector space $V$,
we can construct a \pin\ bundle on a manifold, and define a pinor as a
section of such a bundle. The same is true for the \spin\ group and
spinors. The essence of bundle
theory\footnote{We use the notation of Choquet-Bruhat et al.
\cite{Yellow} which is fairly standard.} is patching together
trivial bundles
\[
(\mbox{Manifold patch}) \times (\mbox{Typical fibre})
= U_i \times V
\]
on the overlap of two manifold patches, $U_i \cap U_j$. The fiber
$V_x$ at a point $x$ in the manifold consists of all the pinors at
this point. A map from the fiber at $x$ to the typical fibre
\[
\phit_{i,x} : V_x \rightarrow V \; , \qquad x \in U_i
\]
defines the coordinates of a pinor $\Psi(x)$ for $x \in U_i$. But if
$x \in U_i \cap U_j$, the map
\[
\phit_{j,x} : V_x \rightarrow V \; , \qquad x \in U_j
\]
defines (probably different) coordinates
for the same pinor $\Psi(x)$.
The patching is done by consistently choosing the maps
\[
\phit_{i,x} \circ \phitinv_{j,x} \; : \; V \rightarrow V
\]
which relate the coordinates of $\Psi(x)$ for $x \in U_i$ and $x \in
U_j$. These maps are the transition functions $g_{ij}(x)$ which act on
$V$ by the chosen representation on $\rho$ of the \pin\ group on $V$
\[
g_{ij}(x)= \phit_{i,x} \circ \phitinv_{j,x}
\]
The consistency condition is
\[
g_{ik}(x) \, g_{kj}(x) = g_{ij}(x) \; .
\]
\  \\

\begin{center}
{\em Pinor coordinates}
\end{center}
See \cite{Yellow} p.\ 415. The subtleties involved in defining the
coordinates of a pinor arise from the fact that there is no unique
choice of a transformation $\Lambda_L$ corresponding to a given
Lorentz transformation $L$. Recall that if one wishes to define a
vector $v$ in a $d$-dimensional vector space
by its coordinates, one says that $v$ is an equivalence class of pairs
$(v_i, \rho_i)$ with $v_i \in \Rset^d$ and $\rho_i$ a linear frame in
$v$, with the equivalence relation
\[
(u_i, \rho_i) \simeq (u_j, \rho_j)
\]
if and only if
\[
u_i = L \, u_j \; , \qquad \rho_j = L \,   \rho_i \; , \qquad L \in
GL(d)
\; .
\]
Similarly the coordinates of a pinor $\Psi$ can be defined by an
equivalence class of triples $(\Psi_i, \rho_i, \Lambda_i)$ with
$\Psi_i \in \Cset^4$, $\rho_i$ an orthonormal frame, and $\Lambda_i
\in \pin(1,3)$, with the equivalence relation
\[
\Psi_i = \Lambda_L \Psi_j\; , \quad
\rho_j = L \rho_i \; , \quad
\Lambda_L = (\Lambda_L)_i \,   (\Lambda_L^{-1})_j
\]
The four complex components of $\Psi(x)$ in the \pin\ frame
$(\rho_i,\Lambda_i)$ are the four complex numbers $\Psi_i(x)$.
\ \\ \ \\
In a similar fashion,
the components of a copinor are defined by the equivalence class
$(\Psi_i, \rho_i, \Lambda_i)$ with the equivalence relation
\[
\Psi_i = \Psi_j \Lambda_L^{-1} \; , \quad
\rho_j = L \rho_i \; , \quad
\Lambda_L = (\Lambda_L)_i \,   (\Lambda_L^{-1})_j \; .
\]
\ \\
\begin{center}
{\em Dirac adjoint in Pin(1,3) \/}
\end{center}
Equipped with the definition of a pinor as an equivalence class of
triples $(\Psi_{(i)}, \rho_{(i)}, \Lambda_{(i)})$ we can extend the
definition of Dirac adjoint (\ref{eq:Diracadj}) from spinors to pinors
\cite[p.\ 36]{Blue}. Let $a(\Lambda)$ be a representation of \pin(1,3)
in $\Zset_2 = \{ 1,-1 \}$, such that
\[
\ba{rcll}
a(\Lambda) &=& 1  & \qquad \mbox{for $\Lambda$ covering orthochronous Lorentz
transformations} \\
a(\Lambda) &=& -1  & \qquad {\rm otherwise} \\
\ea 
\]
Then 
\[
\bar{\Psi}_{(i)} = a(\Lambda) \Psi^{\dagger}_{(i)} \Gamma_0
\]
is such that
\[
\bar{\Psi}_{(i)} = \bar{\Psi}_{(j)} \Lambda^{-1}
\; {\rm when} \; \Psi_{(i)} = \Lambda  \Psi_{(j)}
\]
\begin{pf}
When $\psi\mapsto \Lambda\psi$, then $\psi^{\dagger}
 \mapsto \psi^{\dagger}\Lambda^{\dagger}$ and
\be
\bar{\psi} \; \mapsto \;
a   \psi^{\dagger}\Lambda^{\dagger} \, \Gamma_0 \; .
\ee
When is $\Lambda^{\dagger}\, \Gamma_0=\Gamma_0 \Lambda^{-1}$?
Equivalently, when is $\Gamma_0^{-1} \Lambda^{\dagger} \,   \Gamma_0
\Lambda = \id$ ?
\ \\ \ \\
We can check that
${\Gamma_0}^{-1}\Lambda^{\dagger}\, \Gamma_0 \Lambda$ commutes with
all generators $\Gamma_{\alpha}$ in the basis of the \pin\ group, 
therefore it is a multiple of the unit matrix,
\[
\Gamma_0^{-1} \Lambda^{\dagger} \Gamma_0 \Lambda
= a(\Lambda) \id_4 \; ;
\]
by taking the determinant of both sides one obtains
\[
a^4(\Lambda)=1 \; .
\]
Since $a(\Lambda)$ takes discrete values, it is constant for $\Lambda$ in any
one connected component of the \pin\ group. We check that
$a(\Lambda)=1$ for $\Lambda=\id$ and
$\Lambda=\Gamma_0=\Gamma_0^{\dagger}$; hence
\[
a(\Lambda)=1
\qquad \mbox{for
$\Lambda \in$ components of \pin\ group labelled $\id$ and $P$,}
\]
in other words for $\Lambda$ covering the two components of the
orthochronous Lorentz transformations. We check that
\[
a(\Lambda)=-1   \qquad \mbox{otherwise.}
\]
\end{pf}
\ \\

\begin{center}
{\em Copinors, Dirac adjoints in Pin(3,1) \/}
\end{center}
Following the same arguments as in the case of \pin(1,3), one defines
the Dirac adjoint
\[
\bar{\hat{\psi}} := a \hat{\psi}^{\dagger}\hat{\Gamma}_4
\]
where $a$ is determined from the fact that
\[
\hat{\Gamma}_4^{-1}\hat{\Lambda}^{\dagger}
\hat{\Gamma}_4 \hat{\Lambda} = a(\hat{\Lambda}) \id_4 \; .
\]
With $\hat{\Gamma}_4^{\dagger} = -\hat{\Gamma}_4 $
and $\hat{\Gamma}_4^2=-1$, we find that\\
\ \\
\begin{tabular}{ccrl}
\qquad $a(\hat{\Lambda})$ & $=$ & $1$ \qquad & for $\hat{\Lambda}$ in
the   component of \pin(3,1) labelled
$\id$ or $P$ \\
\qquad$a(\hat{\Lambda})$ & $=$ & $-1$ \qquad & otherwise. \\
\end{tabular}\\
\ \\
For Dirac adjoints on hyperbolic manifolds, see for instance
\cite[p.\ 36]{Blue}.
\ \\

\begin{center}
{\em Pin structures \/}
\end{center}
We are now in a position to define a {\em pin structure}. Let
${\mathcal H}$ : \pin\ group $\rightarrow$ Lorentz group be the 2-to-1
homomorphism defined by
\[
\Lambda_L \Gamma_{\alpha} \Lambda_L^{-1} = \Gamma_{\beta} L^{\beta}{}_{\alpha}
\; ; \qquad {\mathcal H}(\Lambda_L)=L \; .
\]
A pin structure over a (pseudo-)riemannian manifold $M$ of signature
$(t,s)$ is a bundle of \pin\ frames over $M$ together with its
projection ${\mathcal H}$ over a given bundle of Lorentz frames over
$M$. Two different \pin\ structures correspond to two different
prescriptions for patching the pieces of the bundle of \pin\ groups
(as a double cover of the Lorentz bundle) at the overlap of two
patches on $M$ (\eg\ \cite[p.\ 152]{Blue}).
\ \\

\begin{center}
{\em Fermi fields on topologically nontrivial manifolds}
\end{center}
The transition functions of a
\pin(1,3) bundle are elements of \pin(1,3); the transition functions of a
\pin(3,1) bundle are elements of \pin(3,1). The difference between
\pin(1,3) and \pin(3,1) for pinors defined on a topologically 
nontrivial manifold is
spectacular as we shall see shortly. But one should not conclude that
the difference is topological: it is a group difference with
topological implications which are fairly easy to display, and which
were indeed the first ones to be analyzed.
In chronological order,
\begin{itemize}
\item
Obstructions to the construction of \spin\
and \pin\ bundles (\cite{Blue} p{.} 134). The criteria for obstruction
are the nontriviality of some $n$-Stiefel-Whitney classes $w_n$. For
example,
\begin{itemize}
\item
A \pin(2,0)-bundle can be constructed if $w_2$ is trivial
\item
A \pin(0,2)-bundle can be constructed if $w_2+w_1 \cup w_1$ is trivial
\end{itemize}
\item
In supersymmetric Polyakov path
integrals the contributing 2-surfaces depend on the choice of the \pin\
group.   \cite{Carlip}
\item
Quantized fermionic currents \cite{DeWittDeWitt} on
\[
\Rset \mbox{(time)} \times (\Rset \times \mbox{Klein Bottle})
\equiv \Rset^2 \times {\mathbb K}^2 \; . \quad
\]
The Klein bottle alone would
have been sufficient for displaying the difference between the \pin\
groups, but it was convenient to use earlier works done on 3 space, 1
time manifolds \cite{HartIsham}.
\end{itemize}
The Klein bottle ${\mathbb K}^2$ is an interesting manifold for displaying the
difference between the \pin\ groups for the following reasons:
\begin{itemize}
\item
${\mathbb K}^2$ is not orientable (the first Stiefel-Whitney class
$w_1$ is not trivial), Thus a \pin\ bundle, if it exists, is not
reducible to a \spin\ bundle. The Klein bottle forces one to construct
Fermi fields with nontrivial transformation laws under space inversion
on at least one of the overlaps of the coordinate patches.
\item
The Klein bottle admits both kinds of pinor fields, since both
$w_2$ and $w_1 \cup w_1$ are trivial.
\end{itemize}
We review briefly the results. In particular, we explain
 which types of currents
(scalar, vector, ...) can exist on $\Rset^2 \times {\mathbb K}^2$ once
a \pin\ group is chosen, and we review the explicit expectation values of those
currents.
\ \\ \ \\
To obtain a topology $\Rset^2 \times {\mathbb K}^2$ we identify, in a
cartesian coordinate system, the points
\[
(x^0, x^1, x^2, x^3) \quad \mbox{with} \quad
(x^0, x^1, x^2+ma, x^3+(-1)^mx^3+nb)
\]
for all integers $m$, $n$.
\ \\ \ \\
Schematically, one expresses the vacuum expectation values of all
fermionic bilinears $\langle \bar{\Psi}(x) A \Psi(x) \rangle$ in terms
of vacuum expectation values of the chronologically ordered product
$\langle \bar{\Psi}(x) \Psi(x) \rangle$, \ie\ in terms of the Feynman
-Green function $G(x,x')$. The Feynman-Green's function $G$ is expressed in
terms of the Feynman-Green function ${\mathcal G}$ of the Klein-Gordon
operator. For a massless field
\[
G \propto \Gamma^{\alpha} \partial_{\alpha} {\mathcal G} \; .
\]
${\mathcal G}$ is infinite at the coincidence point $x=x'$. Therefore
we subtract the term which equals the Minkowski Feynman-Green
function. This is a cheap and easy way to renormalize, 
but it is valid in this case.
\ \\ \ \\
We find that
${\mathcal G}$, or rather ${\mathcal G}$ renormalized to eliminate the
infinity at the coincidence point, is in the case of \pin(1,3)
\begin{eqnarray*}
{\mathcal G}_{\rm ren}&=&
{i \over (2\pi)^2} \left[
\sum_{m\neq 0,n\neq 0}(-1)^m \left(-(x^0-x'^0)^2+(x^1-x'^1)^2+
(x^2-x'^2+2ma)^2 \right. \right. \\
&+& \left. (x^3-x'^3+nb)^2\right)^{-1}
+ \sum_{m,n} (-1)^m (-i\Gamma_0\Gamma_1\Gamma_2) \left(-(x^0-x'^0)^2 \right. \\
&+& \left. \left. (x^1-x'^1)^2+
(x^2-x'^2+(2m+1)a)^2+(x^3+x'^3+nb)^2\right))^{-1} \right] \; .
\end{eqnarray*}
The sum has been split into two sums, one with the contributions of
$2m$, and one with the contributions of $2m+1$ because of the factor
$(-1)^m$ affecting $x^3$. The ``renormalization'' consists in removing
the $n=0$, $m=0$ term from the first term since this term is, as in
the Minkowski case, infinite at $x=x'$. 
The remainder goes to zero as the Klein bottle becomes large ($a, b
\rightarrow \infty$); ${\mathcal G}$ should properly be treated as a
distribution, but ${\mathcal G}$ is a distribution equivalent to a function.
The term
$\Gamma_0\Gamma_1\Gamma_2$ implements on the fermion field the
periodic reversal of the $x^3$ coordinate.
\ \\ \ \\
For \pin(3,1),
\begin{eqnarray*}
{\mathcal G}_{\rm ren}&=&
{1 \over (2\pi)^2}\left[
\sum_{m\neq 0,n\neq 0}\left(-(x^0-x'^0)^2+(x^1-x'^1)^2+
(x^2-x'^2+2ma)^2 \right. \right. \\
&+& \left. (x^3-x'^3+nb)^2\right)^{-1}
+ \sum_{m,n}\Gamma_0\Gamma_1\Gamma_2 \left(-(x^0-x'^0)^2 \right. \\
&+& \left. \left. (x^1-x'^1)^2+
(x^2-x'^2+(2m+1)a)^2+(x^3+x'^3+nb)^2\right)^{-1}\right] \; .
\end{eqnarray*}
The bilinear fermionic expectation values are expressed in terms of
${\mathcal G}$ by
\[
\langle \bar{\Psi} A \Psi \rangle \, \propto \,
\bigl.{\rm tr}(A \Gamma^{\alpha} \partial_{\alpha} {\mathcal
G})\bigr|_{x=x'}   \qquad \mbox{for \pin(1,3)}
\]
and a similar expression for \pin(3,1). In both cases the derivatives
with respect to $x^0$ and $x^1$ vanish at the coincidence points,
but
\begin{itemize}
\item[]
for \pin(1,3) the derivative w.r.t. $x^3$ vanishes,
\item[]
for \pin(3,1) the derivative w.r.t. $x^2$ vanishes.
\end{itemize}
It follows that
\begin{itemize}
\item[]
for \pin(1,3) the only nonvanishing current is a {\em tensor} with
$A = [\Gamma_0,\Gamma_1]$,
\item[]
for \pin(3,1) the only nonvanishing current is a {\em pseudoscalar} with
$A = \Gamma_5$.
\end{itemize}
We refer to \cite{DeWittDeWitt} for the explicit expressions of the currents
and their graphs. The two cases are totally different. Currents are
observables and in principle one could measure them.
\ \\ \ \\
While a spacetime with a Klein bottle topology, if it exists, would
be difficult to probe, one could imagine solid-state systems for which
the configuration space would be periodic like a Klein bottle.
\ \\ \ \\
Pending such a situation, we searched for other observable differences
in the \pin\ groups. This work, begun by two of us (SJG and EK) has
been continued by MB.

\subsection{Bundle reduction}
\label{sec:reduction}
We recall briefly the essence of bundle reduction. Consider a
principal \pin\ bundle over a manifold $M$ (\ie\ a bundle whose
typical fiber is the \pin\ group) and a principal \spin\ bundle over
the same manifold; or simply a $G$-bundle and an
$H$-bundle, where $H$ is a subgroup of $G$.
\ \\ \ \\
Let the principal $G$-bundle be labeled $(P,M,\pi,G)$, $\pi:P
\rightarrow M$; and let the principal $H$-bundle be labeled $(P_H, M,
\pi_H, H)$. One says that the $G$-bundle is reducible to the
$H$-bundle if $P_H \subset P$, $\pi_H = \pi\bigr|_{P_H}$.
Alternatively: the $G$-bundle is reducible to the $H$-bundle
if the $G$-bundle admits a family of local
trivializations with $H$-valued transition functions. One says: the
structure group $G$ is reducible to $H$.
\ \\ \ \\
A useful criterion: the $G$-bundle is reducible to an $H$-bundle if
and only if the bundle $P \, \backslash \, H$ (typical fibre $G
\, \backslash \,   H$, associated to
$P$ by the canonical left action of $G$ on $G \, \backslash \, H$) admits a
cross
section.
\ \\ \ \\
An example of a reducible bundle: the structure group $GL(n,\Rset)$ of
the tangent bundle of the differentiable mainfold $\Rset^n$ is
reducible to the identity. In other words, the tangent bundle is
reducible to a trivial bundle. This does not mean that the action of
$GL(n,\Rset)$ on the tangent bundle is without interest.
\ \\ \ \\
A vector bundle with typical fiber $V$ is said to be associated to a
principal bundle $G$, if the transition functions act on $V$ by a
representation of $G$ on $V$. Pinors are sections of vector bundles
associated to a principal \pin\ bundle. For brevity we shall say
``pinors are sections of a \pin-bundle''. The properties of a
principal $G$-bundle induce corresponding properties on its
associated bundles, such as reducibility.
\[\fbox{\begin{tabular}{l}
{\small Massless pinors are
sections of \pin\ bundles reducible to \spin\ bundles,} \\
{\small Massive pinors are sections of \pin\ bundles
{\em not\/} reducible to \spin\ bundles.}
\end{tabular}}
\]
Consider a \pin\ bundle reducible to a \spin\ bundle, and an object,
say a Lagrangian defined on the \pin\ bundle;
let the inclusion map
\[
i: \mbox{\spin\ bundle} \rightarrow \mbox{\pin\ bundle} \; .
\]
The pullback $i^*$ maps forms on the \pin\ bundle to forms on the
\spin\ bundle; it maps the \pin\ bundle Lagrangian into a \spin\ bundle
Lagrangian --- which is likely to be different from the Lagrangian
obtained by replacing pinors by spinors in the original
Lagrangian. For example,
symmetry breaking is responsible for introducing mass terms in a
Lagrangian. Bundle reduction, the mathematical expression of symmetry
breaking, yields the mass terms by pulling back the original
Lagrangian into the subbundle.
\ \\ \ \\
An obvious investigation is to apply bundle reduction to a massless
neutrino Lagrangian defined on a \pin\ bundle in order to determine
its pull back on a \spin\ bundle. But
we are temporarily putting this project aside since this paper has been
a long time on the drawing board and we wish to bring it to a closure.
\newpage

\section{Search for Observable Differences}
\label{sec:search}
In section \ref{sec:search}, we investigate what the observable
consequences of the mathematical issues discussed in section
\ref{sec:Pin13} are. We are cautiously optimistic of finding experiments
which can be used to select one \pin\ group over the other for a given
particle. We have ruled out certain setups which seem attractive at
first glance; we present them nevertheless because their failures are
instructive. There are several promising ideas but it is too early to
assess their chances of success. There is one iron-clad identification:
the neutrino exchanged in neutrinoless double beta decay is a
\pin(1,3) particle. As means for selecting a \pin\ group we examine
experiments involving parity
in section \ref{sec:parityexp},
time reversal in section \ref{sec:timerev} and charge conjugation in
section \ref{sec:chargeconj}.
\ \\ \ \\
One conclusion which is easy to see is that \pin(3,1)
fermions cannot interact with \pin(1,3) fermions via terms
$\bar{\psi}M\hat{\psi}$ where $M$ is some matrix, because, as mentioned
by Berestetskii et al \cite{BLP} this term acquires an $i$ under
parity transformations and damps the exponential of the action.
Of course,
\pin(3,1) and \pin(1,3) fermions {\em could} interact via 
nonrenormalizable four-fermion terms
$\bar{\psi}M\psi \, \bar{\hat{\psi}} N \hat{\psi}$ for some matrices $M$
and $N$, but we do not consider such terms.

\subsection{Computing observables with Pin(1,3) and Pin(3,1)}
\label{sec:observables}

One of the fundamental quantities one
calculates in particle physics is the
scattering cross section, or alternatively decay rate.
It is almost always computed using pinors
from \pin(1,3). However, when using
\pin(3,1), there are some changes that could potentially
affect observables.
\ \\

\begin{center}
{\em Trace theorems\/}
\end{center}
Traces of $4n+2$ gamma matrices, where $n = 0,1,2, \ldots$,
are equal between the two \pin\ groups. Traces of $4n+4$ gamma
matrices differ by a sign between the two groups. Therefore, linear
combinations of the following traces, for instance, are different:
\be
\begin{array}{rcl}
{\rm tr}(\Gamma_{\mu}\Gamma_{\nu})&=&
4 \eta_{\mu\nu} \\
{\rm tr}(\hat{\Gamma}_{\mu}\hat{\Gamma}_{\nu})&=&
4\hat{\eta}_{\mu\nu}= -4 \eta_{\mu\nu} \\
{\rm tr}(\Gamma_{\mu}\Gamma_{\nu}\Gamma_{\rho}\Gamma_{\sigma})&=&
4 \left(\eta_{\mu\nu}\eta_{\rho\sigma}-
\eta_{\mu\rho}\eta_{\nu\sigma}+\eta_{\mu\sigma}\eta_{\nu\rho}
\right) \\
{\rm
tr}(\hat{\Gamma}_{\mu}\hat{\Gamma}_{\nu}\hat{\Gamma}_{\rho}\hat{\Gamma}_{\sigma})&=&
4 \left(\hat{\eta}_{\mu\nu}\hat{\eta}_{\rho\sigma}-
\hat{\eta}_{\mu\rho}\hat{\eta}_{\nu\sigma}+\hat{\eta}_{\mu\sigma}\hat{\eta}_{\nu\rho}
\right) \\
&=& 4 \left({\eta}_{\mu\nu}\eta_{\rho\sigma}-
\eta_{\mu\rho}\eta_{\nu\sigma}+\eta_{\mu\sigma}\eta_{\nu\rho}
\right)
\end{array}
\ee
For example, if $A$ is proportional to $\tr \, \Gamma_{\mu}\Gamma_{\nu}$
and $B$ is proportional to $\tr
\, \Gamma_{\mu}\Gamma_{\nu}\Gamma_{\rho}\Gamma_{\sigma}$, then $A+B$ in
\pin(1,3) corresponds to $-\hat{A}+\hat{B}$ in \pin(3,1).
\ \\
\begin{center}
{\em Spin sums\/}
\end{center}
When computing unpolarized cross sections, one needs to sum over spin
states. Let $\Psi(x)$ in \pin(1,3) be defined by (\ref{eq:Psidef}) and
$\hat{\Psi}(x)$ in
\pin(3,1) be defined similarly. With $\sla p = \Gamma^{\mu} p_{\mu}$,
or $\sla p = \hat{\Gamma}^{\mu} p_{\mu}$ as the case may be, we have
\[
\begin{array}{rclcrcl}
\sum_s {u}\, (p,s)\bar{u}(p,s)&=& \sla p + m \qquad &&
\sum_{s} {\hat{u}}\, (p,s)\bar{\hat{u}}(p,s)&=& - i\sla p + m \\ [1mm]
\sum_{s}{v}\, (p,s)\bar{v}(p,s)&=& \sla p - m &&
 \sum_{s} {\hat{v}}\, (p,s)\bar{\hat{v}}(p,s)&=&
-i\sla p - m \; ,
\end{array}
\]
if we use the normalizations
\[
\begin{array}{rclcrcl}
\bar{u}\, (p,r)u(p,s)&=& 2m\delta_{rs} \qquad &&
\bar{\hat{u}}\, (p,r)\hat{u}(p,s)&=& 2m\delta_{rs} \\
\bar{v}\, (p,r)v(p,s)&=& -2m\delta_{rs}&&
\bar{\hat{v}}\, (p,r)\hat{v}(p,s)&=&
-2m\delta_{rs} \; .
\end{array}
\]
This is shown in Appendix \ref{app:coll}.
\label{page:spinsums}

\subsection{Parity and the Particle Data Group publications}
In section \ref{sec:pinrepq} we defined the intrinsic parity of a
quantum field $\Psi$ as $\eta / \lambda$, where the phase
$\eta$ comes from the definition (\ref{eq:unitary}) of the
unitary operator $U_P$ acting on the field operators,
and the phase $\lambda$ is the parity eigenvalue
of the pinor $u(p,s)$ (or $\hat{u}(p,s)$) in
eq. (\ref{eq:uveigen}).
\ \\ \ \\
Here in section \ref{sec:search}, $P$ stands for $P(3)$,
reversal of the three space axes. See eq.\ (\ref{eq:lambdatable}) in
section \ref{sec:pingroup} for the calculation of $\Lambda_P \in
\pin(1,3)$ and $\hat{\Lambda}_P \in \pin(3,1)$. We recall
\[
\Lambda_P^2=\id \; , \qquad \hat{\Lambda}_P^2 = -\id \; .
\]
The phase $\eta$ is usually a matter
of convention, but as was shown in section \ref{sec:pinrepq},
$\eta$ {\em must be imaginary} for a Majorana particle (a particle which is
its own antiparticle). Thus there is at least one way of
restricting the choice of phase $\eta$.
In the Particle Data Group (PDG) publications,
intrinsic parity is always real, so for us $\eta /
\lambda = \pm 1$. The \pin\ group used in PDG
publications is \pin(1,3), and we infer that $\eta=\pm 1$ corresponds
to the PDG convention.
\ \\ \ \\
As can be seen in eq. (\ref{eq:uveigen}),
the eigenvalue $\lambda$ of the parity operator only takes on the
values $+1$ and $-i$ unless we change {\em pin structure} (see
sec. \ref{sec:bundles} for a discussion of pin structures, see also
eqs. (\ref{eq:etalambda}) and (\ref{eq:etalambdahat}) and the remark
thereafter),
which is only necessary on
spaces with nontrivial topology (also section \ref{sec:bundles}).
\ \\ \ \\
One reason many physicists discard parities $ \pm i$
is the intuitive, but, in the case of
fermions, faulty argument that two successive reflections bring us
back to the original state: fermions change sign under $2\pi$
rotations. Recall that this is true for fermions of {\em both} pin
groups -- see eqs. (\ref{eq:lambdatable}) and (\ref{eq:sqrtable}). As
pointed out in the book
by Bjorken and Drell \cite{Bjorken}, ``four reflections
return the spinor to itself in analogy with a rotation through $4\pi$
radians''. Indeed $(\eta/\lambda)^4=1$ and $(\eta /\hat{\lambda})^4=1$.
\ \\ \ \\
Finally, it is clear that intrinsic parity is a ``relative'' concept,
\ie\ one needs to define some particle to have, say, parity $+1$
to fix the number for another particle transforming under $\pin(1,3)$.
In the PDG publications, three (composite) particles are chosen as
``reference particles'' and parities of other particles are determined
by comparison with one of the three reference particles\footnote{There
is   no logical necessity for having {\em three} reference particles,
other than a convenience for analyzing experimental data
within the bounds of the possibly approximate conservation laws
of baryon number, lepton number and strangeness (or, as Weinberg
proposes \cite{Wein}, electric charge, since strangeness
conservation is now known to be approximate).}.
\ \\ \ \\
Since we have intrinsic parity as $\eta/\lambda$, defining a reference
parity still leaves some freedom in specifying $\eta$ and $\lambda$
unless there are extra conditions such as that for a Majorana
particle. Examples of this freedom are
given in section \ref{sec:parityexp}, for example in determining the
intrinsic parity of a pion.
The PDG defines
\begin{center}
\begin{tabular}{|c|c|} \hline
\multicolumn{2}{|c|}{\bf Reference Particles} \\ \hline\hline
 Particle & Intrinsic Parity
\\ \hline\hline
Proton & +1 \\ \hline
Neutron & +1   \\ \hline
$\Lambda$ & +1   \\ \hline
\end{tabular}
\end{center}
These particles are not {\em elementary} particles, but the
distinction seems to be insignificant; one finds
experimentally that these composite particles have well-defined
intrinsic parities, so they are just as good reference particles as
any other. Furthermore, as far as the present set of
elementary particles goes, confined quarks would be difficult to have as
experimental references, and leptons are written as
Weyl fermions in the Standard
Model and as such cannot be acted upon by the parity operator (see
\ref{sec:pinreps}).
\ \\
\begin{center}
{\em Parity conservation} 
\end{center}
In Appendix \ref{app:coll}, we briefly review 
how the observed angular distribution of scattered
particles is used for concluding whether or not parity is conserved.
When it is, one arrives at the expression for
conservation of parity:
\label{page:paritycons}
\be
(-1)^{\ell_{\rm i}}\eta_a \eta_b = (-1)^{\ell_{\rm f}} \eta_c \eta_d
\label{eq:conspar}
\ee
which says that $\eta$ is (multiplicatively) conserved in a
parity-conserving interaction $H_{\rm int}$, \ie\ if $H_{\rm int}$
commutes with $U_P$. We can use
(\ref{eq:conspar})
to determine one unknown $\eta$, for example, using previously
known or defined parities.
\begin{rem}
When representing a particle by a classical field $\psi$, the
eigenvalues $\lambda$ of the parity operator $\Lambda_P$ identify the
relevant \pin\ group. When representing a particle by a Fock state
built by creation operators, parity is not identified by $\lambda$
but by $\eta/\lambda$. Notice that $\eta$ is the quantity which
appears in the conservation law.
\end{rem}
Clearly the fact that intrinsic parity is {\it
multiplicatively} rather than additively conserved is irrelevant, we
could redefine $\eta$ to be the exponential of another symbol.
\ \\ \ \\
In the decay of Co${}^{60}$, electrons are predominantly emitted in
a certain direction, therefore eq.\ (\ref{eq:symmetryM}) 
in the appendix is not satisfied,
and the interaction is said to {\it
violate} parity (conservation).

\subsection{Determining parity experimentally}
\label{sec:parityexp}
There are two different broad approaches to determining intrinsic
parity experimentally: by using selection rules, or by
studying decay rates, cross sections and
polarizations.
\ \\
\begin{center}
{\em Selection rules: Pion decay}
\end{center}
The textbook example \cite{Frauenfelder,Wein} of determining
intrinsic parity by a selection
rule is the negative
pion ($\pi^-$). It is reviewed in Appendix \ref{app:coll},
here we just give the result:
\[
\eta_{\pi} \eta_{d}   = (-1)\eta_{n}\eta_n
\]
where $d$ is the deuteron captured by the pion,
which then decays to two neutrons $n$.
We study how the determination of the pion's parity 
proceeds in \pin\ group language.
\label{page:pion}
\ \\ \ \\
First, the deuteron and the pion have integer spins, so
they cannot have imaginary $\lambda$ values.
We show in the positronium example below that the parity of an
$s$-wave bound state of two fermions $a$ and $b$ is
$\eta_a\eta_b$ (not just for positronium).
Then the intrinsic parities of the pion and deuteron are
$\eta_{\pi}$ and $\eta_d=\eta_a\eta_b$, respectively.
If we assume the neutrons are \pin(1,3) particles, then
$\lambda_p=\lambda_n=+1$ so $\eta_p=\eta_n=+1$ by the reference parities.
With these assumptions we find
\[
\eta_{\pi}=-{\eta_n^2 \over \eta_d}=-1 \; .
\]
If we assume the neutrons are \pin(3,1) particles,
then $\hat{\lambda}_p=\hat{\lambda}_n=-i$ so $\eta_p=\eta_n=-i$ and
we obtain the same result.

From the explicit discussion in the appendix,
we see that even this comparatively simple argument
relies on input from various sources (orbital state of deuteron and
$\pi^- d$ atom as a whole, fact that interaction is parity-conserving).
The only three principles we invoked
were angular momentum conservation, Fermi statistics
and conservation of intrinsic parity. All of these
are independent of the choice of
\pin\ group. On general grounds we can therefore
expect this experiment to be incapable of detecting a difference
between the two \pin\ groups, but
it is somewhat instructive.
\ \\

\begin{figure}[h]
   \begin{center}
   \vspace{5mm}
     \resizebox{5cm}{!}{\includegraphics{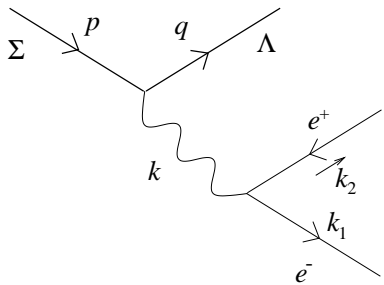}}
     \vspace{2mm}
   \end{center}
\caption{Tree-level QED diagram for
$\Sigma^0 \rightarrow\Lambda^0 + e^+ + e^-$.}
\label{fig:diagram}
\end{figure}

\begin{center}
{\em Selection rules: Three-fermion decay\/}
\end{center}
Another example of a ``selection rule'' type argument
can be found in the book by Sternberg \cite{Sternberg}. There it
is claimed that the following argument {\em can\/} determine the difference
between the two \pin\ groups.
\ \\ \ \\
It is mentioned that a fermion cannot decay into three fermions through a
parity-conserving interaction in the
Pin group for which $\lambda$, the eigenvalue of $\Lambda_P$,
is imaginary (in our conventions, this is \pin(3,1)), because
\[
(\pm 1)^3= \pm 1 \; \mbox{   whereas   } \; (\pm i)^3=-(\pm i)
\]
There are two arguments that show why the conclusion ``a \pin(3,1)
fermion cannot decay into three fermions'' is too hasty. First, since
intrinsic parity is $\eta/\lambda$, where $\eta$ is the phase in the
definition of $U_P$ and the quantity which appears in the
parity conservation law,
intrinsic parity is not directly related to \pin\
group through $\eta$ {\em unless\/} there is an extra requirement
on $\eta$, such as the Majorana condition. If $\eta$ can be chosen
real or imaginary
by convention, we cannot determine the \pin\ group in this way.
\ \\ \ \\
Second, just like a fermion with $\eta=+1$ can
decay into three fermions of $\eta=-1$, $-1$ and $+1$, for
example, a
fermion of $\eta=+i$ can decay into three
fermions with $\eta=+i$, $+i$ and $-i$. We cannot infer that
three-fermion decay is always forbidden merely from $i^3=-i$.
\ \\
\begin{center}
{\em Selection rules: Positronium}
\end{center}
\label{page:positron}
There is a beautiful experiment, first proposed by Wheeler
\cite{Wheeler}, to verify experimentally the relation (\ref{eq:etaconj})
for the phase of the parity operator from section \ref{sec:pinrepq}.
We review the experiment in Appendix \ref{app:coll} for completeness.
\ \\
\begin{center}
{\em Decay rates; cross sections\/}
\end{center}
There is a plethora of different accelerator experiments which are
capable of determining the intrinsic parity of a particle. Actual examples
include but are not limited to polarized target experiments, production
experiments and electromagnetic decays. The methods for
studying parity are sometimes similar to those used in determining
spin, but there is no theoretical reason that we know of why the two
should be related.
\ \\ \ \\
We choose to concentrate on one particular experiment for
definiteness. We have chosen the beautiful $\Sigma^0$-parity
Steinberger experiment \cite{Steinberger} from 1965. One could ask why we
have chosen to analyze such an old experiment, given the immense progress
that has been made in experimental particle physics during the last
three decades. However, once a discrete attribute
such as the intrinsic parity of a particle
(or resonance) is determined to good accuracy,
it is of course unattractive for experimentalists to construct a
dedicated experiment to measure it again. The attribute might be
measured as a ``byproduct'' of other experiments, but such
determinations would, by the same token,
be less straightforward for us to analyze here.
Since parities of most particles
were already measured in the 1960s, one finds that most of these
dedicated experiments were done around 1965 or before.
\ \\ \ \\
The experiment revolves around the electromagnetic decay
\[
\Sigma^0 \quad \rightarrow \quad \Lambda^0 + e^+ + e^-
\]
where the parity of the $\Lambda^0$ is chosen as one of the reference
parities.
\ \\ \ \\
The simple idea, put forward by Feinberg \cite{Feinberg}, is to
measure the branching ratio for the above decay relative to the
main decay mode $\Sigma^0 \rightarrow \gamma \gamma$.
The QED prediction is
different for the hypotheses $\Sigma^0$-parity $+1$ and
$\Sigma^0$-parity $-1$ (we prefer to not use the terms
``odd'' and ``even'' due to the possible existence
of four parities). We shall show very briefly how this difference
arises.
Let us fix $\eta$, the phase in $U_P$, to be $\eta=1$ for now.
The tree-level diagram is shown in fig.\ \ref{fig:diagram}.
Using standard notation \cite{Peskin}
we write down the matrix elements for the two hypotheses.
Since we do not have an {\em a priori} $\Lambda\Sigma$ vertex,
we write down all possible bilinears, and determine the coefficients
of each experimentally. It turns out that the dominant contribution
is the tensor or pseudotensor:
\begin{eqnarray*}
{\mathcal M}_+&=&e^2 {iF \over M} \left(\bar{u}_{\Lambda}(q)
\Gamma_{\mu\nu}k^{\nu}u_{\Sigma}(p)\right){1 \over k^2} \left(
\bar{u}(k_1)\Gamma^{\mu} v(k_2)\right) \\ [1mm]
{\mathcal M}_-&=&e^2 {iF \over M} \left(\bar{u}_{\Lambda}(q)
\Gamma_5 \Gamma_{\mu\nu}k^{\nu}u_{\Sigma}(p)\right){1 \over k^2} \left(
\bar{u}(k_1)\Gamma^{\mu} v(k_2)\right)
\end{eqnarray*}
Here $F$ is a form factor, and
we have written $\Gamma^{\mu \nu}={i \over 2}
[\Gamma^{\mu}, \Gamma^{\nu}]$ and used the average mass
$M=\half(M_{\Lambda}+M_{\Sigma})$. There are also other terms contributing
to the diagram, but the form factor $F$ is sufficiently
large for terms with other form factors to be neglected.
\ \\ \ \\
The idea is that if,
for example, ${\mathcal M}_-$ is the correct matrix element, we
can shift the $\Gamma_5$ to the right and
include it in $u_{\Sigma}$. This means
that $\lambda_{\Sigma}$, the eigenvalue of $\Lambda_P$ for the
$\Sigma$ particle, switches sign
due to $\Lambda_P \Gamma_5 = - \Gamma_5 \Lambda_P$. Thus in the case
of ${\mathcal M}_-$
the relative parity of $\Lambda$ and $\Sigma$ would be $-1$.
\ \\ \ \\
Since there is only one diagram (in this approximation), we
immediately see that any phase will eventually be canceled when we
take the absolute value squared. However, 
we summarize in Appendix \ref{app:coll}
how this is manifested using the rules from section
\ref{sec:observables}, since a similar calculation may prove important
in other settings.
\label{page:sigma}

\subsection{Interference, reversing magnetic fields, reflection}
\label{sec:interference}
\begin{figure}[h]
   \begin{center}
     \resizebox{8cm}{!}{\includegraphics{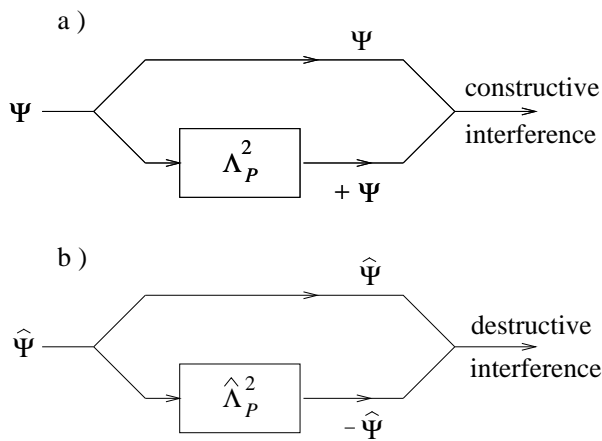}}
     \caption{A type of experiment which should give
different results for the two types of pinors.}
   \label{fig:boxes}
     \vspace{3mm}
   \end{center}
\end{figure}

\begin{figure}[h]
   \begin{center}
       \vspace{3mm}
       \resizebox{10cm}{!}{\includegraphics{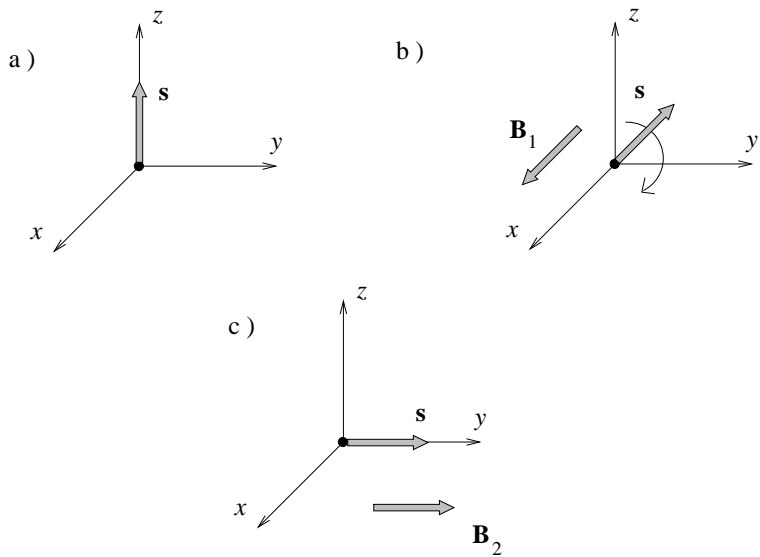}}
   \caption{A $z$-polarized electron (fig. a) is placed for a while in a
magnetic field ${\bf B}_1$ along the $x$-axis 
causing its spin to precess
around the $x$-axis (fig. b); later on 
a magnetic field ${\bf B}_{2}$ in the
$y$-direction is switched on (fig. c). One could also consider an electrically
neutral spin 1/2 particle in a potential
$\mu {\bf s} \cdot {\bf B} = \mu \epsilon_{ijk} \sigma^{ij}B^{k} \;
(i, j, k \in \{1, 2, 3\})$ where $\mu$ is the gyromagnetic ratio
and $\sigma^{ij}$ the spin angular momentum operator.}
   \label{fig:magnetic}
     \vspace{3mm}
   \end{center}
\end{figure}

In \pin(1,3) and \pin(3,1), two successive parity transformations
are given, respectively, by $\Lambda _P^2 =   \id$ and
$\hat{\Lambda}_P^2 = -\id$.   Hence if one could construct an
experiment corresponding to fig. \ref{fig:boxes},
one might be able to differentiate
between the two types of pinor particles.   A beam must
somehow be split into two beams, one of which is inverted twice while
the other is left unaffected.   Then the two beams must be brought
back together and allowed to interfere.   Where \pin(1,3) particles
interfere constructively (fig. a), \pin(3,1) particles will interfere
destructively (fig. b).
\newpage
\begin{center}
{\em Reversing magnetic fields}
\end{center}
Not knowing how to construct a space reversal apparatus (the boxes in
fig. \ref{fig:boxes}) we considered the following experiment:
\ \\ \ \\
A particle beam is split in two parts: One part passes through a magnetic
field
in the $x$ direction followed by a magnetic field in the
$y$ direction as in fig. \ref{fig:magnetic}, the other part passes through
a magnetic field in the $x$ direction followed by a magnetic
field in the $-y$ direction. The two parts of the beam
are then recombined and allowed to interfere.
An explicit calculation by one of us (EK) showed that the interference
of the two parts of the beam having experienced the two different magnetic
field configurations is the same for a \pin(3,1) beam and for a \pin(1,3)
beam.
\ \\ \ \\
The explicit calculation consists in comparing the
transition amplitudes for Pin (3,1) and Pin (1,3) electrons moving
under the conditions described in Fig. 3. The system evolves
according to the Dirac equation which reads
\begin{eqnarray*}
(i\Gamma^{\alpha} (\partial_{\alpha} + iqA_{\alpha}) - m) \psi   &=& 0 \;
\; \; \mbox{for Pin(1,3) particles} \\
(\hat{\Gamma}^{\alpha} ( \partial_{\alpha} + iqA_{\alpha}) -   m )
\hat{\psi} &=& 0 \; \; \; \mbox{for Pin (3,1) particles}
\end{eqnarray*}
Since $\hat{\Gamma}^{\alpha}$ may be represented as the matrices
$-i\Gamma^{\alpha}$, the equations are the same.
Two equivalent initial states under the same evolution remain
equivalent throughout all time. Thus,
both parts of the beam in the Pin (3,1) case
evolve exactly the same way as their respective counterparts in the Pin
(1,3) case, the interference patterns produced in either case are
identical. The
same will hold true for any configuration
in which the matter field is required to change continuously.
\ \\

\begin{figure}[h]
   \begin{center}
       \vspace{3mm}
       \resizebox{5cm}{!}{\includegraphics{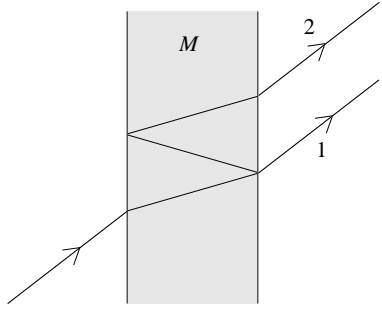}}
   \caption{A reflection experiment intended to find a difference
      between two types of pinors.}
   \label{fig:medium}
     \vspace{3mm}
   \end{center}
\end{figure}

\begin{center}
{\em Reflection\/}
\end{center}
Since the previous   experiment turned out not to give a
parity transformation, EK went on to study
the interference between two parts of a fermion beam passing through some
medium $M$ as shown in fig. \ref{fig:medium}.
The part which follows path 1 is transmitted
directly through the medium, whereas the part which follows path 2 is
reflected
twice before passing through.
Such an experiment could be
achieved by passing neutrons through a magnetic crystal.
\ \\ \ \\
Although the idea of reflection from a surface
might bring to mind the idea of parity
transformation, the reflections involved here have nothing to do with the
parity
transformation $\Lambda_{P}$ (or $\hat{\Lambda}_{P}$).   In other words,
although
the neutrons following path 2 are reflected twice, they do not undergo any
parity
transformations as called for in fig. \ref{fig:boxes}.
Hence this setup also fails to realize
fig. \ref{fig:boxes}.
\begin{figure}[h]
   \begin{center}
       \vspace{3mm}
       \resizebox{5cm}{!}{\includegraphics{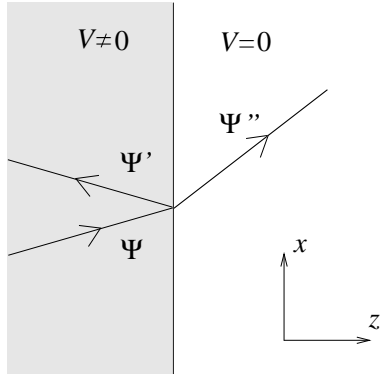}}
   \caption{Reflection and transmission of a pinor particle at an
          interface.}
   \label{fig:interface}
     \vspace{3mm}
   \end{center}
\end{figure}
That reflection at a boundary does not produce any parity transformation is
best
seen by just solving the problem of reflection and transmission of a plane
wave
between two media, one with a free-particle Dirac Hamiltonian and one with a
Hamiltonian consisting of the free-particle part plus a potential $V$,
as shown schematically in
fig. \ref{fig:interface}.
The solution of this problem can be found in several books
\cite{Flugge}. It is done by matching coefficients of solutions and
will not bring in any parity transformations; again the requirement of
continuity is the stumbling block.
\ \\ \ \\
Thus reflection at a boundary cannot be used to distinguish between
particles of different Pin groups either.
\ \\ \ \\
\subsection{Time reversal and Kramers' degeneracy}
\label{sec:timerev}
In quantum mechanics, if one requires the Hamiltonian
to be invariant under time reversal,
then the time reversal operator is antiunitary.
Indeed the Hamiltonian $H$ generates time evolution:
\[
H|\, t\, \rangle = i\partial_t \, |\, t \, \rangle \; ;
\]
$H$ is invariant under time reversal if there exists an antiunitary
operator $A_T$ such that
\[
A_T H^* A_T^{-1} = H
\]
with the asterisk denoting complex conjugate; then $A_T|\,
t\, \rangle^*$
satisfies the time reversed equation.
\ \\ \ \\
Kramers has shown (see \eg\ \cite[p.\ 81]{Wein} 
\cite[p.\ 281]{Sakurai} or \cite[p.\ 408]{Merz};
see also \cite[p.\ 601]{Avron}) that, given an {\em antiunitary}
time reversal operator $A_T$ defined as in eq.\ (\ref{eq:AT}) with
the matrix ${\mathcal A}_T$ satisfying
\[
{\mathcal A}_T {\mathcal A}_T^*=-\id
\]
and such that it commutes with the Hamiltonian $H$ of the system, an
eigenstate $(|n\rangle)$ of $H$ and the time reversed eigenstate
$A_T(|n\rangle^{*})$ are two different states with the same energy.
This degeneracy can be removed by adding an interaction which does not
commute with $A_T$.
\ \\ \ \\
Can Kramers' theorem provide a method for distinguishing
$\Psi$-particles (with $\Lambda^2_T=\id$) from
$\hat\Psi$-particles (with $\hat\Lambda^2_T=-\id$)?
We have established the relationship
between the time reversal operator ${\mathcal A}_T$ and the unitary
time reversal operator $\Lambda_T$ in section \ref{sec:Pin13}, namely
\[
{\mathcal A}_T {\mathcal C}^{-1} = \Lambda_T \; .
\]
Kramers' theorem cannot be used to distinguish the two \pin\ groups
(even though $\Lambda_T^2 = \id$ and $\hat{\Lambda}_T^2 = -\id$):
\ \\ \ \\
Let the operators ${\mathcal A}_T$ and $\hat{{\mathcal
A}}_T$ correspond to $\Lambda_T$ and $\hat{\Lambda}_T$, respectively.
\[
\Lambda_T = {\mathcal A}_T {\mathcal C}^{-1} \;
\qquad
\hat{\Lambda}_T = \hat{{\mathcal A}}_T \hat{\mathcal C}^{-1}
\]
where
\[
{\mathcal C} \Gamma^*_{\alpha} {\mathcal C}^{-1} = -
\Gamma_{\alpha} \quad \mbox{and} \quad
\hat{\mathcal C} \hat{\Gamma}^*_{\alpha} \hat{\mathcal C}^{-1} =
\hat{\Gamma}_{\alpha} \; .
\]
A double time reversal is not produced by ${\mathcal A}_T^2$ but by
${\mathcal A}_T {\mathcal A}^*_T$ since, according to the defining
equation $\psi'(Tx)={\mathcal A}_T \psi^*(x)$, and the field operator
$\Psi$ transforms according to equation (\ref{eq:AT}) as
\[
A_T A_T \Psi(x) A_T^{-1} A_T^{-1} = \zeta {\mathcal A}_T
(\zeta {\mathcal A}_T \Psi^*(x))^* = {\mathcal A}_T {\mathcal A}_T^*
\Psi(x) \; .
\]
Whereas $\Lambda^2_T \neq \hat{\Lambda}^2_T$, the double antiunitary
time reversal shows no such difference:
\[
{\mathcal A}_T{\mathcal A}_T^*=
\hat{\mathcal A}_T \hat{\mathcal A}_T^* = -\id \; .
\]
Both $\Psi$- and $\hat{\Psi}$-particles can be used to construct
degenerate time-reversed pairs. It follows that Kramer's degeneracy
cannot be used to distinguish $\Psi$-particles and $\hat{\Psi}$-particles.
\newpage

\subsection{Charge Conjugation; Positronium; Neutrinoless Double Beta Decay}
\label{sec:chargeconj}
\begin{figure}[h]
   \begin{center}
       \vspace{3mm}
       \resizebox{7cm}{!}{\includegraphics{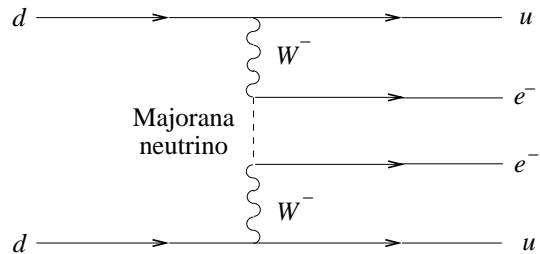}}
       \vspace{1mm}
   \caption{Quark diagram for neutrinoless double beta decay
($2n\to 2p+2e^-$ in ${\rm^{82}Se}\to{\rm^{82}Kr}+2e^-$).}
   \label{fig:quark1}
   \end{center}
\end{figure}

\begin{figure}[h]
   \begin{center}
       \vspace{1mm}
       \resizebox{7cm}{!}{\includegraphics{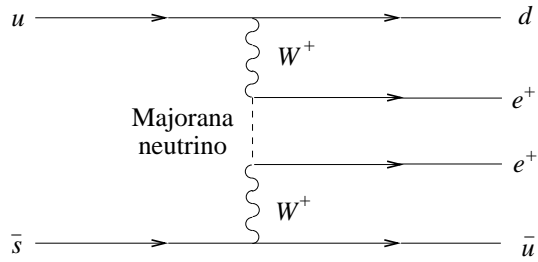}}
       \vspace{1mm}
   \caption{We note also another type of neutrinoless double beta
       decay: Quark diagram for $K^+\to\pi^-+2e^+$.}
   \label{fig:quark2}
  \end{center}
\end{figure}

\begin{center}
{\em Positronium\/}
\end{center}
Once more, positronium proves to be a useful test bed for discrete
symmetries. In Appendix \ref{app:coll}
we briefly review from our perspective
the textbook example of how charge conjugation decides
the lifetimes of two different positronium states.
\label{page:posit2}
\ \\
\begin{center}
{\em Neutrinoless double beta decay\/}
\end{center}
Double beta decay occurs usually with the emission of two neutrinos.
However, if the neutrino associated with a beta decay is reabsorbed to
produce a second beta decay, then no neutrino is emitted, and the
process is called neutrinoless double beta decay.   Diagrams of neutrinoless
double beta decay in two different reactions are given in fig.
\ref{fig:quark1} and fig. \ref{fig:quark2}.
If the same neutrino is emitted and absorbed, it has to be a particle
identical with its antiparticle, \ie\ 
{\em it has to be a Majorana
particle}. (Recall that a Majorana particle is one for which
$\psi^{\rm c}=\psi$).
The possible observation of neutrinoless double beta decay
has been analyzed by Klapdor-Kleingrothaus
\cite{Klapdor}.
\ \\ \ \\
The discussion in section \ref{sec:pinrepq} indicates that
a Majorana particle can only be a \pin(3,1) particle. Thus
the existence of Majorana neutrinos, if confirmed, would have some
implication for the topology of the universe, namely that the universe
is a manifold which can serve as a base for a \pin(3,1)-bundle.

\newpage

\section{The Pin group in $s$ space, $t$ time dimensions}

In order to analyze the properties of the two Pin groups in arbitrary
dimensions, we first review and simplify
a few topics, treated in detail in
\cite{Yellow,Blue} and \cite{CDGwo}. This section is
organized as follows.
\ \\
\begin{enumerate}
\item[\ref{sec:difference}]
The difference between $s+t$ even and $s+t$ odd
\item[\ref{sec:chirality}]
Chirality
\item[\ref{sec:mod8}]
Construction of the gamma matrices. Periodicity modulo 8.
\item[\ref{sec:groups}]
Conjugate and complex gamma matrices.
\item[\ref{sec:sequence}]
The short exact sequence $\id
\rightarrow \spin(t,s) \rightarrow \pin(t,s) \rightarrow
{\mathbb Z}_2 \rightarrow 0$
\item[\ref{sec:susy}]
Grassman (superclassical) pinor fields
\item[\ref{sec:strings}]
String theory and spin structures
\end{enumerate}
\ \\
In this section we will use the $P(1)$ parity transformation, which
reverses only one axis instead of three.
\newpage

\subsection{The difference between $s+t$ even and $s+t$ odd}
\label{sec:difference}
In this section $s+t=d=2p$ and $s'+t'=d+1=2p+1$.
In brief:
\begin{itemize}
\item[] $t+s=2p=d$
\begin{itemize}
\item[]
\begin{itemize}
   \item[$\bullet$]
   Only one irreducible faithful representation of the gamma matrices
   \item[$\bullet$]
   The center of Pin$(t,s)$ is ${\mathbb R}^+_{\times}\id$
   (multiples of the unit element)
\end{itemize}
\end{itemize}
\vspace{3mm}
\item[] $t'+s'=2p+1=d+1$
\begin{itemize}
\item[]
\begin{itemize}
   \item[$\bullet$]
   Two inequivalent irreducible faithful representations of the gamma matrices
   \item[$\bullet$]
   The center of Pin$(t',s')$ is ${\mathbb R}^+_{\times}\id \cup 
      {\mathbb R}^+_{\times}
     \Gamma_{d+2}$
   \item[$\bullet$]
   The map $\pin(t',s') \rightarrow O(t',s')$ is not surjective
\end{itemize}
\end{itemize}
\end{itemize}
A key element in the proofs of some of the above statements is the
construction of the generators for $\pin(t',s')$ given the generators
$\{\Gamma_{\alpha}\}$ for $\pin(t,s)$.
\ \\ \ \\
We begin with
$s+t=d=2p$. The algebra over the reals generated by the (possibly
complex) $2^p \times 2^p$ matrices ($\Gamma_{\alpha}$) is a faithful
representation of the Clifford algebra ${\mathcal C}(t,s)$. This
representation is unique, modulo similarity transformations, and irreducible.
\ \\ \ \\
The center (set of elements which commute with all elements) of
$\pin(t,s)$ is ${\mathbb R}^+_{\times}\id$.
\ \\ \ \\
We now consider the case
$s'+t'=d+1=2p+1$, with either $s'=s+1$ or $t'=t+1$.
Set
\be
 \Gamma_{d+1}:= \Gamma_1 \Gamma_2 ... \Gamma_d
\qquad \mbox{ and similarly for $\hat{\Gamma}_{d+1}$}
\qquad \mbox{($d$ even)}.
\ee
We note
\be
   (\Gamma_{d+1})^2=(-1)^{s+p} \id_{2^p} \; , \qquad
 (\hat{\Gamma}_{d+1})^2=(-1)^{t+p} \id_{2^p} \; , \qquad d=2p.
\ee
We also note that $\Gamma_{d+1}$ anticommutes with all $\Gamma_{\alpha} \in
\pin(t,s)$, and that a similar statement holds for $\hat{\Gamma}_{d+1}$.
Therefore we can use $k\Gamma_{d+1}$, where $k$ is a phase,
to construct a basis for
$\pin(t+1,s)$ and for $\pin(t,s+1)$ as follows; similar construction
of $\pin(s+1,t)$ and of $\pin(s,t+1)$
can be done using $\hat{\Gamma}_{d+1}$.
\ \\ \ \\
For
\be
k^2=(-1)^{s+p}, \qquad (k\Gamma_{d+1})^2=\id
\ee
the set of anticommuting matrices $\{ \Gamma_{\alpha},
k\Gamma_{d+1}\}$ generates $\pin(t+1,s)$. The two choices
$k=\pm(-1)^{(s+p)/2}$ provide two inequivalent representations of
the group $\pin(t+1,s)$.
\ \\ \ \\
For
\be
k^2=(-1)^{s+p+1}, \qquad (k\Gamma_{d+1})^2=-\id
\ee
the set of anticommuting matrices   $\{ \Gamma_{\alpha},
k\Gamma_{d+1}\}$ generates $\pin(t,s+1)$. The two choices
$k=\pm i(-1)^{(s+p)/2}$ provide two inequivalent representations of
$\pin(t,s+1)$. Similar results hold for $\pin(s+1,t)$ and $\pin(s,t+1)$.
\ \\ \ \\
For $s'+t'$ odd,
the center of $\pin(t',s')$ consists of ${\mathbb R}^+_{\times} \id$
and ${\mathbb R}^+_{\times}\Gamma_{d+2}$.
Indeed the product
\be
\Gamma_{d+2}=
\Gamma_1 \Gamma_2 ... \Gamma_d \Gamma_{d+1}
\qquad \mbox{with} \qquad   \Gamma_{d+1}=\Gamma_1
\Gamma_2 ... \Gamma_d
\ee
commutes with all elements in $\{\Gamma_{\alpha}, \Gamma_{d+1}\}$.
\ \\ \ \\
For $s'+t'$ odd, the map $\pin(t',s') \rightarrow O(t',s')$ is not
surjective. Namely,
there is no element in $\pin(t',s')$ which maps into
the element of $O(t',s')$ which reverses the axes. Indeed let
$(PT)=\diag(-1,...-1)$, then there is no $\Lambda_{PT}$
satisfying
\be
\Lambda_{PT} \Gamma_{\alpha} \Lambda_{PT}^{-1} =
\Gamma_{\beta} {(PT)^{\beta}}_{\alpha} =
- \Gamma_{\alpha} \; ,
\qquad \mbox{for all $\Gamma_{\alpha}$}
\ee
since this would imply
\[
\Lambda_{PT} \Gamma_{d+2} +
\Gamma_{d+2}\Lambda_{PT} =0 \; ,
\]
but $\Gamma_{d+2}$ commutes with all elements in
$\pin(t',s')$, and $\Lambda_{PT} \Gamma_{d+2} \neq 0$
since $\Lambda_{PT}$ and $\Gamma_{d+2}$ are invertible.
\begin{center}
{\em The twisted map}
\end{center}
To eliminate some of the differences between $d$ odd and $d$ even, one
can introduce a map, sometimes called the twisted map,
\[
\tilde{\mathcal H}: \pin(t,s) \rightarrow O(t,s) \; ,
\]
surjective in all dimensions, as follows.
The Clifford algebra is a graded algebra
\be
{\mathcal C}(t,s)=
{\mathcal C}_+(t,s)+
{\mathcal C}_-(t,s)
\ee
where ${\mathcal C}_+$ is generated by even products of
elements of the basis, and ${\mathcal C}_-$ is generated by odd
products. Let
\be
\begin{array}{cccl}
\alpha(\Lambda_+)&=&\Lambda_+ & \qquad \mbox{ for }\Lambda_+ \in
{\mathcal C}_+(t,s) \\
\alpha(\Lambda_-)&=&-\Lambda_- & \qquad \mbox{ for }\Lambda_- \in
{\mathcal C}_-(t,s)
\end{array}
\ee
The map $\tilde{\mathcal H}$,
defined by
\be
\alpha(\Lambda_L)\Gamma_{\alpha} \Lambda_L^{-1}=
\Gamma_{\beta} {L^{\beta}}_{\alpha}
\label{eq:twistedmap}
\ee
is surjective in all dimensions.
\ \\ \ \\
We note also that $\tilde{{\mathcal H}}(\Gamma_{\alpha})$ reverses the
$\alpha$-axis. The twisted map $\tilde{\mathcal H}$ seems desirable,
but eq.\ (\ref{eq:twistedmap}) is not a similarity transformation and
the invariance of the Dirac equation under Lorentz transformations
requires the similarity transformation $\Lambda_L \Gamma_{\alpha}
\Lambda^{-1} = \Gamma_{\beta} {L^{\beta}}_{\alpha}$. Attempts to find
maps \cite{Karoubi}
\[
\rho:\pin(t,s) \rightarrow \pin(t,s)
\]
to recover a similarity transformation, \ie\ $\rho$ such that
\[
\alpha(\Lambda_L)\Gamma_{\alpha} \Lambda^{-1}_L=
\rho(\Lambda)
{\Gamma_{\alpha}} (\rho(\Lambda))^{-1} \; ,
\label{eq:twist}
\]
obviously fail in odd dimensions, and are very awkward in even
dimensions. We shall not work with the twisted map.

\subsection{Chirality}
\label{sec:chirality}
In this paragraph
the matrix $\Gamma_{d+1}=\Gamma_1\Gamma_2 ... \Gamma_d$ is used to
define chirality for $d$ even.

$\Gamma_{d+1}$ is a linear
operator on the space $S$ of spinors. We recall
\be
(\Gamma_{d+1})^2=(-1)^{s+p} \id_{2^p} \; , \qquad
(\hat{\Gamma}_{d+1})^2=(-1)^{t+p} \id_{2^p} \; .
\label{eq:gamma2}
\ee
The eigenvalue equation
\[
\Gamma_{d+1}\psi= \alpha \psi
\]
implies
\[
(\Gamma_{d+1})^2\psi=\alpha^2 \psi = (-1)^{s+p} \psi
\qquad \mbox{ by (\ref{eq:gamma2}) }
\]
hence
\[
\alpha^2=(-1)^{s+p} \; .
\]
Thus
\bea
\alpha &=& \pm 1 \qquad \mbox{ if $s+p$ is even} \\
\alpha &=& \pm i \qquad \mbox{ if $s+p$ is odd (as in section
\ref{sec:pinreps})}
\eea
One denotes by $S_+$ the eigenspace with eigenvalue 1 or $i$, and by
$S_-$ the eigenspace with eigenvalue $-1$ or $-i$.
\[
S=S_+ \oplus S_-
\]
Therefore the projection matrices
\be
{\mathcal P}_{\pm}=
\left\{
\ba{ll}
\half (\id_{2^p} \pm \Gamma_{d+1})\; , &   \mbox{$s+p$ even} \\
\half (\id_{2^p} \pm i\Gamma_{d+1}) \; ,   & \mbox{$s+p$ odd}
\ea
\right.
\ee
project a $2^p$-component pinor into two $2^{p-1}$-component spinors.
\ \\ \ \\
$\Gamma_{d+1}$ for $s+p$ even, and $i\Gamma_{d+1}$ for $s+p$ odd, are
called {\em chirality operators\/}.
\ \\ \ \\
A chiral basis is a basis adapted to the
splitting $S=S_+ \oplus S_-$.
In a chiral basis
\[
\Gamma_{d+1}=
\left(
\begin{array}{cc}
-\id_{2^{p-1}} & 0 \\
0 & \id_{2^{p-1}}
\end{array}
\right) \; .
\]
Polchinski \cite[app.\ B]{Polchinski} gives a recursion construction of a
chiral basis in terms of the Pauli matrices $\{\sigma_i\}$.
\begin{rem}
For $s+t$ even,
 $\Gamma_{\alpha}\Gamma_{d+1}=-\Gamma_{\alpha}\Gamma_{d+1}$, hence
the matrix $\Gamma_{d+1}$ is a solution $\Lambda_L$ of
\[
\Lambda_L \Gamma_{\alpha} \Lambda_L^{-1} = -\Gamma_{\alpha}
\]
which implies $(L)=\diag(-1,...,-1)$. Since the dimension of
space-time is even, this Lorentz transformation does not change the
handedness of the system of coordinates.
\end{rem}

\subsection{Construction of gamma matrices.
Periodicity modulo 8. }
\label{sec:mod8}
Two useful mathematical references are
the books by Gilbert and Murray \cite{Gilbert} and by
Porteous \cite{Porteous}.
\ \\ \ \\
The study of $\pin(t,s)$ for arbitrary $t$ and $s$ is considerably
simplified by the fact that the groups depend not on $s$ and $t$ but
on $|s-t|$ modulo 8. Moreover, $\pin(t,s)$ and $\pin(s,t)$ are
isomorphic for $|s-t|=0$ modulo 4.
\ \\ \ \\
The proof uses the isomorphism
\be
M_4({\mathbb R}) \simeq {\Hset} \otimes_{\mathbb R} {\Hset}
\label{eq:isoMH}
\ee
where $M_4({\mathbb R})$ is the algebra of real $4\times 4$ matrices
and ${\Hset}$ the quaternion algebra.
This isomorphism is not
trivial. It has been questioned on the grounds that the left hand side
is real and the right hand side seems to be complex, given that a well
known two-dimensional representation of the quaternion basis consists
of the matrix $\id_2$
together with $i$ times the Pauli matrices which cannot be all
imaginary.
This argument for the complexity of the quaternion algebra is
obviously meaningless since complex representations of real algebras
abound, as can be seen, for instance, in this paper.
We prove the isomorphism in appendix
\ref{app:iso} because it is not easily available
to non-specialists.
\ \\ \ \\
In paragraph \ref{sec:difference}, we constructed $\pin(t+1,s)$ and
$\pin(t,s+1)$ by adding $k\Gamma_{d+1}$, with different values of
$k^2$, to the basis of $\pin(t,s)$ with $t+s=d=2p$.
Now we combine $t+s=2p$ and
$t'+s'=d'$ arbitrary. By tensoring the Clifford algebras ${\mathcal
   C}(t,s)$ and ${\mathcal C}(t',s')$ one can obtain either one of
the two Clifford algebras
${\mathcal C}(t+t',s+s')$ or ${\mathcal C}(t+s',s+t')$, depending on
the sign of $k^2$ in $k\Gamma_{d+1}$.
\begin{eqnarray}
   {\mathcal C}(t,s) \otimes_k {\mathcal C}(t',s') &=&
   {\mathcal C}(t+t',s+s') \qquad \mbox{for } k^2=1
   \label{eq:k21} \\
   {\mathcal C}(t,s) \otimes_k {\mathcal C}(t',s') &=&
   {\mathcal C}(t+s',s+t') \qquad \mbox{for } k^2=-1
   \label{eq:k2m1}
\end{eqnarray}
Let $\{\id, \Gamma_{\alpha}\}$ be a basis of ${\mathcal C}(t,s)$ and
$\{\id', \Gamma_{\alpha'}\}$ be a basis of ${\mathcal C}(t',s')$, then
the $d+d'$ elements $\{\Gamma_{\alpha} \otimes \id',
k\Gamma_{d+1} \otimes \Gamma_{\alpha'}\}$ form a basis for their tensor
product.
Henceforth we abbreviate $\otimes_k$ to $\otimes$.
\begin{pf}
Since $\Gamma_{\alpha}$ anticommutes with
$\Gamma_{d+1}$, the elements in this basis anticommute
pairwise. Their squares are
\begin{eqnarray*}
   (\Gamma_{\alpha} \otimes \id')^2 &=&
   (\Gamma_{\alpha})^2 \otimes \id' =
   (\id \otimes \id')\eta_{\alpha \alpha} \\
   (k\Gamma_{d+1} \otimes \Gamma_{\alpha'})^2 &=&
   k^2(\id \otimes \id')\eta_{\alpha' \alpha'}
   \; ;
\end{eqnarray*}
the sign of $k^2$ determines the combination
$t+t'$ or $t+s'$ in the tensor product
${\mathcal C}(t,s) \otimes {\mathcal C}(t',s')$.
\qed
\end{pf}
\ \\ \ \\
In order to prove the periodicity of the Clifford algebra modulo 8, we
prove
\begin{eqnarray}
{\mathcal C}(0,s+8) &\simeq& M_{16}(\Rset) \otimes {\mathcal C}(0,s)
 \label{eq:period1} \\
{\mathcal C}(t,s) &\simeq& M_{2^t}(\Rset) \otimes {\mathcal C}(0,s-t)
\qquad \mbox{for $s>t$}
 \label{eq:period2}
\end{eqnarray}
When tensoring Clifford algebras we can use either (\ref{eq:k21}) or
(\ref{eq:k2m1}). Using (\ref{eq:k21}) is easier but using
(\ref{eq:k2m1}) brings out interesting results. In brief, if we use
(\ref{eq:k21})
\bea
{\mathcal C}(0,s+8) &\simeq& {\mathcal C}(0,2) \otimes {\mathcal C}(0,s+6) \\
     &\simeq& \left(\otimes {\mathcal C}(0,2)\right)^4 \otimes {\mathcal
C}(0,s) \\
   &\simeq& M_{16}(\Rset) \otimes {\mathcal C}(0,s) \qquad \qquad
\mbox{since ${\mathcal C}(0,2) \simeq M_2(\Rset)$}
\eea
But if we use (\ref{eq:k2m1}) we bring out the quaternionic algebras
since ${\mathcal C}(2,0)$ is isomorphic to ${\Hset}$; we have then
enough information to construct the classification table. Using
(\ref{eq:k2m1}) we obtain
\bea
{\mathcal C}(0,s+8) &\simeq& {\mathcal C}(0,2) \otimes {\mathcal C}(s+6,0) \\
     &\simeq& {\mathcal C}(0,2) \otimes {\Hset} \otimes {\mathcal
C}(0,s+4) \simeq \ldots \\
   &\simeq& M_{2}(\Rset) \otimes
{\Hset} \otimes M_{2}(\Rset)\otimes{\Hset}
\otimes {\mathcal C}(0,s)
\eea
The proof of (\ref{eq:period2}) is analogous.
With $k^2=1$,
\bea
{\mathcal C}(t,s) &\simeq& (\otimes{\mathcal C}(1,1))^t
\otimes {\mathcal C}(0,s-t) \\
&=& M_{2^t}(\Rset) \otimes {\mathcal C}(0,s-t) \; .
\eea
\begin{exmp}
Constructing gamma matrices from Pauli matrices.
\[
   \sigma_1 =
   \left( \begin{array}{cc} \; 0 \; & \; 1 \; \\
       1 & 0 \end{array} \right)
   \; , \quad
 \sigma_2 =
     \left( \begin{array}{cc} \; 0 \; & -i \\
         i & 0 \end{array} \right)
   \; , \quad
 \sigma_3 =
     \left( \begin{array}{cc} \; 1 \; & 0 \\   
           0 & -1 \end{array} \right)
     \; , \quad
\]
\[
\sigma_j \sigma_k = i \epsilon_{jkl} \sigma_l
\; , \qquad
(\sigma_i)^2 = \id_2
\; , \qquad
\sigma_1\sigma_2\sigma_3=i\id_2 \; .
\]
In dimensions 2 and 3, the gamma matrices are $2 \times 2$ matrices
and we can write down table \ref{table:gamma}. Let a matrix in
${\mathcal C}$ be written $M=m^0\id + m^i \Gamma_i +
m^{ij}\Gamma_i\Gamma_j + m\Gamma_1\Gamma_2 \cdots \Gamma_d $ ;
$M$ is an element of a real vector space; the isomorphism in the last
column is dictated by the properties of the vector space basis.
\begin{table}[h]
\[
\begin{array}{|c|c|c|c|c|}\hline
\mbox{Clifford alg.} &
\mbox{alg. generators} &
\mbox{vector space basis} &
\, {\rm dim}_{\Rset} &
\mbox{isomorphism} \\ \hline\hline
{\mathcal C}(0,1) & 1, i & 1, i & 2 & \Cset \\ \hline
{\mathcal C}(1,0) & 1, 1' & 1, 1' & 2 & \Rset \oplus \Rset \\ \hline
\hline
{\mathcal C}(0,2) & \id_2, i\sigma_1,
i\sigma_3   &
\id_2, i\sigma_1, i\sigma_2, i\sigma_3 & 4 & \Hset \\ \hline
{\mathcal C}(1,1) & \id_2, (\sigma_1 \mbox{ or } \sigma_3), i\sigma_2 &
\id_2, \sigma_1, i\sigma_2, \sigma_3 & 4 & M_2(\Rset) \\ \hline
{\mathcal C}(2,0) & \id_2, \sigma_1, \sigma_3 &
\id_2, \sigma_1, i\sigma_2, \sigma_3 & 4 & M_2(\Rset) \\ \hline
\hline
{\mathcal C}(0,3) & \id_2, i\sigma_1, i\sigma_2, i\sigma_3
& \id_2 + 7 \mbox{ matrices}& 8 &
\Hset \oplus \Hset
 \\ \hline
{\mathcal C}(1,2) & \id_2, \sigma_2, i\sigma_1, i\sigma_3 &
\id_2 + 7 \mbox{ matrices}   & 8 & \Hset \otimes \Cset \\ \hline
{\mathcal C}(2,1) & \id_2,   \sigma_1, \sigma_3, i\sigma_2&
\id_2 + 7 \mbox{ matrices} &8 &
M_2(\Rset) \oplus M_2(\Rset)   \\ \hline
{\mathcal C}(3,0) & \id_2, \sigma_1, \sigma_2, \sigma_3 &
\id_2 + 7 \mbox{ matrices}   & 8 &
M_2(\Cset)   \\ \hline
\end{array}
\]
\caption{Gamma matrices in 1,2 and 3 dimensions.}
\label{table:gamma}
\end{table}
In dimensions higher than 3, equations (\ref{eq:k21}) and
(\ref{eq:k2m1}) can be used for constructing gamma matrices. We work
out ${\mathcal C}(1,3)$ and ${\mathcal C}(3,1)$ explicitly:
\begin{itemize}
\item[]
${\mathcal C}(1,3) \simeq {\mathcal C}(1,1) \otimes {\mathcal C}(0,2)$
for $k^2=1$, $k=\pm 1$
   \begin{itemize}
   \item[]
   the algebra generators are:
   $\sigma_1 \otimes \id_2$, $i\sigma_2 \otimes \id_2,
   -k\sigma_3 \otimes (i\sigma_1)$, $-k\sigma_3 \otimes (i\sigma_2)$ \\
   consist of 3 real matrices, and 1 imaginary one.
   \end{itemize}
\item[]
${\mathcal C}(1,3) \simeq {\mathcal C}(1,1) \otimes {\mathcal C}(2,0)$
for $k^2=-1$, $k=\pm i$
   \begin{itemize}
   \item[]
   the algebra generators are:
   $\sigma_1 \otimes \id_2$, $i\sigma_2 \otimes \id_2,
   -k\sigma_3 \otimes \sigma_1$, $-k\sigma_3 \otimes \sigma_3$ \\
   consist of 2 real matrices, and 2 imaginary ones.
   \end{itemize}
\item[]
${\mathcal C}(3,1) \simeq {\mathcal C}(1,1) \otimes {\mathcal C}(2,0)$
for $k^2=1$, $k=\pm 1$
   \begin{itemize}
   \item[]
       Changing the value of $k$ in the previous basis yields 4 real
       matrices. This is a Majorana representation.
   \end{itemize}
\item[]
${\mathcal C}(3,1) \simeq {\mathcal C}(1,1) \otimes {\mathcal C}(0,2)$
for $k^2=-1$, $k=\pm i$
   \begin{itemize}
   \item[]
     Changing the value of $k$ in the first basis yields 3 real
matrices, and 1 imaginary one.
   \end{itemize}
\end{itemize}
${\mathcal C}(1,3)$ does not admit a real representation; ${\mathcal
C}(3,1)$ does admit a real representation.
\end{exmp}
\ \\
In section \ref{sec:pingroup}, the label $t$ for time is equal to 1
and the label $s$ for space is equal to 3, therefore ${\mathcal
C}(t,s)$ signals at a glance a metric of signature $(+,-,-,-)$ and
${\mathcal C}(s,t)$ a metric of signature $(+,+,+,-)$.
Here $t$ and $s$ are arbitrary, and we shall use $(m,n)$ rather than
$(t,s)$\footnote{${\mathcal C}(n,m)$ in ref. \cite{Blue} is
${\mathcal C}(m,n)$ in this report.}.
Table \ref{table:iso}
lists the algebra isomorphisms
of ${\mathcal C}(m,n)$ with $d=m+n$ for all possible values of $(m-n)$
mod 8.
The vector space $M_k(\Rset)$ of $k \times k$ real matrices is
abbreviated to $\Rset(k)$ and the space of $k \times k$ quaternionic
matrices (the matrix elements are quaternions) is denoted
$\Hset(k)$. For example the isomorphism
\[
M_4(\Rset) \simeq \Hset \otimes \Hset
\quad \mbox{ is abbreviated } \quad \Rset(4) \simeq \Hset(2) \; .
\]
\begin{table}[h]
\[
\begin{array}{|c|c|c|c|c|}\hline
(m-n) \, {\rm mod}\,   8 & 0 & 1 & 2 & 3 \\ \hline
{\mathcal C}(m,n) & \Rset(2^{d/2}) & \Rset(2^{(d-1)/2})
\oplus \Rset(2^{(d-1)/2}) &   \Rset(2^{d/2}) &
\Cset(2^{(d-1)/2}) \\ \hline \hline
(m-n) \, {\rm mod} \, 8 & 4 & 5 & 6 & 7 \\ \hline
{\mathcal C}(m,n) & {\Hset}(2^{d/2-1}) & {\Hset}(2^{(d-1)/2-1})
\oplus {\Hset}(2^{(d-1)/2-1}) &   {\Hset}(2^{d/2-1}) &
\Cset(2^{(d-1)/2}) \\ \hline
\end{array}
\]
\caption{Algebra isomorphisms}
\label{table:iso}
\end{table}
This table is valid for both $m-n>0$ and $m-n<0$ since a negative
number modulo 8 is equal to a positive number.
\ \\ \ \\
Given $\hat{\Gamma} \in {\mathcal C}(m,n)$, $m>n$, and
$\Gamma \in {\mathcal C}(m,n)$, $m<n$, one can use
$\hat{\Gamma}=i\Gamma$ for verifying, for instance,
${\mathcal C}(0,5)$ given ${\mathcal C}(5,0)=\Hset(2) \oplus \Hset(2)$;
one finds ${\mathcal C}(0,5)=\Hset(2) \oplus i\Hset(2)=\Hset(2)
\otimes \Cset$.
Now $\Hset \otimes \Cset \simeq \Cset(2)$, so
${\mathcal C}(0,5)\simeq \Cset(4)$ which is correct
for $-5=3$ mod 8.
\ \\ \ \\
From table \ref{table:iso} we conclude the following:
\begin{itemize}
\item
For $d$ even, the vector space isomorphisms are either with vector
spaces of real matrices, or vector spaces of quaternionic matrices.
\item
For $n=1$, ${\mathcal C}(m,1)$ admits a real representation for
\bea
m &=& 1,2,3 \mbox{ mod } 8 \\
{\rm i.e.} \quad   d &=& 2,3,4,10,11,12 \mbox{ etc.}
\eea
\item
If ${\mathcal C}(m,1)$ admits a real representation, ${\mathcal C}(1,m)$
admits a purely imaginary one.
\item
${\mathcal C}(m,n)$ and ${\mathcal C}(n,m)$ are not isomorphic
unless $m-n=0$ mod 4.
\end{itemize}
Another technique for identifying the dimensions which admit real
representations consists in assuming all the $\Gamma_{\alpha}$'s real,
and seeing if it leads to a contradiction. For instance, let $d=4$, and
assume $\Gamma_1$, $\Gamma_2$, $\Gamma_3$, $\Gamma_4$ to be
real, with $(\Gamma_j)^2=\id$, $j\in \{1,2,3\}$
and $(\Gamma_4)^2=-\id$. The $\Gamma_j$ are symmetric, and $\Gamma_4$
is antisymmetric. The algebra generated by the $\Gamma_{\alpha}$'s
consists of
\begin{itemize}
\item[]
10 symmetric matrices: $\id$, $\Gamma_j$, $\Gamma_4\Gamma_j$,
$\Gamma_4\Gamma_j\Gamma_k$
\item[]
6 antisymmetric matrices: $\Gamma_4$, $\Gamma_j\Gamma_k$,
$\Gamma_j\Gamma_k\Gamma_l$, $\Gamma_1\Gamma_2\Gamma_3\Gamma_4$
\end{itemize}
These 16 matrices make a basis for $M_4(\Rset)$. There is no
contradiction in having assumed the $\Gamma_{\alpha}$'s to be real.
\ \\

\begin{center}
{\em Onsager construction of gamma matrices\/}
\end{center}
In the proof of (\ref{eq:period1}) and (\ref{eq:period2}) we have
given a construction for a basis of ${\mathcal C}(t+t', s+s')$ and
${\mathcal C}(t+s', s+t')$ given a basis for ${\mathcal C}(t,s)$ and
${\mathcal C}(t',s')$.
It is worth mentioning another construction using the Onsager solution
of the Ising model \cite{Onsager}; the explicit representation using
this construction can be found in \cite{CDGwo}. In particular, one sees
by inspection which matrices are real, and which are imaginary both
for $d$ even and $d$ odd.
\ \\

\begin{center}
{\em Majorana pinors, Weyl-Majorana spinors\/}
\end{center}
A pinor is said to be Majorana if it is real (or purely imaginary). 
If a space $S$ of
Majorana pinors is of even dimension, it can be split into two
eigenspaces of a chirality operator
\[
S=S_+ \oplus S_- \; ,
\]
then each eigenspace is a space of Weyl-Majorana spinors.
\ \\ \ \\
Majorana pinors have been used to avoid confusion in
charge conjugation. When a real representation is not available, one
replaces a $d$-dimensional complex pinor $\psi=\psi_1+i\psi_2$
by a $2d$-dimensional real
pinor
\[
\psi = \left(
\begin{array}{c}
\psi_1 \\
\psi_2
\end{array}
\right)
\qquad \mbox{and} \qquad
\psi^* = \left(
\begin{array}{cc}
\; \id \; & 0 \\
0 & -\id
\end{array}
\right)
\psi \; .
\]
The $2d$-dimensional representation is reducible.

\subsection{Conjugate and complex gamma matrices}
\label{sec:groups}
The set of hermitian conjugate matrices $\{\pm
\Gamma_{\alpha}^{\dagger}\}$, the set of inverse matrices $\{ \pm
\Gamma_{\alpha}^{-1} \}$ and the set of complex conjugate matrices $\{
\Gamma_{\alpha}^*\}$ obey the same algebra as the set $\{ \pm
\Gamma_{\alpha}\}$ and the same normalization
$\Gamma_{\alpha}\Gamma_{\alpha}^{\dagger}=\pm \id$ (no
summation).
\ \\ \ \\
For $d$ even there is only one irreducible faithful representation of
the gamma matrices of dimension $2^{d/2}$. Hence there are similarity
transformations
\be
\begin{array}{rcll}
{\mathcal H}^{-1}_{\pm} \Gamma^{\dagger}_{\alpha}
 {\mathcal H}_{\pm} &=& \pm \Gamma_{\alpha} \qquad
& \mbox{($\Gamma_{\alpha}^{\dagger}$ operates on bras)   \qquad (a)}
\label{eq:gammadagger} \\
{\mathcal C}_{\pm}
\Gamma^*_{\alpha} {\mathcal C}^{-1}_{\pm} &=& \pm \Gamma _{\alpha}\qquad
& \mbox{($\Gamma_{\alpha}^*$ operates on kets) \qquad     (b)}
\label{eq:gammastar}
\end{array}
\ee
We shall not need the similarity transformation of inverse
matrices.
\begin{rem}
The similarity transformation on $\{ \Gamma_{\alpha}^{\dagger}\}$ does
not imply that there is a similarity transformation on products
$\Gamma_{\alpha}^{\dagger}\Gamma_{\beta}^{\dagger}$. Indeed ${\mathcal
H}^{-1} (\Gamma_{\alpha}\Gamma_{\beta})^{\dagger}{\mathcal H} =
{\mathcal
H}^{-1} (\Gamma_{\beta}^{\dagger}\Gamma_{\alpha}^{\dagger}){\mathcal H}
= \Gamma_{\beta}\Gamma_{\alpha} \neq \Gamma_{\alpha} \Gamma_{\beta}$.
This explains the factor $a(\Lambda)$ in the definition of copinor in
\ref{sec:copinor}.
\end{rem}
\ \\
For $d=2p+1$ odd there are two inequivalent irreducible faithful
representations of the gamma matrices of dimension $2^{(d-1)/2}$;
hence there may not exist matrices ${\mathcal H}_{\pm}$ and
${\mathcal C}_{\pm}$ satisfying those similarity transformations ---
in other words, we could have
\be
{\mathcal H}^{-1}_{\pm} \Gamma^{\dagger}_{\alpha}
 {\mathcal H}_{\pm} = (-1)^{\alpha} \Gamma_{\alpha}
\quad \mbox{where} \quad
(-1)^{\alpha}=\left\{
\ba{ll} 1 \quad & \mbox{for $\alpha=0$} \\
0 & \mbox{for $\alpha\in \{ 1,2,3 \}$}
\ea   \right.
\ee
In section \ref{sec:difference} we gave a construction for a basis of
$\pin(t',s')$, $t'+s'=2p+1$, given a basis of $\pin(t,s)$, $t+s=2p$.
In this
construction the first $2p$ elements were the same as in $\pin(t,s)$,
hence they satisfy (\ref{eq:gammastar}a) or (\ref{eq:gammadagger}b) as
the case may be. We need to check only which equation is satisfied by
the new element $k\Gamma_{d+1}$, or $\hat{k} \hat{\Gamma}_{d+1}$.
\ \\ \ \\
We recall (\ref{eq:gamma2})
the properties of $\Gamma_{d+1}$ and $\hat{\Gamma}_{d+1}$:
for $d=2p$, $\Gamma_{d+1}:= \Gamma_1 \Gamma_2
\ldots \Gamma_d$ and similarly for $\hat{\Gamma}_{d+1}$,
\[
\Gamma_{d+1}^2 = (-1)^{s+p} \id_{2^p} \; ,
\qquad \hat{\Gamma}^2_{d+1}=(-1)^{t+p} \id_{2^p}
\]
The square of the new element, $(k\Gamma_{d+1})^2$, tells us
which group we generate (table \ref{table:even}).
\begin{table}[h]
\[
\begin{array}{|r|c|r|r|}\hline
\; (k\Gamma_{d+1})^2 \; & \mbox{Group} & k^2 \; \quad & k\quad \qquad
\\ \hline \hline
\id \qquad & \; \pin(t+1,s) \; &
(-1)^{s+p}   & \pm (-1)^{(s+p)/2} \\ \hline
-\id \qquad & \pin(t,s+1) &
-(-1)^{s+p} & \pm i \, (-1)^{(s+p)/2} \\ \hline
\end{array}
\]
\caption{Constructing a \pin\ group in odd dimensions.}
\label{table:even}
\end{table}
The two choices for $k$ in their respective groups
correspond to two different
representations.
We calculate the transformation under (\ref{eq:gammadagger}a) and
(\ref{eq:gammastar}b) of the new element
$k\Gamma_{d+1}$:
\bea
{\mathcal H}_{\pm}^{-1}(k\Gamma_{d+1}){\mathcal H}_{\pm}
&=& (-1)^t \, \Gamma_{d+1}^{\dagger}
\\
{\mathcal C}_{\pm}(k\Gamma_{d+1}){\mathcal C}^{-1}_{\pm}
&=&
\left\{
\ba{rl}
\Gamma_{d+1}^* & \mbox{if $k^*=k$}\\
-\Gamma_{d+1}^* & \mbox{if $k^*=-k$}\\
\ea \right.
\eea
From this, and the requirement that (\ref{eq:gammadagger}a)
and (\ref{eq:gammastar}b) extend to
$\Gamma_{d+1}$, we read off
\begin{center}
\begin{itemize}
\item[]
if $t$ is even, the only choice is ${\mathcal H}_+$
\item[]
if $t$ is odd, in particular if $t=1$,
the only choice is ${\mathcal H}_-$
\item[]
if $s-t+1$ mod $4=2$, the only choice is ${\mathcal C}_+$
\item[]
if $s-t+1$ mod $4=0$, the only choice is ${\mathcal C}_-$
\end{itemize}
\end{center}
For the corresponding $\pin(t,s)$ transformations one finds
\[
\hat{\mathcal H}_{\pm}={\mathcal H}_{\mp}
\quad \mbox{and} \quad
\hat{\mathcal C}_{\pm}={\mathcal C}_{\mp}
\]
The details of this calculation can be found in \cite{CDGwo}\footnote
{ In that paper $\Gamma_{d+1}$ is called the ``orientation matrix''
$\epsilon$.}, as well as properties of ${\mathcal H}$ and ${\mathcal
C}$.

\subsection{The short exact sequence $\id \rightarrow$ Spin$(t,s)
   \rightarrow $ Pin$(t,s) \rightarrow {\mathbb Z}_2 \rightarrow 0$}
\label{sec:sequence}
We shall prove that a Pin group is a semidirect product of a Spin
group with ${\mathbb Z}_2$, the group consisting of two elements
$\{e,z\}$, where $z^2=e$ :
\begin{eqnarray*}
   \pin(t,s) &=& \spin(t,s) \ltimes {\mathbb Z}_2 \qquad 
   s+t>1 \\
   \pin(s,t) &=& \spin(s,t) \ltimes {\mathbb Z}_2 \qquad 
   s+t>1
\end{eqnarray*}
where $\ltimes$ is defined below.
\spin$(t,s)$ and \spin$(s,t)$ are isomorphic, but a semidirect product
``scrambles'' the elements of its components and, as we know,
\pin$(t,s)$ is not necessarily isomorphic to \pin$(s,t)$. As we have
shown in \ref{sec:mod8}, they are isomorphic only when $s-t=0$ mod
4. The semidirect product construction does not work for \pin(0,1), 
since ${\mathbb Z}_2 \ltimes {\mathbb Z}_2 = 
{\mathbb Z}_2 \times {\mathbb Z}_2$.
\\
\ \\
{\em a)} First we prove general properties of semidirect products,
then we apply them to \spin\ and \pin. Let
\begin{eqnarray*}
   G &=& H \ltimes {\mathbb Z}_2 \\
   \hat{G} &=& \hat{H} \ltimes {\mathbb Z}_2 \\
\end{eqnarray*}
where there is an isomorphism $\phi : H \rightarrow \hat{H}$.
\ \\ \ \\
Let $t: ({\mathbb Z}_2 \times H) \rightarrow H$ be an element
of the   group of automorphisms of $H$ indexed by ${\mathbb Z}_2$.
\[
   t_z : H \rightarrow H
\qquad \mbox{ by } \qquad
h \mapsto t_z(h) = t(z,h)=zhz^{-1} \; .
\]
Note that $z$ is not in $H$ but acts multiplicatively
on $H$. The semidirect product
$G=H \ltimes {\mathbb Z}_2$ is the space of ordered pairs $(h,z)$ and
$(h,e)$, $h \in H$ with the product law
\[
(h_1,z_1)\cdot (h_2,z_2)=(h_1 t(z_1,h_2),z_1 z_2) \; ;
\]
here $z_i$, $i=1,2$,
is either $e$ or $z$; when working with ${\mathbb Z}_2$, it
is difficult to be correct without being pedantic.
\ \\ \ \\
If ${\mathbb Z}_2$
and $H$ commute, $t(z_i,h)=h$ and the semidirect product becomes a
direct product. As the simplest example of semidirect product scrambling,
compare (see section \ref{sec:spinsub} and also appendix
\ref{app:covers})
\[
{\mathbb Z}_3 \times {\mathbb Z}_2
\qquad \mbox{ and } \qquad
{\mathbb Z}_3 \ltimes {\mathbb Z}_2 = D_3
\]
where the dihedral group $D_3$ is the group of symmetries of
a regular triangle.
\ \\ \ \\
Consider the short exact sequence
\[
\id \rightarrow   \spin(t,s)
   \rightarrow \pin(t,s) \rightarrow {\mathbb Z}_2 \rightarrow 0 \; .
\]
We shall show that
\begin{center}
\begin{tabular}{lll}
 \qquad & if $s+t$ is odd, & $O(s,t) = SO(s,t) \times {\mathbb Z}_2$ \\
 \qquad & if $s+t$ is even, & $O(s,t) = SO(s,t) \ltimes {\mathbb Z}_2$
\end{tabular}
\end{center}
Since a direct product is a special case of a semidirect product, we
begin with
\[
   (a,z_i) \in SO(s,t) \ltimes {\mathbb Z}_2
\]
The identification $(a_i,z_i)$ with $a_i z_i =: g_i \in O(s,t)$ makes
sense because $z_i$ is not necessarily in $SO(s,t)$, and because it
makes the definition of the semidirect product consistent with the
group product in $O(s,t)$. Indeed
\begin{eqnarray*}
(a_1, z_1) \cdot (a_2, z_2) &=&
(a_1 z_1 a_2 z_1^{-1}, z_1 z_2) \\
&\simeq& a_1 z_1 a_2 z_1^{-1} z_1 z_2
= g_1 g_2 \; .
\end{eqnarray*}
The difference between $s+t$ odd and $s+t$ even stems from the fact
that if $s+t$ is odd we can choose
\[
{\mathbb Z}_2 = (\id, -\id)
\qquad \mbox{ because } \qquad
-\id \notin SO(s,t) \qquad \mbox{($s+t$ odd).}
\]
${\mathbb Z}_2$ commutes, then, with $SO(s,t)$ and the semidirect
product is simply a direct product.
\ \\ \ \\
If $s+t$ is even, $-\id \in SO(s,t)$ so we cannot use $-\id$ for
$z$ (since we require $z \notin SO(s,t)$). Other options for $z \notin
SO(s,t)$ are reflections. Reflections do not commute with all elements
in $SO(s,t)$ and the semidirect product does not reduce to a direct
product.
\ \\ \ \\
In general, $G=H\times {\mathbb Z}_2$ if $G$ has a central element $z$
of order 2 which is not in $H$. \\
\ \\
{\em b)} Given $G=H\ltimes {\mathbb Z}_2$, $\hat{G}
=\hat{H}\ltimes {\mathbb Z}_2$ and an isomorphism
\[
\phi: H\rightarrow \hat{H} \; ,
\]
we shall prove that there is an isomorphism
\[
\Phi:G \rightarrow \hat{G}
\]
if and only if, for every $h \in H$,
\begin{equation}
   \Phi(h,z_i) = (\phi(h),\hat{z}_i)
\label{eq:Phi}
\end{equation}
where $\hat{z}_i$ is defined by
\begin{equation}
   \hat{z}_i \phi(h) \hat{z}_i^{-1} =
   \phi(z_i h z_i^{-1}) \; .
\label{eq:z_i}
\end{equation}
It may be useful to refer to this diagram:
\[
\begin{array}{ccccc}
   &     & \ltimes \; {\mathbb Z}_2 & & \\ 
   & H & \longrightarrow & G & \\ 
   \phi & \downarrow & & \downarrow & \Phi \\ [1mm]
   & \hat{H} & \longrightarrow & \hat{G} & \\ 
   &     & \ltimes \; {\mathbb Z}_2 & &
\end{array}
\]
\begin{pf}
Let $(h_1,z)$ be identified with $g_1=h_1 z$ and $(h_2,e)$ be
identified with $g_2=h_2$. $\Phi$ is an algebra isomorphism if
$\Phi(g_1) \Phi(g_2)=\Phi(g_1 g_2)$. It is sufficient to choose
$g_1=(h_1, z)$ and $g_2=(h_2, e)=h_2$ for identifying under which
condition $\Phi$ is an isomorphism.
\ \\ \ \\
If the condition (\ref{eq:Phi}) is satisfied
\[
   \Phi((h,z))=(\phi(h),\hat{z}) \; , \qquad \Phi((h,e))=\Phi(h)=\phi(h)
\]
then
\[
\Phi((h_1, z)) \, \Phi(h_2) = (\phi(h_1), \hat{z}) \, \phi(h_2)
\; ,
\qquad \mbox{ identified with } \qquad \phi(h_1) \, \hat{z} \,
\phi(h_2)
\; ,
\]
on the other hand
\[
\begin{array}{rcll}
\Phi((h_1,z) \cdot (h_2,e))
&=&
\Phi(h_1 z h_2 z^{-1}, ze) & \\
&=&
(\phi(h_1 z h_2 z^{-1}), \hat{z}) & \mbox{assuming (\ref{eq:Phi})} \\
&=&
(\phi(h_1)\phi(zh_2z^{-1}), \hat{z}) & \mbox{since $\phi$ is an
   isomorphism} \\
\mbox{identified to} && \phi(h_1) \phi(z h_2 z^{-1})\, \hat{z} \\
&=&
\phi(h_1) \, \hat{z} \, \phi(h_2) & \mbox{by the definition
   (\ref{eq:z_i}) of $\hat{z_i}$} \\
\end{array}
\]
and thus
\begin{equation}
\Phi(g_1) \Phi(g_2) = \Phi(g_1 g_2) \; .
\label{eq:Phimorph}
\end{equation}
We have proven that eq. (\ref{eq:Phi}) implies
eq. (\ref{eq:Phimorph}).
The converse follows by identification.
\qed
\end{pf}

\subsection{Grassman (superclassical) pinor fields}
\label{sec:susy}
In several studies, pinors are sections of a supervector bundle
associated to a principal \pin\ bundle by a representation $(\rho, V)$
of the \pin\ group where the typical fiber $V$ is a supervector
space. A supervector space is a linear space over the supernumbers. (A
linear space is a module for which the ring of operators is a field,
e.g. the real numbers or the complex numbers). Supernumbers are
generated by a Grassman algebra; i.e. the generators of the algebra
$\{\zeta^a\}$ with $a \in \{1,\ldots,N\}$, with possibly $N=\infty$,
anticommute:
\[
\zeta^a \zeta^b = - \zeta^b \zeta^a
\]
and a supernumber $z$ can be expressed in the form
\[
\zeta = \zeta_B + \zeta_S
\]
where $\zeta_B$ is an ordinary complex number and
\[
\zeta_S=\sum^{\infty}_{n=1} c_{a_1 \cdots a_n}
\, \zeta^{a_n} \cdots \zeta^{a_1}
\]
the $c_{a_1 \cdots a_n}$
being complex numbers, completely antisymmetric in the indices.
\ \\ \ \\
It is often said (and we have done so in the past) that
choosing representations of the \pin\ groups on supervector spaces is
desirable for considering classical physics as the limit of quantum
physics. In other words, if the anticommutator of a quantum field at
two different causally related points goes to zero with
Planck's constant $\hbar$, then the classical pinor
field at two different points anticommute when $\hbar=0$. But, as
pointed out by Cartier, the anticommutator
of the fermionic fields in the Lagrangian is not
proportional to $\hbar$. 
Indeed, in QED, the electric current density ${\mathbf
j}$ in terms of the electron field $\Psi$ is (restoring 
$\hbar$ and $c$ in this subsection):
\[
j_{\mu}=ec \, \bar{\Psi} \Gamma_{\mu} \Psi \; .
\]
The physical dimension of the current
\[
J_{\mu}(t)=\int j_{\mu}(x,t) d^3x
\]
is $[J_{\mu}(t)]=eT^{-1}$.
Hence we can compare physical dimensions:
\[
ec\left[
\int \bar{\Psi} \Gamma_{\mu} \Psi d^3 x \right]
= e T^{-1}
\]
which implies that $ [\bar{\Psi} \Gamma_{\mu} \Psi ] = L^{-4}$.
Thus the dimension of the field operator is $[\Psi] = L^{-2}$ 
and so
\[
[\{\Psi(x), \Psi(y)\}]=L^{-4} \; .
\]
The anticommutator does not have the same physical dimension
as $\hbar$, which is
$[\hbar]=M L^2 T^{-1}$.
In order to have the anticommutator proportional to $\hbar$,
it suffices to take the anticommutator of $\sqrt{\hbar}\, \Psi$.
\ \\ \ \\
One important reason for treating classical pinors as supervector
fields is functional integration:
the functional integral needed to construct matrix elements of
operators built with Fermi quantum fields is a functional integral over
a space of functions with values in a Grassman algebra.
In general it is convenient to have quantum fields and the
corresponding classical fields taking their values in the same
algebra. 
\ \\ \ \\
Having discussed
the motivation for classically treating pinors as
superclassical fields we refer the reader to the existing literature
on supermanifolds \cite{Blue,Super} and on the use of superclassical
fields in studies aimed at comparing the two \pin\ groups
\cite{DeWittDeWitt,CDGwo}.

\subsection{String theory and pin structures}
\label{sec:strings}
The following remarks discuss string theory, where the \pin\ groups may
be particularly relevant. 
Pin structures are defined in section \ref{sec:bundles}.
\ \\ \ \\
The concept ``string theory'' now encompasses more objects than
the one-dimensional strings of the original string theories; those
original theories emerge as different limits of modern string theory, or
appear in duality relationships with other theories included in modern
string theory. Of course, the original string theories are still of
interest when viewed as different corners of the parameter space
of modern string theory.
\ \\ \ \\
We note briefly how spin structures enter into the Ramond-Neveu-Schwarz (RNS)
formalism of closed superstrings in ten dimensions
\cite{Ginsparg,Green,Polchinski,Witten};
the extension to pin structures follows the same pattern.
\ \\ \ \\
One {\em a priori} 
problem in superstring theory in ten dimensions is the existence
of a tachyon in the spectrum.
It is solved by the projection on the space of states
known as the Gliozzi-Scherk-Olive (GSO) projection:
\be
P_{\rm GSO}={\scriptstyle {1 \over 2}}(1+(-1)^F)
\label{eq:GSO}
\ee
where $F$ is the fermion number.
This projection takes away the tachyon, and leaves an equal number of
fermions and bosons, as required for a linear realization of supersymmetry.
(It also solves other problems.)
\ \\ \ \\
To show how the GSO projection involves spin
structures, we study a torus diagram, which can represent the creation
and annihilation of a pair of closed strings, as they move in time; 
a one-loop string diagram.
\begin{figure}[h]
   \begin{center}
       \vspace{3mm}
       \resizebox{5cm}{!}{\includegraphics{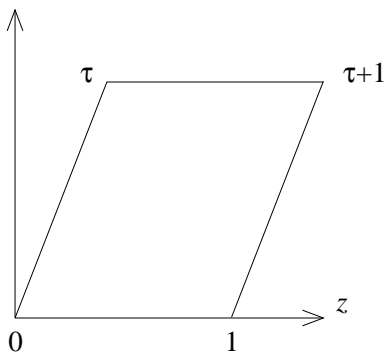}}
   \caption{Opposite sides of the parallelogram are identified, and the
   parallelogram has the topology of a torus.}
   \label{fig:torus}
     \vspace{3mm}
   \end{center}
\end{figure}
If we carry a fermion field around either one of the two
nontrivial cycles of the torus, the spin
structure dictates whether the fermion comes back to itself (periodic) or
changes sign (antiperiodic).
There are two cycles on the torus, so we have four combinations of
``boundary conditions'' for the functional integral,
we label them $(P,P)$, $(P,A)$, $(A,P)$, $(A,A)$. The first letter refers to
periodicity in $z$ (see fig. \ref{fig:torus}).
\ \\ \ \\
For functional integrals of a single fermion
we find, denoting by $\tr_A$ the trace in the
antiperiodic sector,
\bea
(P,P) &=& q^{-1/48} \, \tr_P \, (-1)^F q^{L_0} \\
(P,A) &=& q^{-1/48} \, \tr_P \,   q^{L_0} \\
(A,P) &=& q^{-1/48} \, \tr_A \,   (-1)^F q^{L_0} \\
(A,A) &=& q^{-1/48} \, \tr_A \,     q^{L_0} \\
\eea
where $L_0$ is the normal-ordered Hamiltonian,
and $q=\exp(2\pi i\tau)$ where $\tau$ is the modular parameter on the
torus. (We are not interested in the details here, just the $(-1)^F$
factors.)
The general principle of modular invariance
can be used as a guide for combining these four amplitudes. Here
we simply add the four amplitudes; this
amounts to inserting a factor $(1+(-1)^F)$ in both $\tr_A$
and $\tr_P$. Inserting this factor is
identical (up to the factor $\half$) to performing a GSO projection
(\ref{eq:GSO}).
Thus adding functional integral
contributions from each spin structure
({\em summing over spin structures}), is a prescription
which gives useful results,
at least in weakly coupled string theory at the one-loop level.
\ \\ \ \\
There is no reason to limit the above discussion to spin
structures. In string theory one considers unoriented string
diagrams (such as the Klein bottle) in addition to the torus diagram
discussed above.
The full Lorentz group is certainly relevant, and hence the pin
structures. Criteria for the existence of pin structures on orientable
and non-orientable manifolds with metrics of arbitrary signatures can
be found in Karoubi \cite{Karoubi}. 
The first thing one runs into is the criterion for isomorphicity:
$s-t=0$ mod 4. On a Minkowski string worldsheet,
there is evidently only \pin(1,1). 
On a Euclidean worldsheet the
\pin\ groups are different: \pin(0,2) and \pin(2,0).
Beyond the worldsheet, there are higher-dimensional hypersurfaces
in string theory to which fermions may be restricted.
Of these, checking the criterion we see that
the two  \pin\ groups are isomorphic 
only for 5+1-dimensional hypersurfaces and in 9+1-dimensional
spacetime, alternatively 4-dimensional or 8-dimensional 
Euclidean hypersurfaces.
In all other cases directly relevant to string theory (spatial dimensions
2,3,4,6,7,8 and 10 of Minkowski space or 2,3,5,6,7,9 and
11 dimensions of Euclidean space) the \pin\ groups are {\it not}
isomorphic.
\ \\ \ \\
There are already existing attempts in this direction in the
literature. Chamblin \cite{Chamblin} has
mentioned one way of selecting pin structures in string theory. 
In a note on the 3D Ising model as a string theory \cite{Distler},
Distler pointed out that fermions used in open string theory make sense
with \pin(0,2) structure but not with \pin(2,0) structure, since only
\pin(0,2) structure can be defined on any 2-manifold.
The discussion takes place within his approach to the 3D Ising model, which
in the continuum limit is equivalent to a certain unoriented string theory.
The implication of the non-existence of pin structures on some
2-dimensional surfaces has been worked out for the
Polyakov path integral of the NSR superstring action  \cite{Carlip}.
Finally, a paper by Dasgupta, Gaberdiel and Green
discussed \pin\ groups in relation to breaking of $O(16)$
to $SO(16)$ symmetry in string theory \cite{Dasgupta}.

\section{Conclusion}
\subsection{Some facts}
In 3+1 dimensions, there are two \pin\ groups,
\pin(1,3) and \pin(3,1), which come into play in the analysis of time
or space reversal. In principle the existence of two \pin\ groups
provides a finer classification of fermions than one \pin\ group. Such
a classification is useful only if one can design experiments which
distinguish the two types of fermions. Many promising experimental
setups give, for one reason or another, identical results for both
types of fermions. These negative results are reported here because
they are instructive. Two notable positive results show that the
existence of two \pin\ groups is relevant to physics:
\begin{itemize}
\item
In a neutrinoless double beta decay, the neutrino emitted and
reabsorbed in the course of the interaction can only be described in
terms of \pin(3,1).
\item
If a space is topologically nontrivial, the vacuum expectation values
of Fermi currents defined on this space can be totally different when
described in terms of \pin(1,3) and \pin(3,1).
\item
Only \pin(0,2) can be used in open string theory \cite{Distler}. The
same conclusion applies to a 3D Ising model which
is in the continuum limit
equivalent to a certain unoriented string theory.
\end{itemize}

\subsection{A tutorial}
The \pin\ groups are technically useful; they provide a simple
framework for the study of fermions, in the context of the full
Lorentz group.
\ \\

\begin{center}
{\em Parity}
\end{center}
The parity operator operates on the space of {\em pinors}.
It cannot be defined
on the space of {\em spinors} (Weyl fermions)
for the following reason: the parity operator consists of an odd
number of gamma matrices, whereas the elements of the \spin\ group
consist of even numbers of gamma matrices. When there is no parity
operator, there is no parity eigenspinor, therefore
no parity eigenvalue can be assigned to a Weyl
fermion. One often hears that no parity is assigned to Weyl fermions
``because weak interactions do not conserve parity''
but to say that an interaction does not
conserve parity implies that a parity can be assigned to the initial
state and to the final state.
The statement is meaningless because the same word ``parity'' is used
for two different concepts: ``intrinsic parity'' of a
fermion (as
in section \ref{sec:pinrepq})
and the ``parity
non-conservation of an interaction'' (as in section \ref{sec:pinreps}).
\ \\ \ \\
The {\em square} of the parity operator does operate
on the space of spinors. In \pin(1,3), $\Lambda^2_{P(3)}=+\id$, and in
\pin(3,1), $\hat{\Lambda}^2_{P(3)}=-\id$, hence it is meaningful to say if a
Weyl fermion belongs to a subgroup of \pin(1,3) or \pin(3,1).
\ \\

\begin{center}
{\em Time reversal}
\end{center}
There are two definitions of the time
reversal operator: a unitary one and an antiunitary one, which serve
different purposes.
The unitary one is in the toolbox of the Lorentz group. The
antiunitary one is used in motion reversal, an expression which Wigner
credits to L\"{u}ders \cite[p.\ 54]{Wigner2}. Invariance under antiunitary
time reversal is required so that quantum systems are free of negative
energy states.
\ \\

\begin{center}
{\em Charge conjugation}
\end{center}
Charge conjugation of pinors
has nothing to do with the Lorentz group, nor does
antiunitary time reversal; but the ``$C\!PT$'' transformation on
pinors correspons simply to
an orientation preserving Lorentz transformation (transformation of
determinant 1).
\ \\

\begin{center}
{\em Wigner's classification and classification by Pin groups \/}
\end{center}
To prefer one classification over another is in part a matter of
taste, and in part a matter of its intended use. We prefer the 
classification by \pin\ groups  because it is a straightforward
consequence of the use of the full Lorentz group in physics. Wigner
begins with $SL(2,\Cset)$ which is isomorphic, but not identical,
to the covering group $\spin^{\uparrow}(3,1) \subset \spin(3,1)$. 
Then he combines it with reflections and constructs four different
convering groups, but needs to discard representations which are not
physically admissible. He raises the question of a ``whole group'' but
notes that it is not uniquely defined in the context of his
classification. We regret that his work precedes the identification of
the two \pin\ groups. He would have made use of this fact in a more
perceptive fashion than we have done so far.
\ \\

\begin{center}
{\em Fock space and one-particle states \/}
\end{center}
Operators on a Fock space and operators on a
space of pinors are different objects.
In section \ref{sec:pinreps} we present \pin\ group operators
acting on the space of unquantized pinor fields (classical fields)
$\psi$. In section \ref{sec:pinrepq} we define charge conjugation,
space and time reversal operators on quantum fields $\Psi$.  The
relationship between $\psi$ and $\Psi$ can be found in the dictionary
of notation  (section \ref{sec:notation}). Equations
(\ref{eq:AT}), (\ref{eq:UC}) and (\ref{eq:UP}) provide the bridges
between operators on Fock space and operators on a space of
pinors. These equations are necessary in the analysis of the \pin\
groups in the quantum field theory of particle physics.

\subsection{Avenues to explore}
This report is limited to the case where the inversion operators $U_P$
and $U_T$ take one-particle states into other one-particle states of
the same species. Inversions may act in a more complicated way than
this on degenerate multiplets of one-particle states. This possibility
was first suggested by Wigner \cite{Wigner2}. Weinberg \cite{Wein}
explored generalized versions of the inversion operators, in which
finite matrices appear in place of the inversion phases, but without
making some of Wigner's limiting assumptions. In the ``Collected
References'' of appendix \ref{app:ref}, under the heading ``Carruther's
Theorem'' we list basic references on the subject, in particular works
of Moussa and Stora, which can be used as a starting point for
investigating the role of the {\em two} \pin\ groups in the case of
degenerate multiplets of one-particle states.
Indeed, in section \ref{sec:pinreps} we learn to distinguish the \pin\
groups; in section \ref{sec:pinrepq} we introduce the phases
associated to projective representations of quantum field operators.
\ \\ \ \\
Other investigations, such as the following, could reveal differences
between $\pin(t,s)$ and $\pin(s,t)$ fermions:
\begin{itemize}
\item
Time or space reversal in the complex environment of atomic and
molecular physics, dipole moments, etc.
\item
Topologically nontrivial configuration spaces.
\item
To first order, decay rates and cross sections computed in this
report do not depend on the choice of \pin\ group, but given the trace
and spin sum differences, it is not excluded that higher order
contributions would be different.
\end{itemize}

\newpage
\section{Acknowledgements}

The first version was written by two of us (CD and SJG) in 1991 and
then kept on the
backburner while we analyzed situations in
which one could observe experimentally the differences between the two
Pin groups. EK joined us and worked out in detail
the section on interference
\ref{sec:interference}. Retrospectively, one can
argue that the answers are obvious, but only an explicit calculation
can be convincing, because the issues are subtle and the signs
dictated by the choice of groups enter at various stages of the
calculation.
\ \\ \ \\
MB undertook the major project of attempting to
make this work meaningful to
experimental physicists. He investigated selection rules in positronium
decay, three-fermion decay, positronium and decay rates, in particular
$\Sigma^0$ decay.
\ \\ \ \\
The new team (MB and CD) has completely rewritten the previous
version.
\ \\ \ \\
In the course of nearly a decade, many colleagues have commented on
this work. We thank them for their interest, but the list of their
names would necessarily be incomplete, and serve little purpose other
than name dropping. Special thanks are due to Steven Carlip whose
suggestions vastly improved the first draft, to Yuval Ne'eman
for his support during the adventures of the second version,
and to Raymond Stora who read it seriously and noticed a number of
issues requiring improvement.
\ \\ \ \\
MB wishes to thank the Sweden-America Foundation for financial support.

\newpage
\appendix
\noindent 
{\bf {\Large  Appendix}}
\renewcommand{\thesection}{\Alph{section}}
\renewcommand{\theequation}{\Alph{section}.\arabic{equation}}
\setcounter{equation}{0}

\section{Induced transformations}
We recall briefly the transformation laws induced by a Lorentz
transformation $L$ of spacetime.
\begin{figure}[h]
   \begin{center}
       \resizebox{5cm}{!}{\includegraphics{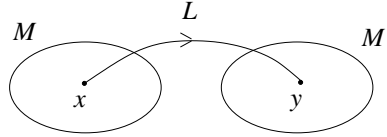}}
   \end{center}
   \caption{Lorentz transformation $L$.}
   \label{fig:induced}
\end{figure}
Let $(M,g)\equiv M^{1,3} \equiv M $ be a spacetime manifold with
metric $g$ of signature $\eta=(1,-1,-1,-1)$. Let $L$ map $(M,g)$ into
itself. Let $T_x M$ and $T_y M$ be the tangent spaces to $M$ at $x$
and $y$ respectively, and $T^*_x M$ and $T_y^* M$ be their dual spaces
(spaces of linear maps on the tangent spaces). Let $V(x) \in T_x M$
and $\omega(x) \in T_x^* M$, $V(x)$ is a contravariant vector,
$\omega(x)$ is a covariant vector. In terms of components, the duality
is
\[
\langle \, \omega(x), V(x) \, \rangle =
\sum_{\alpha} \omega_{\alpha}(x) \, V^{\alpha}(x) \equiv
\omega_{\alpha}(x)\,   V^{\alpha}(x) \; .
\]
Since $L$ is a linear map, its derivative mapping $L'(x)$ is $L$
itself, but is now a mapping from $T_x M$ to $T_y M$.
\ \\

\begin{figure}[h]
   \begin{center}
       \resizebox{5cm}{!}{\includegraphics{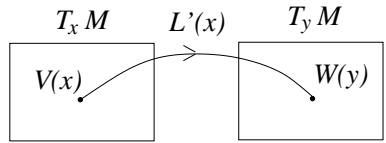}}
   \end{center}
   \caption{Derivative mapping $L'(x)$.}
   \label{fig:derivative}
\end{figure}

\be
\ba{rcl}
W(y) &=& L \, V(x) \\
W(Lx) &=& L \, V(x) \; .
\ea
\label{eq:Wmap}
\ee
In terms of components
\[
W^{\alpha}(Lx) = {L^{\alpha}}_{\beta} \, V^{\beta}(x) \; .
\]
The duality is used to determine the transformation properties of
elements of the dual spaces. Let $\theta(y) \in T_y^* M$, then we
define $\tilde{L}(y) : \theta(y) \mapsto \omega(x)$ by
\[
\langle \,   \omega(x) , V(x) \, \rangle =
\langle \, \theta(y) , W(y) \, \rangle
\]
Since $L'(x)$ is independent of $x$, then $\tilde{L}(y)$ is
independent of $y$.

\begin{figure}[h]
   \begin{center}
       \resizebox{5cm}{!}{\includegraphics{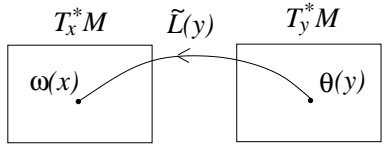}}
   \end{center}
   \caption{Mapping $\tilde{L}(y)$ between dual
       spaces. $\tilde{L}(y)=\tilde{L}$.}
   \label{fig:dual}
\end{figure}
\[
\langle \, (\tilde{L}\theta)(x) , V(x)\,   \rangle =
\langle \, \theta(y) , (LV)(y) \, \rangle
\]
\be
\omega(x) = \tilde{L} \, \theta(Lx)
\ee
In terms of components
\[
\omega_{\beta}(x) = \theta_{\alpha}(Lx)\, { L^{\alpha}}_{\beta} \; .
\]
The same pattern applies to contravariant pinors
(or just ``pinors'') and covariant pinors (or ``copinors'').
Hence, under a Lorentz transformation $L$, a pinor
$\psi(x)$ becomes $\psi'(Lx)$ such
that
\[
\psi'(Lx) = \Lambda_L \psi(x) \; ,
\]
a copinor $\bar{\psi}'(Lx)$ becomes $\bar{\psi}(x)$ such that
\[
\bar{\psi}(x) = \tilde{\Lambda}_L \bar{\psi}'(Lx) \; .
\]

\newpage

\section{The isomorphism
$M_4({\mathbb R}) \simeq {\Hset} \otimes {\Hset}$}
\label{app:iso}
\setcounter{equation}{0}
Let $M_4({\mathbb R})$ be the space of real
$4 \times 4$ matrices, and ${\Hset}$ be the quaternion algebra.
Let $(1,i,j,k)$ be its basis with $i^2=j^2=k^2=-1$ and
$ij=k$, $jk=i$, $ki=j$. Let $(a^1, a^2, a^3, a^4)$ be the coordinates
of $\alpha \in {\Hset}$ and $(b^1, b^2, b^3, b^4)$   the
coordinates of $\beta \in \Hset$.
\[
\alpha=a^1+a^2i+a^3j+a^4k
\]
\[
\alpha^{\dagger}=a_1-a_2i-a_3j-a_4k \; .
\]
As a vector space, $\Hset$ is a real 4-dimensional vector
space. Let
\[
I:{\Hset} \rightarrow V^4 \qquad \mbox{by}
\qquad
I\,(a^1+a^2i+a^3j+a^4k)=
\left(
   \begin{array}{c}
       a^1 \\ a^2 \\ a^3 \\ a^4
   \end{array}
\right)
\]
Let $\alpha \otimes \beta$ act linearly on $\Hset$ by
\[
(\alpha \otimes \beta)(\gamma) = (\alpha_L \beta_R)(\gamma)
:= \alpha \gamma \beta^{\dagger} \qquad \gamma \in \Hset
\]
We map $\Hset \otimes \Hset$ to the
space of $4 \times 4$ real matrices $M_4(\Rset)$ by
\[
f: \Hset \otimes \Hset \rightarrow
M_4(\Rset) \qquad \mbox{by}
\]
\[
f(\alpha \otimes \beta)=M(\alpha, \beta)
\quad \mbox{where} \quad
M(\alpha , \beta) I(\gamma) = I(\alpha \gamma \beta^{\dagger})
\quad \mbox{for all } \gamma \in \Hset
\]
It is straightforward to prove that $f$ is an algebra isomorphism.
One can construct explicitly the matrix
$M(\alpha,\beta)$ for a pair of basis
elements. Each matrix $M(\alpha,\beta)$ thus obtained can be written
as a tensor product of a pair of matrices from the set
$\{\id_2, \sigma_1, i\sigma_2, \sigma_3\}$
(see example below).
\ \\ \ \\
We recall the definitions of tensor products of algebras, and tensor
products of matrices. Let $\{e_i\}$ and $\{e_{\alpha}\}$ be bases for
the real algebras $A$ and $B$. Let $c=c^{i\alpha} e_i \otimes
e_{\alpha}$ and $d=d^{j\beta} e_j \otimes
e_{\beta}$, then
\[
cd = c^{i \alpha} d^{j \beta} (e_i e_j \otimes e_{\alpha} e_{\beta})
\]
Let $a=(a^i_j)$ and $b=(b^{\alpha}_{\beta})$, then
\[
(a \otimes b)^I_J = a^i_j b^{\alpha}_{\beta}
\]
Here $I=(i,\alpha)$, $J=(j,\beta)$. There are two obvious choices for
ordering the pairs. We choose
\[
(a \otimes b)=\left(
\begin{array}{cc}
a^1_1(b) & a^1_2(b) \\[2mm]
a^2_1(b) & a^2_2(b) \\
\end{array}
\right)
\]
To prove that $M(\alpha, \beta)$ is a real matrix, we construct
$M(\alpha, \beta)$ for all the elements in a basis of $\Hset \otimes
\Hset$, then extend the result by linearity. Let the basis for $\Hset
\otimes \Hset$ consist of $\id \otimes \id$, $\id \otimes i$, $\id
\otimes j$, $\id \otimes k$, $i \otimes \id$, $i \otimes i$, etc. Let
$\gamma=a + bi +   cj + dk$, then
\[
M(1,i)I(\gamma)=I(1\gamma(-i))=I(b-ai-dj+ck)
\]
Therefore
\bea
M(1,i)&=&\left(
\begin{array}{cccc}
0 & \; 1 \; & \; 0 \; & 0 \\
-1 & 0 & 0 & 0 \\
0 & 0 & 0 & -1 \\
0 & 0 & 1 & 0
\end{array}
\right)
= \sigma_3 \otimes i\sigma_2 \qquad \mbox{(on the r.h.s. $i=\sqrt{-1}$).}
\eea
A similar calculation for all the elements in the basis of $\Hset
\otimes \Hset$ shows that $M(\alpha, \beta) \in M_4(\Rset)$.

\newpage

\section{Other double covers of the Lorentz group}
\label{app:covers}
\setcounter{equation}{0}
After having identified \pin(3,1) and \pin(1,3) as two inequivalent
covers of the Lorentz group, we have to review briefly the other
double covers of the Lorentz group.
\ \\ \ \\
We can look at $\pin(s,t)$ and $\pin(t,s)$ as extension of
$O(s,t)$ (equivalently $O(t,s)$) by $\Zset_2$ in the short exact
sequence
\[
1 \rightarrow \Zset_2 \rightarrow G \rightarrow O(s,t)
\rightarrow 1
\qquad \Zset_2 = \left\{1, -1\right\}
\]
Two extensions $G$ and $G\,'$ are equivalent if and only if $G$ and $G\,'$
are isomorphic {\em and} the two sequences
\[
\begin{array}{ccccccccc}
1 &\rightarrow &\Zset_2& \rightarrow &G &\rightarrow & O (s,t)
&\rightarrow& 1\\
&   &\downarrow& \circ & \wr\!\wr & \circ & \downarrow \\[1mm]
1 &\rightarrow &\Zset_2& \rightarrow &G\,' & \rightarrow & O (s,t)
&\rightarrow &1
\end{array}
\]
are made of two commutative diagrams: the two maps
$\Zset_2 \rightarrow G \rightarrow G\,'$ and
$\Zset_2
\rightarrow
\Zset_2 \rightarrow G\,'$ are identical (up to
isomorphisms) --- and the
same property for the other
diagram.
\ \\ \ \\
There are eight double covers of the Lorentz group called Pin$^{abc}$
by Dabrowski \cite{Dabrowski} and characterized by
\[
\Lambda^{2}_P = a, \quad \Lambda^{2}_T = b, \quad (\Lambda_P
\Lambda_T)^{2} = c \; , \qquad a, b, c \in \Zset_2 \; .
\]
In the fourth column of table \ref{table:covers}, we give the names of the
corresponding finite groups with elements $\pm \id, \pm \Lambda_P,
\pm \Lambda_T$. The dihedral group $D_n$ is the group of symmetries of
an $n$-sided regular polygon.

\begin{table}[h]
\[
\begin{array}{|c|c|c|l|c|}
\hline
\Lambda^{2}_P& \Lambda^{2}_T&  (\Lambda_P\Lambda_T)^2 &
{\rm Group} & \Lambda_T\Lambda_P = \pm \Lambda_P\Lambda_T \\ 
\hline  \hline 
+1   & +1   & +1 & \Zset_2\oplus \Zset_2\oplus
\Zset_2 & +
\\ \hline 
+1 & -1 & -1 & \Zset_2\oplus \Zset_4 & +\\  \hline 
-1 & +1 & -1 & \Zset_2\oplus \Zset_4 & +\\ \hline 
-1 & -1 & +1 & \Zset_2\oplus \Zset_4 & +\\ \hline 
-1 & -1 & -1 & {\rm quaternion} & -\\ \hline 
-1 & +1 & +1 & \mbox{dihedral generating Pin (1,3)} & -\\ \hline 
+1 & -1 & +1 & \mbox{dihedral generating Pin (3,1)} & -\\ \hline 
+1 & +1 & -1 & {\rm dihedral} & - \\ \hline
\end{array}
\]
\caption{The eight double covers of the Lorentz group.}
\label{table:covers}
\end{table}
The following requirements identify the Pin groups, called
cliffordian by Dabrowski \cite{Dabrowski}:
\begin{eqnarray*}
\Lambda_T\Lambda_P &=& - \Lambda_P\Lambda_T\\
\Lambda_P^2 &\ne& \Lambda^2_T
\end{eqnarray*}
By relating $\Lambda_P, \, \Lambda_T \in \pin (s,t)$ to
$P,\, T \in O(s,t)$ respectively (or $\hat{\Lambda}_P,
\hat{\Lambda}_T$ to $P, T$) we give the explicit pin structure
(see section \ref{sec:bundles}).
There is an extensive literature (see Appendix \ref{app:ref}) on pin
structures, not only for \pin\ groups covering Lorentz groups, but also
for \pin\ groups covering $O(s,t)$ with arbitrary values of $s$ and $t$.

\section{Collected calculations}
\label{app:coll}
\setcounter{equation}{0}
In this Appendix, we collect for reference some calculations that were
either too long to have in the main body of the paper, or fairly standard
and only slightly generalized to accommodate the two \pin\ groups. For
each calculation we refer to the page where the relevant discussion
can be found.
\ \\

\begin{center}
{\it Spin sums} (p.\ \pageref{page:spinsums})
\end{center}
We compute the spin sums, using $u(p,s)$
as an example. The Dirac equation
$(i\Gamma^{\alpha}\partial_{\alpha}-m)\psi(x)=0$ gives
\be
( \sla p - m )\, u(p,s)=0 \; .
\label{eq:Diracp}
\ee
With the given normalization, $u$ is an eigenpinor of the spin sum
with eigenvalue $2m$; indeed
\bea
\left(\sum_{r}{u}\, (p,r)\bar{u}(p,r)\right) u(p,s)
&=& \sum_{r} u(p,r) (2m \delta_{rs}) \\
&=& 2m \, u(p,s) \; .
\eea
We can rewrite $2mu$ using eq.\ (\ref{eq:Diracp}) as
\bea
2m u(p,s) &=& (m+m)u(p,s) \\
&=& (\sla p +m) u(p,s) \; .
\eea
Thus we have the
given spin sum for $u$.
\ \\

\begin{center}
{\em Parity conservation} (p.\ \pageref{page:paritycons})
\end{center}
We review briefly how the observed angular distribution of scattered
particles is used for concluding whether or not parity is conserved.
To be specific, we study scattering of two fermions into two fermions.
The argument is reviewed for a second-order 
contribution to the $S$-matrix, but we could of course
also have considered the full $S$-matrix. Let
\[
\langle {\bf p'}, {\bf k'}| {\scriptstyle
\int} d^4 x \, H_{\rm int} \;
{\scriptstyle \int} d^4 y \, H_{\rm int}
|{\bf p}, {\bf k}\rangle =:
{\mathcal M}({\bf p}, {\bf k}, {\bf p'}, {\bf k'}) \; .
\]
The ``in'' and ``out'' states
$|{\bf p}, {\bf k}\rangle$ and $\langle {\bf p'}, {\bf k'}|$
each have spin labels suppressed.
The evolution of free states into states
of the interacting theory is not relevant to the scattering process,
so we take these states to be free states
as is usual, in other words we only consider amputated Feynman diagrams.
\ \\ \ \\
The external states are:
\[
|{\bf p},s \rangle \sim a^{\dagger}({\bf p},s)|0\rangle \; .
\]
Now recall $U_P \, a({\bf p},s) U_P^{-1}=\eta_a a(-{\bf p},s)$
and $\psi({\bf p},s)=\Lambda_P\psi(-{\bf p},s)$
with the four-momentum being $p=(p_0, {\bf p})$. Thus,
 we have under parity for our two-particle states
\be
\ba{rcll}
U_P|{\bf p}, {\bf k}\rangle &\sim&
\eta_a^* \eta_b^* \, |-{\bf p}, -{\bf k}\rangle
& \quad   \mbox{provided } U_P | 0 \rangle = | 0 \rangle \\
\langle {\bf p'}, {\bf k'}| U_P^{\dagger} &\sim&
\langle -{\bf p'}, -{\bf k'}|\,   \eta_c \eta_d
& \quad \mbox{provided } \langle 0 | U_P^{\dagger}= \langle 0 | \; .
\ea
\ee
Now, if the operator $U_P$ commutes with
the Hamiltonian, a parity transformation simply induces the
following change in the $S$-matrix contribution:
\[
\langle {\bf -p'}, {\bf -k'}| \eta_c \eta_d \, {\scriptstyle
\int} d^4 x \, H_{\rm int} \;
{\scriptstyle \int} d^4 x \, H_{\rm int}
\, \eta^*_a \eta^*_b |{\bf -p}, {\bf -k}\rangle \; .
\]
Under a parity transformation we now have
\[
{\mathcal M}({\bf p}, {\bf k}, {\bf p'}, {\bf k'})
 \mapsto \eta_c \eta_d \eta^*_a \eta^*_b
\, {\mathcal M}({\bf -p}, {\bf -k}, {\bf -p'}, {\bf -k'})
\]
If the matrix element has some symmetry under inversion, we
use this symmetry for reexpressing the
right hand side to deduce a conservation rule for
the intrinsic parities. For instance, when we decompose ${\mathcal M}
({\bf -p}, {\bf -k}, {\bf -p'}, {\bf -k'})$
into partial waves (spherical harmonics) labelled by ${\ell}$, the
matrix element acquires a $(-1)^{\ell}$ due to the parity of
the spherical harmonics $Y_{\ell m}$:
\[
Y_{\ell m}(-\hat{q})=(-1)^{\ell}\, Y_{\ell m}(\hat{q}) \; ;
\]
with $\hat{q}$ the unit momentum transfer $\hat{q}=({\bf p}- 
{\bf p}')/|{\bf p}-{\bf p}'|$;
this relation can be used in a relativistic theory as well as in
non-relativistic quantum mechanics.
Thus we can write
\begin{eqnarray}
{\mathcal M}({\bf p}, {\bf k}, {\bf p'}, {\bf k'})
 &\longrightarrow&   \eta_c \eta_d \eta^*_a \eta^*_b
\, {\mathcal M}({\bf -p}, {\bf -k}, {\bf -p'}, {\bf -k'}) \\
 &=&   \eta_c \eta_d \eta^*_a \eta^*_b
\, (-1)^{\ell_{\rm i}}
(-1)^{\ell_{\rm f}} {\mathcal M}({\bf p}, {\bf k}, {\bf p'}, {\bf k'})
\label{eq:symmetryM}
\end{eqnarray}
or
\[
(-1)^{\ell_{\rm i}}(-1)^{\ell_{\rm f}}\eta^*_d \eta^*_c \eta_a \eta_b
=1 \; ,
\]
or equivalently,
\be
(-1)^{\ell_{\rm i}}\eta_a \eta_b = (-1)^{\ell_{\rm f}} \eta_c \eta_d \;.
\ee
\ \\

\begin{center}
{\it Pion decay} (p.\ \pageref{page:pion})
\end{center}
Consider pion capture by a deuteron followed by emission of two
neutrons:
\[
\pi^- + d   \rightarrow n + n
\]
In this capture, the pionic ``atom'' is known \cite{Chinowsky}
to be in the $\ell=0$
ground state. The pion has spin 0 and the deuteron spin 1,
so the initial state has total angular momentum $j=1$.
\ \\ \ \\
There are two principles we can use for deducing the orbital
angular momentum of the right-hand side
\begin{itemize}
\item
Angular momentum conservation $(j=1)$
\item
Antisymmetry of neutrons under exchange
\end{itemize}
\ \\ \ \\
The first yields the following possibilities for orbital quantum
number $\ell$ of the neutrons, and total neutron spin $s$, consistent
with $j=1$:
\begin{center}
\begin{tabular}{lcccc}
{\em a)} & $\ell=1 $   & & $s=0 $ & \\ 
{\em b)} & $\ell=0 $   & & $s=1 $ & \\
{\em c)} & $\ell=1 $   & & $s=1 $ & \\
{\em d)} & $\ell=2 $   & & $s=1 $ &
\end{tabular}
\end{center}
The wave function of two neutrons $n_1$ and $n_2$ (total spin $s$) in
an $\ell$-state satisfies
\[
\psi(n_1, n_2) = (-1)^{\ell+s+1}\psi(n_2, n_1) \; . \\
\]
Since it is required that
\[
\psi(n_1, n_2) = - \psi(n_2,n_1) \; ,
\]
the only option in the above table is {\em c)}.
\ \\ \ \\
Now that we have $\ell$ for the two-neutron final state, it is a trivial
matter to calculate the intrinsic parity of the pion. The orbital
contribution is $(-1)^{\ell}=(-1)$, so the $\eta$-phases of the
initial and final states are related by
eq. (\ref{eq:conspar}) since intrinsic parity is conserved:
\[
\eta_{\pi} \eta_{d}   = (-1)\eta_{n}\eta_n \; .
\]
\ \\

\begin{center}
{\em Selection rules: Positronium} (p.\ \pageref{page:positron})
\end{center}
The experiment revolves around positronium, the Coulomb bound state of an
electron and a positron. To analyze it, we will use a common approach
to bound states \cite{Peskin} which uses some nonrelativistic quantum
mechanics in conjunction with quantum field theory.
\ \\ \ \\
A bound state is created by letting the operator
\[
B=\int {d^3 {\bf p} \over (2\pi)^3 }\sum_{s_{\rm e}, s_{\rm p}}
\psi({\bf p}, s_{\rm e}, s_{\rm p}) \, a^{\dagger}_{{\bf p}, s_{\rm e}}
b^{\dagger}_{-{\bf p}, s_{\rm p}}
\]
act on the vacuum state. Here, $\psi$ is the Schr\"{o}dinger
wavefunction obtained from solving the nonrelativistic
Schr\"{o}dinger equation in a Coulomb potential. The electron spin is
$s_{\rm e}$ and the positron spin is $s_{\rm p}$, and we are
working in the
bound state CM system, hence the opposite momenta ${\bf p}$ and $-{\bf p}$.
\ \\ \ \\
To find the parity of the system, we compute the action
of parity on the bound state operator $B$, using equations
(\ref{eq:Uona}) and (\ref{eq:Uonb}) from the section on
intrinsic parity in \ref{sec:pinrepq}:
\bea
U_P B \, U_P^{-1}&=&\int {d^3 {\bf p} \over (2\pi)^3}
 \sum_{s_{\rm e}, s_{\rm p}}
\psi({\bf p}, s_{\rm e}, s_{\rm p}) \eta^*_a
a^{\dagger}_{-{\bf p}, s_{\rm e}}
\eta_b^* b^{\dagger}_{{\bf p}, s_{\rm p}} \\
&=&(\eta_a \eta_b)^* \int {d^3 {\bf p} \over (2\pi)^3}
\sum_{s_{\rm e}, s_{\rm p}}
\psi(-{\bf p}, s_{\rm e}, s_{\rm p}) a^{\dagger}_{{\bf p}, s_{\rm e}}
b^{\dagger}_{-{\bf p}, s_{\rm p}}\\
&=&(\eta_a \eta_b) \int {d^3 {\bf p} \over (2\pi)^3}
\sum_{s_{\rm e}, s_{\rm p}}
(-1)^{\ell}\psi({\bf p}, s_{\rm e}, s_{\rm p})
a^{\dagger}_{{\bf p}, s_{\rm e}}
b^{\dagger}_{-{\bf p}, s_{\rm p}} \\
&=& \eta_a \eta_b (-1)^{\ell} B
\eea
How can we measure this phase $\eta_a \eta_b (-1)^{\ell}$?
The amplitude for annihilation into two photons is
\[
{\mathcal M}(B \rightarrow 2\gamma) =
\int {d^3 {\bf p} \over (2\pi)^3} \sum_{s_{\rm e}, s_{\rm p}}
\psi({\bf p}, s_{\rm e}, s_{\rm p}) \,
{\mathcal M}({\bf p}, s_{\rm e}, -{\bf p}, s_{\rm p}
\rightarrow 2\gamma)
\]
where the matrix element ${\mathcal M}({\bf p}, s_{\rm e},
-{\bf p}, s_{\rm p}
\rightarrow 2\gamma)$ is the ordinary field theory amplitude for a
free electron and positron of momenta ${\bf p}$ and $-{\bf p}$ and
spins $s_{\rm e}$, $s_{\rm p}$.
\ \\ \ \\
Nonrelativistically, we may think of this matrix element as being
composed of a wavefunction overlap integral
$\int \psi_B^* \psi_{2\gamma} \, d^3 x$, and we can draw conclusions
about $\psi_{2\gamma}$ from the photon part of the tree-level QED
amplitude. In particular, since each photon vertex introduces a
(transverse) polarization vector $e^{\mu}$, and the only other
vector available is one photon momentum ${\bf k}$, we can only form
the following scalar or pseudoscalar combinations:
\bea
\psi^+_{2\gamma} \; &\propto& \; {\bf e}_1 \cdot {\bf e}_2 \\
\psi^-_{2\gamma} \; &\propto& \; {\bf k} \cdot
({\bf e}_1 \times {\bf e}_2)
\eea
If we denote by $\phi$
the angle between ${\bf e}_1$ and ${\bf e}_2$,
it is clear that the probability of the photons coming out polarized
at $\phi=90^{\circ}$ is zero in the first (even) case and nonzero in the
second (odd) case. Wu and Shaknov \cite{WuShaknov} performed the experiment
with a ${}^{64}{\rm Cu}$ positron source and found
\[
{ {\rm rate}(\phi = 90^{\circ}) \over {\rm rate}(\phi = 0^{\circ})} =
2.04 \pm 0.08 \; .
\]
If we write the wavefunction of the two-photon state as a product
of spatial and spin wavefunctions
$\psi({\rm space}) \psi({\rm spin})$, 
the spatial part is odd under
inversion. Since parity is conserved in QED, this means that
experiment dictates that for $s$-wave positronium (which is
predominantly the case),
\[
\eta_a \eta_b   = -1 \; .
\]
Thus the theoretical result (\ref{eq:etaconj}) is on firm experimental ground.
\ \\
\newpage

\begin{center}
{\it Cross sections of} $\; \Sigma^0 \rightarrow \Lambda^0 +
e^+ + e^-$  (p.\ \pageref{page:sigma})
\end{center}
We square and sum 
the given matrix elements
over polarizations, which introduces traces over the
gamma matrices using the spin sums from
section \ref{sec:observables}. \\
$\Sigma$-$\Lambda$ trace in $|{\mathcal M}_+|^2$:
\[
[{\rm tr}(\sla q \, \Gamma_{\mu\nu} \sla p \, \Gamma_{\rho \sigma})
+M_{\Sigma}M_{\Lambda}{\rm tr}(\Gamma_{\mu\nu}\Gamma_{\rho
\sigma})] k^{\mu} k^{\sigma}
\]
$\Sigma$-$\Lambda$ trace in $|{\mathcal M}_-|^2$:
\bea
-&[&{\rm tr}(\sla q \, \Gamma_5 \Gamma_{\mu\nu} \sla p \, \Gamma_5
\Gamma_{\rho \sigma})
+M_{\Sigma}M_{\Lambda}{\rm tr}(\Gamma_5 \Gamma_{\mu\nu} \Gamma_5
\Gamma_{\rho
\sigma})] k^{\mu} k^{\sigma} = \\ [1mm]
&\lbrack& {\rm tr}(\sla q \, \Gamma_{\mu\nu} \sla p \,
\Gamma_{\rho \sigma})
-M_{\Sigma}M_{\Lambda}{\rm tr}(\Gamma_{\mu\nu}
\Gamma_{\rho
\sigma})] k^{\mu} k^{\sigma}
\eea
(both of these are to be contracted with the electron-positron trace).
We now see where the difference in the prediction comes in: a sign
change in the $M_{\Sigma} M_{\Lambda}$ term. We can
study the decay rate as a function of the invariant mass of the
electron-positron pair, and compare it to experiment. The
Steinberger experiment yields a curve which to good accuracy
agrees with the hypothesis $\Sigma^0$-parity $+1$.
\ \\ \ \\
Now for the main question: would this prediction change if we
allowed for four different hypotheses $(\pm 1, \pm i)$ for the $\Sigma^0$
parity? That is, what if $\Sigma^0$ transformed under $\pin(3,1)$
instead of the previously assumed $\pin(1,3)$?
\ \\ \ \\
For all particles in \pin(1,3), using the rules from
section \ref{sec:observables}, we compute the squared matrix element:
\begin{eqnarray*}
{\scriptstyle {1 \over 4}}\sum_{s s' r r'} |{\mathcal M}_+|^2 &\;
\propto \; &
\left[ {\rm tr}\left(
\sla q \, \Gamma_{\mu\nu}
\sla p \, \Gamma_{\rho \sigma}\right)+M_{\Sigma}M_{\Lambda}{\rm
tr}(\Gamma_{\mu\nu}\Gamma_{\rho
\sigma})\right] \times \\
&& k^{\mu} k^{\sigma}   \left[\tr \left(
\sla k_1\Gamma^{\nu}
\sla k_2 \Gamma^{\rho}\right)-
m_e^2 \, \tr \left(
\Gamma^{\nu}\Gamma^{\rho}\right)\right]
\end{eqnarray*}
\begin{eqnarray*}
{\scriptstyle {1 \over 4}}\sum_{s s' r r'} |{\mathcal M}_-|^2
&\; \propto \; &
\left[{\rm tr}\left(
\sla q \, \Gamma_{\mu\nu}
\sla p \, \Gamma_{\rho \sigma}\right)-M_{\Sigma}M_{\Lambda}{\rm
tr}(\Gamma_{\mu\nu} \Gamma_{\rho \sigma})\right] \times \\
&& k^{\mu} k^{\sigma}   \left[\tr \left(
\sla k_1\Gamma^{\nu}
\sla k_2 \Gamma^{\rho}\right)-
m_e^2 \, \tr \left(
\Gamma^{\nu}\Gamma^{\rho}\right)\right]
\end{eqnarray*}
Similarly, we compute for \pin(3,1):
\begin{eqnarray*}
{\scriptstyle {1 \over 4}}\sum_{s s' r r'} |\hat{{\mathcal M}}_+|^2
&\; \propto \; &
\left[ {\rm tr}\left(
\sla q \, \hat{\Gamma}_{\mu\nu}
\sla p \, \hat{\Gamma}_{\rho \sigma}\right) -
M_{\Sigma}M_{\Lambda}{\rm tr}(\hat{\Gamma}_{\mu\nu}
\hat{\Gamma}_{\rho \sigma})\right] \times \\
&& k^{\mu} k^{\sigma}   \left[ -\tr \left(
\sla k_1\hat{\Gamma}^{\nu}
\sla k_2 \hat{\Gamma}^{\rho}\right)-
m_e^2 \, {\rm tr}\left(
\hat{\Gamma}^{\nu}\hat{\Gamma}^{\rho}\right)\right] \\
&& \; \propto \;
{\scriptstyle {1 \over 4}}\sum_{s s' r r'} |{\mathcal M}_+|^2
\end{eqnarray*}
\begin{eqnarray*}
{\scriptstyle {1 \over 4}}\sum_{s s' r r'} |\hat{{\mathcal M}}_-|^2
&\; \propto \; &
\left[ {\rm tr}\left(
\sla q \, \hat{\Gamma}_{\mu\nu}
\sla p \, \hat{\Gamma}_{\rho \sigma}\right)
+ M_{\Sigma}M_{\Lambda}{\rm tr}(\hat{\Gamma}_{\mu\nu}\hat{\Gamma}_{\rho
\sigma})\right] \times \\
&& k^{\mu} k^{\sigma}   \left[-\tr \left(
\sla k_1\hat{\Gamma}^{\nu}
\sla k_2 \hat{\Gamma}^{\rho}\right)-
m_e^2 \, {\rm tr}\left(
\hat{\Gamma}^{\nu}\hat{\Gamma}^{\rho}\right)\right] \\
&& \; \propto \;
{\scriptstyle {1 \over 4}}\sum_{s s' r r'} |{\mathcal M}_-|^2
\end{eqnarray*}
where the constants of proportionality are everywhere the same.
Thus we have shown how the difference in decay rates
between $\pin(1,3)$ and
$\pin(3,1)$ particles disappears in this calculation.
\ \\ 

\begin{center}
{\it Positronium} (p.\ \pageref{page:posit2})
\end{center}
We recall from section \ref{sec:parityexp}
that the bound state operator is
\[
B=\int {d^3 {\bf p} \over (2\pi)^3 }\sum_{s_{\rm e}, s_{\rm p}}
\psi({\bf p}, s_{\rm e}, s_{\rm p}) \, a^{\dagger}_{{\bf p}, s_{\rm e}}
b^{\dagger}_{-{\bf p}, s_{\rm p}}
\]
so the action of $U_C$ on the bound state operator is
\bea
U_C B \, U_C^{-1}&=&\int {d^3 {\bf p} \over (2\pi)^3}
\sum_{s_{\rm e}, s_{\rm p}}
\psi({\bf p}, s_{\rm e}, s_{\rm p}) \xi^*_a
b^{\dagger}_{{\bf p}, s_{\rm e}}
\xi_b^* a^{\dagger}_{-{\bf p}, s_{\rm p}} \\
&=&-(\xi_a \xi_b)^* \int {d^3 {\bf p} \over (2\pi)^3} \sum_{s_{\rm e},
s_{\rm p}}
\psi({\bf p}, s_{\rm e}, s_{\rm p}) a^{\dagger}_{-{\bf p}, s_{\rm p}}
b^{\dagger}_{{\bf p}, s_{\rm e} }\\
&=&(\xi_a \xi_b) \int {d^3 {\bf p} \over (2\pi)^3}
\sum_{s_{\rm e}, s_{\rm p}}
(-1)^{\ell+s+1}\psi({\bf p}, s_{\rm e}, s_{\rm p})
a^{\dagger}_{{\bf p}, s_{\rm e}}
b^{\dagger}_{-{\bf p}, s_{\rm p}} \\
&=& \xi_a \xi_b \, (-1)^{\ell+s+1} B
\eea
We see that since $C$-parity, unlike $P$-parity, depends on the total
positronium spin $s$, there will be different selection rules for the
two spin states $s=0$ (known as ``para''-positronium) and $s=1$
(``ortho''-positronium).
\ \\ \ \\
To obtain these selection rules, we consider the $C$-parity of the
final state of photons. Two photons have even $C$-parity whereas three
photons have odd parity.
\bea
&& \mbox{$C$-parity of two photons: $(-1)^2=+1$} \\
&& \mbox{$C$-parity of three photons: $(-1)^3=-1$}
\eea
Consider $\ell=0$. Decay to three photons is suppressed by a factor of
order $\alpha$ (the fine-structure constant) compared to the amplitude
for two photons, and we find
\begin{center}
\begin{tabular}{|l|c|l|c|}\hline
Spin state & $C$-parity & exp. half-life & indicates decay mode \\ \hline\hline
Para ($s=0$) & $+\xi_a \xi_b $ & $\simeq 10^{-10}$ \, {\rm s} &
2 photons \\ \hline
Ortho ($s=1$) & $-\xi_a \xi_b $ & $\simeq 10^{-7}$ \, {\rm s} &
3 photons \\ \hline
\end{tabular}
\end{center}
Thus we see agreement with experiment provided $\xi_a \xi_b=1$, which
verifies equation (\ref{eq:xiconj}) of section \ref{sec:pinrepq}.
\ \\

\newpage

\section{Collected references}
\label{app:ref}

So that the reader does not have to look up many different
references, we have
used --- when possible --- three basic references; {\em Analysis, Manifolds
and Physics Part I: Basics} and {\em Part II: 92 Applications} by
Y. Choquet-Bruhat and C. DeWitt-Morette \cite{Yellow,Blue},
and {\em Introduction to Quantum Field Theory} by M.E. Peskin and
D.V. Schroeder \cite{Peskin}.
The word \pin\ appears only in
{\em Part II: 92 Applications}
(8 entries in the index). Needless to say, the
word \spin\ appears in all three references.
Note that in this report
we use the Pauli matrices commonly used in physics (see
section \ref{sec:notation}), not the ones used in the first two
references \cite{Yellow,Blue}).
\ \\ \ \\
We would also like to mention some related work that was not
directly used for this report, but which may be of interest
to the reader depending on his or her specific interest in the \pin\
groups.

\begin{itemize}
\item General discussion of the mathematics of the two \pin\ groups
and/or $C\!PT$
\cite{Coque1,Coque2,Gursey,Houtappel,Jost2,Michel,Roman,%
Sternberg2,Sternberg3,Umezawa,Wick2,Wigner2}
\item Superselection rules
\cite{Streater,Streater2,Strocchi,Wightman} (see also
\cite{Aharanov,Aharanov2,Badurek,Wick})
\item Pin structures
\cite{Alty,Chamblin2,Chamblin3,Yellowref,Hawking}
\item Spinors on non-trivial manifolds
\cite{Barut,Friedman,Grinstein,Isham}
\item Solution of the Dirac equation in electromagnetic fields
\cite{Barut3,Barut2}
\item $C\!PT$ theorem
\cite{Dyson,Jost,Luders,Zumino,Pauli,Stapp}
\item Carruther's theorem
\cite{Carruthers1,Carruthers2,Fabri,Lee3,Moussa,MoussaStora,%
MoussaStora2,Steinmann,ZuminoZwanziger}
\item $C\!PT$ and cosmology
\cite{Sakharov1,Sakharov2,Sakharov3}
\item Solar neutrinos
\cite{Bahcall}
\item Interference and $C\!PT$ in neutron physics
\cite{Werner}
\item Majorana neutrinos and double lepton decay
\cite{Dassie,Kayser,Littenberg,Majorana,Tomoda}
\item Non-trivial manifolds in condensed matter physics
\cite{Mermin,Michel2,Toulouse}
\item Phase factor observation
\cite{Bernstein,Rauch,Werner2}
\item Space and time reversal in atomic and molecular processes
\cite{Bouchiat,Bouchiat2,Macpherson,Meekhof}
\end{itemize}

\end{document}